\documentclass{aastex631}
\newcommand{\lhaaso}{LHAASO J0621+3755}
\newcommand{\lhaasocat}{1LHAASO J0622+3754}
\newcommand{\psr}{PSR J0622+3749}
\newcommand{\fermi}{\textit{Fermi}-LAT}
\newcommand{\fgl}{4FGL J0622.2+3749}
\newcommand{\xmm}{\textit{XMM-Newton}}
\newcommand{\nustar}{\textit{NuSTAR}}
\newcommand{\degree}{^{\circ}}

\def\simgt{\lower.5ex\hbox{$\; \buildrel > \over \sim \;$}}
\def\simlt{\lower.5ex\hbox{$\; \buildrel < \over \sim \;$}}
\usepackage{soul}
\graphicspath{{./}{figures/}{figures/veritas/}{figures/model/}{figures/SpecAnalysis/}{figures/timing/}}
\begin{document}

\title{Multiwavelength observation of a candidate pulsar halo \lhaaso{} \\ and the first X-ray detection of \psr{}}

\author[0000-0002-9021-6192]{C.~B.~Adams}\affiliation{Physics Department, Columbia University, New York, NY 10027, USA}
\author{A.~Archer}\affiliation{Department of Physics and Astronomy, DePauw University, Greencastle, IN 46135-0037, USA}
\author[0000-0002-3886-3739]{P.~Bangale}\affiliation{Department of Physics and Astronomy and the Bartol Research Institute, University of Delaware, Newark, DE 19716, USA}
\author[0000-0002-9675-7328]{J.~T.~Bartkoske}\affiliation{Department of Physics and Astronomy, University of Utah, Salt Lake City, UT 84112, USA}
\author[0000-0003-2098-170X]{W.~Benbow}\affiliation{Center for Astrophysics $|$ Harvard \& Smithsonian, Cambridge, MA 02138, USA}
\author[0000-0001-6391-9661]{J.~H.~Buckley}\affiliation{Department of Physics, Washington University, St. Louis, MO 63130, USA}
\author[0009-0001-5719-936X]{Y.~Chen}\affiliation{Department of Physics and Astronomy, University of California, Los Angeles, CA 90095, USA}
\author{J.~L.~Christiansen}\affiliation{Physics Department, California Polytechnic State University, San Luis Obispo, CA 94307, USA}
\author{A.~J.~Chromey}\affiliation{Center for Astrophysics $|$ Harvard \& Smithsonian, Cambridge, MA 02138, USA}
\author[0000-0003-1716-4119]{A.~Duerr}\affiliation{Department of Physics and Astronomy, University of Utah, Salt Lake City, UT 84112, USA}
\author[0000-0002-1853-863X]{M.~Errando}\affiliation{Department of Physics, Washington University, St. Louis, MO 63130, USA}
\author{M.~Escobar~Godoy}\affiliation{Santa Cruz Institute for Particle Physics and Department of Physics, University of California, Santa Cruz, CA 95064, USA}
\author[0000-0002-5068-7344]{A.~Falcone}\affiliation{Department of Astronomy and Astrophysics, 525 Davey Lab, Pennsylvania State University, University Park, PA 16802, USA}
\author{S.~Feldman}\affiliation{Department of Physics and Astronomy, University of California, Los Angeles, CA 90095, USA}
\author[0000-0001-6674-4238]{Q.~Feng}\affiliation{Department of Physics and Astronomy, University of Utah, Salt Lake City, UT 84112, USA}
\author[0000-0002-1067-8558]{L.~Fortson}\affiliation{School of Physics and Astronomy, University of Minnesota, Minneapolis, MN 55455, USA}
\author[0000-0003-1614-1273]{A.~Furniss}\affiliation{Santa Cruz Institute for Particle Physics and Department of Physics, University of California, Santa Cruz, CA 95064, USA}
\author[0000-0002-0109-4737]{W.~Hanlon}\affiliation{Center for Astrophysics $|$ Harvard \& Smithsonian, Cambridge, MA 02138, USA}
\author[0000-0003-3878-1677]{O.~Hervet}\affiliation{Santa Cruz Institute for Particle Physics and Department of Physics, University of California, Santa Cruz, CA 95064, USA}
\author[0000-0001-6951-2299]{C.~E.~Hinrichs}\affiliation{Center for Astrophysics $|$ Harvard \& Smithsonian, Cambridge, MA 02138, USA and Department of Physics and Astronomy, Dartmouth College, 6127 Wilder Laboratory, Hanover, NH 03755 USA}
\author[0000-0002-6833-0474]{J.~Holder}\affiliation{Department of Physics and Astronomy and the Bartol Research Institute, University of Delaware, Newark, DE 19716, USA}
\author[0000-0002-1432-7771]{T.~B.~Humensky}\affiliation{Department of Physics, University of Maryland, College Park, MD, USA and NASA GSFC, Greenbelt, MD 20771, USA}
\author[0000-0002-1089-1754]{W.~Jin}\affiliation{Department of Physics and Astronomy, University of California, Los Angeles, CA 90095, USA}
\author[0009-0008-2688-0815]{M.~N.~Johnson}\affiliation{Santa Cruz Institute for Particle Physics and Department of Physics, University of California, Santa Cruz, CA 95064, USA}
\author[0000-0002-3638-0637]{P.~Kaaret}\affiliation{Department of Physics and Astronomy, University of Iowa, Van Allen Hall, Iowa City, IA 52242, USA}
\author{M.~Kertzman}\affiliation{Department of Physics and Astronomy, DePauw University, Greencastle, IN 46135-0037, USA}
\author{M.~Kherlakian}\affiliation{Fakult\"at f\"ur Physik \& Astronomie, Ruhr-Universit\"at Bochum, D-44780 Bochum, Germany}
\author[0000-0003-4785-0101]{D.~Kieda}\affiliation{Department of Physics and Astronomy, University of Utah, Salt Lake City, UT 84112, USA}
\author[0000-0002-4260-9186]{T.~K.~Kleiner}\affiliation{DESY, Platanenallee 6, 15738 Zeuthen, Germany}
\author[0000-0002-4289-7106]{N.~Korzoun}\affiliation{Department of Physics and Astronomy and the Bartol Research Institute, University of Delaware, Newark, DE 19716, USA}
\author{F.~Krennrich}\affiliation{Department of Physics and Astronomy, Iowa State University, Ames, IA 50011, USA}
\author[0000-0002-5167-1221]{S.~Kumar}\affiliation{Department of Physics, University of Maryland, College Park, MD, USA }
\author{S.~Kundu}\affiliation{Department of Physics and Astronomy, University of Alabama, Tuscaloosa, AL 35487, USA}
\author[0000-0003-4641-4201]{M.~J.~Lang}\affiliation{School of Natural Sciences, University of Galway, University Road, Galway, H91 TK33, Ireland}
\author[0000-0003-3802-1619]{M.~Lundy}\affiliation{Physics Department, McGill University, Montreal, QC H3A 2T8, Canada}
\author[0000-0001-9868-4700]{G.~Maier}\affiliation{DESY, Platanenallee 6, 15738 Zeuthen, Germany}
\author[0000-0001-7106-8502]{M.~J.~Millard}\affiliation{Department of Physics and Astronomy, University of Iowa, Van Allen Hall, Iowa City, IA 52242, USA}
\author{J.~Millis}\affiliation{Department of Physics and Astronomy, Ball State University, Muncie, IN 47306, USA and Department of Physics, Anderson University, 1100 East 5th Street, Anderson, IN 46012}
\author[0000-0001-5937-446X]{C.~L.~Mooney}\affiliation{Department of Physics and Astronomy and the Bartol Research Institute, University of Delaware, Newark, DE 19716, USA}
\author[0000-0002-1499-2667]{P.~Moriarty}\affiliation{School of Natural Sciences, University of Galway, University Road, Galway, H91 TK33, Ireland}
\author[0000-0002-3223-0754]{R.~Mukherjee}\affiliation{Department of Physics and Astronomy, Barnard College, Columbia University, NY 10027, USA}
\author[0000-0002-6121-3443]{W.~Ning}\affiliation{Department of Physics and Astronomy, University of California, Los Angeles, CA 90095, USA}
\author[0000-0002-4837-5253]{R.~A.~Ong}\affiliation{Department of Physics and Astronomy, University of California, Los Angeles, CA 90095, USA}
\author[0000-0003-3820-0887]{A.~Pandey}\affiliation{Department of Physics and Astronomy, University of Utah, Salt Lake City, UT 84112, USA}
\author[0000-0001-7861-1707]{M.~Pohl}\affiliation{Institute of Physics and Astronomy, University of Potsdam, 14476 Potsdam-Golm, Germany and DESY, Platanenallee 6, 15738 Zeuthen, Germany}
\author[0000-0002-0529-1973]{E.~Pueschel}\affiliation{Fakult\"at f\"ur Physik \& Astronomie, Ruhr-Universit\"at Bochum, D-44780 Bochum, Germany}
\author[0000-0002-4855-2694]{J.~Quinn}\affiliation{School of Physics, University College Dublin, Belfield, Dublin 4, Ireland}
\author{P.~L.~Rabinowitz}\affiliation{Department of Physics, Washington University, St. Louis, MO 63130, USA}
\author[0000-0002-5351-3323]{K.~Ragan}\affiliation{Physics Department, McGill University, Montreal, QC H3A 2T8, Canada}
\author{P.~T.~Reynolds}\affiliation{Department of Physical Sciences, Munster Technological University, Bishopstown, Cork, T12 P928, Ireland}
\author[0000-0002-7523-7366]{D.~Ribeiro}\affiliation{School of Physics and Astronomy, University of Minnesota, Minneapolis, MN 55455, USA}
\author{L.~Rizk}\affiliation{Physics Department, McGill University, Montreal, QC H3A 2T8, Canada}
\author{E.~Roache}\affiliation{Center for Astrophysics $|$ Harvard \& Smithsonian, Cambridge, MA 02138, USA}
\author[0000-0003-1387-8915]{I.~Sadeh}\affiliation{DESY, Platanenallee 6, 15738 Zeuthen, Germany}
\author[0000-0002-3171-5039]{L.~Saha}\affiliation{Center for Astrophysics $|$ Harvard \& Smithsonian, Cambridge, MA 02138, USA}
\author{G.~H.~Sembroski}\affiliation{Department of Physics and Astronomy, Purdue University, West Lafayette, IN 47907, USA}
\author[0000-0002-9856-989X]{R.~Shang}\affiliation{Department of Physics and Astronomy, Barnard College, Columbia University, NY 10027, USA}
\author[0000-0003-3407-9936]{M.~Splettstoesser}\affiliation{Santa Cruz Institute for Particle Physics and Department of Physics, University of California, Santa Cruz, CA 95064, USA}
\author[0000-0002-9852-2469]{D.~Tak}\affiliation{SNU Astronomy Research Center, Seoul National University, Seoul 08826, Republic of Korea.}
\author{A.~K.~Talluri}\affiliation{School of Physics and Astronomy, University of Minnesota, Minneapolis, MN 55455, USA}
\author{J.~V.~Tucci}\affiliation{Department of Physics, Indiana University Indianapolis, Indianapolis, Indiana 46202, USA}
\author[0000-0002-8090-6528]{J.~Valverde}\affiliation{Department of Physics, University of Maryland, Baltimore County, Baltimore MD 21250, USA and NASA GSFC, Greenbelt, MD 20771, USA}
\author[0000-0003-2740-9714]{D.~A.~Williams}\affiliation{Santa Cruz Institute for Particle Physics and Department of Physics, University of California, Santa Cruz, CA 95064, USA}
\author[0000-0002-2730-2733]{S.~L.~Wong}\affiliation{Physics Department, McGill University, Montreal, QC H3A 2T8, Canada}
\author[0009-0001-6471-1405]{J.~Woo}\affiliation{Columbia Astrophysics Laboratory, 550 West 120th Street, New York, NY 10027, USA}
\affiliation{Department of Physics, Columbia University, 538 West 120th Street, New York, NY 10027, USA}

\collaboration{65}{(VERITAS collaboration)}

\author[0009-0008-0402-1243]{Jon Kwong}
\affiliation{Department of Physics, Columbia University, 538 West 120th Street, New York, NY 10027, USA}
\author[0000-0002-9709-5389]{Kaya Mori}
\affiliation{Columbia Astrophysics Laboratory, 550 West 120th Street, New York, NY 10027, USA}
\author[0000-0002-3681-145X]{Charles J. Hailey}
\affiliation{Columbia Astrophysics Laboratory, 550 West 120th Street, New York, NY 10027, USA}
\affiliation{Department of Physics, Columbia University, 538 West 120th Street, New York, NY 10027, USA}
\author[0000-0001-6189-7665]{Samar Safi-Harb}
\affiliation{Department of Physics and Astronomy, University of Manitoba, Winnipeg, MB R3T 2N2, Canada}
\author[0000-0002-2967-790X]{Shuo Zhang}
\affiliation{Department of Physics and Astronomy, Michigan State University, East Lansing, MI 48824, USA}
\author[0000-0001-7209-9204]{Naomi Tsuji}
\affiliation{Faculty of Science, Kanagawa University, 3-27-1 Rokukakubashi, Kanagawa-ku, Yokohama-shi, Kanagawa 221-8686, Japan}
\collaboration{6}{(\xmm{} collaboration)}

\author[0000-0003-0268-463X]{Silvia Manconi}
\affiliation{Sorbonne Universit\'e \& Laboratoire de Physique Th\'eorique et Hautes \'Energies (LPTHE),
CNRS, 4 Place Jussieu, Paris, France}
\affiliation{Laboratoire d'Annecy-le-Vieux de
Physique Théorique (LAPTh), CNRS, USMB, F-74940 Annecy, France}
\author[0000-0002-3754-3960]{Fiorenza Donato}
\affiliation{Dipartimento di Fisica, Università di Torino, and INFN, Sezione di Torino, Via P. Giuria 1, Torino, Italy, 
and Theoretical Physics Department, CERN, 1211 Geneva 23, Switzerland}
\author[0000-0003-2759-5625]{Mattia Di Mauro}
\affiliation{INFN, Sezione di Torino, Via P. Giuria 1, Torino, Italy}

\collaboration{3}{}

\correspondingauthor{Jooyun Woo}
\email{jw3855@columbia.edu}
\correspondingauthor{Silvia Manconi}
\email{silvia.manconi@lpthe.jussieu.fr}
\correspondingauthor{Jon Kwong}
\email{jk4768@columbia.edu}
\correspondingauthor{Anne Duerr}
\email{a.duerr@utah.edu}



\begin{abstract}

Pulsar halos are regions around middle-aged pulsars extending out to tens of parsecs. The large extent of the halos and well-defined central cosmic-ray accelerators make this new class of Galactic sources an ideal laboratory for studying cosmic-ray transport. LHAASO J0621+3755 is a candidate pulsar halo associated with the middle-aged gamma-ray pulsar PSR J0622+3749. We observed LHAASO J0621+3755 with VERITAS and \xmm{} in the TeV and X-ray bands, respectively. For this work, we developed a novel background estimation technique for imaging atmospheric Cherenkov telescope observations of such extended sources. No halo emission was detected with VERITAS (0.3--10 TeV) or \xmm{} (2--7 keV) within $1\degree$ and $10'$ around PSR J0622+3749, respectively. Combined with the LHAASO-KM2A and \fermi{} data, VERITAS flux upper limits establish a spectral break at $\sim$1--10 TeV, a unique feature compared with Geminga, the most studied pulsar halo. We model the gamma-ray spectrum and LHAASO-KM2A surface brightness as inverse Compton emission and find suppressed diffusion around the pulsar, similar to Geminga. A smaller diffusion suppression zone and harder electron injection spectrum than Geminga are necessary to reproduce the spectral cutoff. A magnetic field $\leq1\ \mu$G is required by our XMM-Newton observation and synchrotron spectral modeling, consistent with Geminga. Our findings support slower diffusion and lower magnetic field around pulsar halos than the Galactic averages, hinting at magnetohydrodynamic turbulence around pulsars. Additionally, we report the detection of an X-ray point source spatially coincident with PSR J0622+3749, whose periodicity is consistent with the gamma-ray spin period of 333.2 ms. The soft spectrum of this source suggests a thermal origin.

\end{abstract}



\section{Introduction} \label{sec:intro}

Pulsar halos \citep{Giacinti2020:pulsar-halo} are regions around middle-aged (characteristic age $\tau>100$ kyr) pulsars extending out to a few tens of parsecs, typically found as very-high-energy (VHE, E $>100$ GeV) gamma-ray sources. These halos are believed to be the last evolutionary stage of pulsar wind nebulae (PWNe), when relativistic electrons and positrons (collectively referred to as electrons hereafter) escape from the central PWN, diffuse in the interstellar medium and scatter interstellar photons up to TeV energies through inverse Compton (IC) scattering. Since the first pulsar halos around the nearby pulsars Geminga and Monogem were discovered by Milagro Gamma-Ray Observatory \citep{MilagroGalacticSources} and studied in detail by the High-Altitude Water Cherenkov (HAWC) observatory \citep{HAWC_Halos_and_Positron_Flux}, more than a dozen other VHE sources have been categorized as candidate pulsar halos (\citet{3hwc}, \citet{lhaasoj0621}, \citet{j0359}, \citet{j0248}). The majority of these sources are spatially coincident with middle-aged pulsars with spin-down luminosities of the order of $10^{34}$ erg s$^{-1}$. Detecting the lower-energy IC emission of these candidate pulsar halos in the \textit{Fermi} Large Area Telescope (\fermi{}) data has been unsuccessful except for the case of Geminga \citep{DiMauro:2019yvh}. While the electrons within the halos are expected to produce extended, low-surface-brightness synchrotron emission in the interstellar magnetic field, no such emission has been found from radio to X-ray energies (e.g., \citet{Manconi:2024wlq}).

Pulsar halos have been studied extensively as an ideal probe for the cosmic-ray (CR) transport mechanism near energetic CR accelerators. Numerous theoretical CR diffusion models have successfully reproduced the gamma-ray spectra and morphologies of different halos (e.g., \cite{Schroer2023:suppresed-diffusion,Fang:2021qon,Bao:2021hey,DiMauro:2019hwn}). Such models\footnote{An exception is the ballistic diffusion model \citep{Recchia:2021kty}, which explains the observed gamma-ray emission from Geminga, Monogem, and \lhaaso{} without invoking suppressed diffusion.} consistently suggest diffusion coefficients within pulsar halos 2--3 orders of magnitude lower than the Galactic average inferred from the secondary-to-primary CR ratio \citep{bcratio}. The origin of diffusion inhibition within pulsar halos is unknown. Among the proposed origins are the parent supernova remnant's shocks or magnetohydrodynamic turbulence generated by the CRs themselves. Further details on the recent developments in observation and theory of pulsar halos can be found in \citet{amato_halo_review,Fang:2022fof,Liu:2022hqf}.

\lhaaso{} is a candidate pulsar halo detected by the Large High Altitude Air Shower Observatory (LHAASO). The LHAASO Kilometer Square Array (KM2A) detected this extended (2D Gaussian $\sigma=0.40\degree\pm0.07\degree$) VHE source up to 158 TeV \citep{lhaasoj0621}. Its counterpart in the first LHAASO catalog \citep{lhaasocat} is \lhaasocat{}, an extended source also detected by the LHAASO Water Cherenkov Detector Array (WCDA) in the 1--25 TeV energy range (2D Gaussian $\sigma=0.46\degree\pm0.03\degree$). \lhaaso{} is spatially coincident with a Geminga-like (Table \ref{tab:pulsars}) pulsar \psr{} (angular separation $0.11\degree\pm0.12\degree$). \psr{} is a gamma-ray pulsar detected by \fermi{} (\fgl{} in the latest \fermi{} point source catalog \citep{4fgl-dr4}); see Table \ref{tab:pulsars}. The pulsar is detected below 10 GeV during the on-pulse phase, while it is not detected during the off-pulse phase \citep{pulsar-gating}. No multiwavelength counterparts of the pulsar or its halo have been found in the radio, infrared, and X-ray bands \citep{fermipsr,lhaasoj0621,pulsar-gating}. 

Multiwavelength observation plays a key role in understanding the physical nature of pulsar halos. VHE observation using imaging atmospheric Cherenkov telescopes (IACTs) can map the spatial distribution of a halo’s gamma-ray emission with a superior angular resolution ($<0.1\degree$), thereby placing a tight constraint on the diffusion coefficient within the halo. X-ray observation can uniquely determine the magnetic field within a halo, which is one of the key factors characterizing CR diffusion. In this work, we present the VHE observation of \lhaaso{} with the Very Energetic Radiation Imaging Telescope Array System (VERITAS) in \S \ref{sec:veritas} and present a novel background estimation technique for observations of extended sources in \S \ref{app:gammapy}. 
We present the X-ray observation of \lhaaso{} with \xmm{} and report the first X-ray detection of \psr{} in \S \ref{sec:xmm}. We model the observed X-ray and gamma-ray emission as synchrotron and IC emission of energetic electrons injected by \psr{} (\S \ref{sec:modeling}). We discuss the multiwavelength aspect of \lhaaso{} as a pulsar halo and the interpretation of the X-ray spectrum of \psr{} in \S \ref{sec:discussion}. Finally, we summarize our work and propose future observation of \lhaaso{} and other candidate pulsar halos in \S \ref{sec:summary}.

\begin{deluxetable*}{ccccccccccc}
\tablecaption{\psr{}, Geminga, and Monogem pulsar properties. \label{tab:pulsars}}
\tablehead{
 \colhead{Name} & \colhead{$P$} & \colhead{$\dot{P}$} & \colhead{$\tau$} & \colhead{$\dot{E}$} & \colhead{$B_S^{\dagger}$} & \colhead{$d$} \\
 \cline{8-10}
 & \colhead{(ms)} & \colhead{($10^{-14}$ s s$^{-1}$)} & \colhead{(kyr)} & \colhead{($10^{34}$ erg s$^{-1}$)} & \colhead{($10^{12}$ G)} & \colhead{(kpc)}
}
\startdata
\psr{} & 333.2 & 2.542 & 207.7 & 2.713 & 2.945 & 1.6?$^{\ddagger}$ \\
Geminga & 237.1 & 1.097 & 342.4 & 3.250 & 1.632 & $0.25^{+0.23}_{-0.08}$ \\
Monogem & 384.9 & 5.496 & 111.0 & 3.804 & 4.654 & $0.29\pm0.03$ \\
\enddata
    \tablecomments{$^{\dagger}$Minimum magnetic field strength at the neutron star surface, assuming the $\dot{E}$ is converted into magnetic dipole radiation.}
    \tablecomments{$^{\ddagger}$``Pseudo-distance" derived from the phenomenological correlation between pulsar gamma-ray luminosity and distance \citep{fermipsr}. This distance was adopted for the pulsar halo modeling in this work. Distance estimate by parallax, dispersion measure, or other methods is unavailable. We present distance estimates by X-ray spectral analysis in \S \ref{subsec:psr_spec}.}
    \tablecomments{References: \cite{geminga-distance} for the Geminga distance, \cite{Brisken2003:monogem-distance} for the Monogem distance, \cite{3pc} for everything else.}
\end{deluxetable*}

\section{VERITAS analysis} \label{sec:veritas}

VERITAS is a ground-based gamma-ray observatory located in Amado, Arizona, consisting of an array of four imaging atmospheric Cherenkov telescopes sensitive in the 85 GeV to $>$30 TeV energy range. Each telescope is equipped with a 12m optical reflector and 499-pixel photomultiplier tube camera, providing a field of view (FoV) of $3.5\degree$ in diameter and angular resolution of $0.08\degree$ (68\% containment radius) at 1 TeV \citep{veritas}.

VERITAS observed \lhaaso{} in 2022 and 2023 for 40 hours at a mean elevation of $72\degree$. Observations were taken in a ``wobble mode" \citep{Fomin1994:wobble} in which the telescopes were pointed $0.7\degree$ offset from the centroid of \lhaaso{}. We used Eventdisplay v490.2 \citep{Eventdisplay,Eventdisplay_v490p2}, a standard VERITAS data analysis pipeline, for event reconstruction and gamma-hadron separation. For gamma-hadron separation, we applied a cut on the air shower image parameters (mean scaled length and mean scaled width) predetermined for a Crab-like spectrum using boosted decision trees \citep{bdt}. The gamma-like events, as well as full-enclosure instrument response functions (effective area and point spread function), were written into DL3 files for analysis in Gammapy \citep{gammapy,gammapy_v1.1}. The source region was set to a circular region with a radius of $1\degree$ corresponding to the 86\% containment radius of \lhaasocat{}. Such a large source extent, compared with VERITAS's FoV (radius of $1.75\degree$), poses challenges in background estimation. We developed a code to generate a 3D acceptance map for the observations and apply the FoV technique \citep{background} for background estimation. The FoV technique estimates the background of the entire FoV by scaling the acceptance in each energy and spatial bin by the observed count rate in the source-free region. Therefore, accurate telescope acceptance estimation is crucial. The energy and spatial dependence of the telescope acceptance are mainly determined by observing conditions such as elevation and azimuth. The details of acceptance map generation and background estimation using Gammapy are described in \S \ref{app:gammapy}. The analysis was cross-checked with an independent analysis using Eventdisplay v490.2 for event reconstruction and the low-rank perturbation method for background estimation \citep{matrix}.

\begin{figure}
    \centering
    \includegraphics[width=0.48\linewidth]{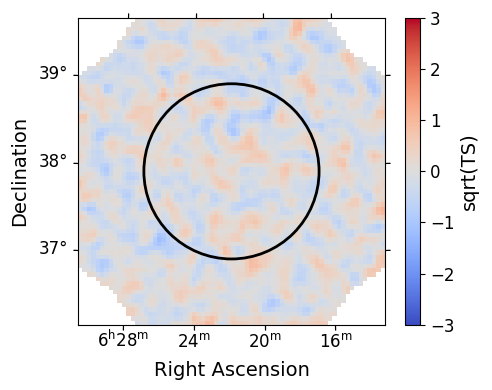}
    \includegraphics[width=0.51\linewidth]{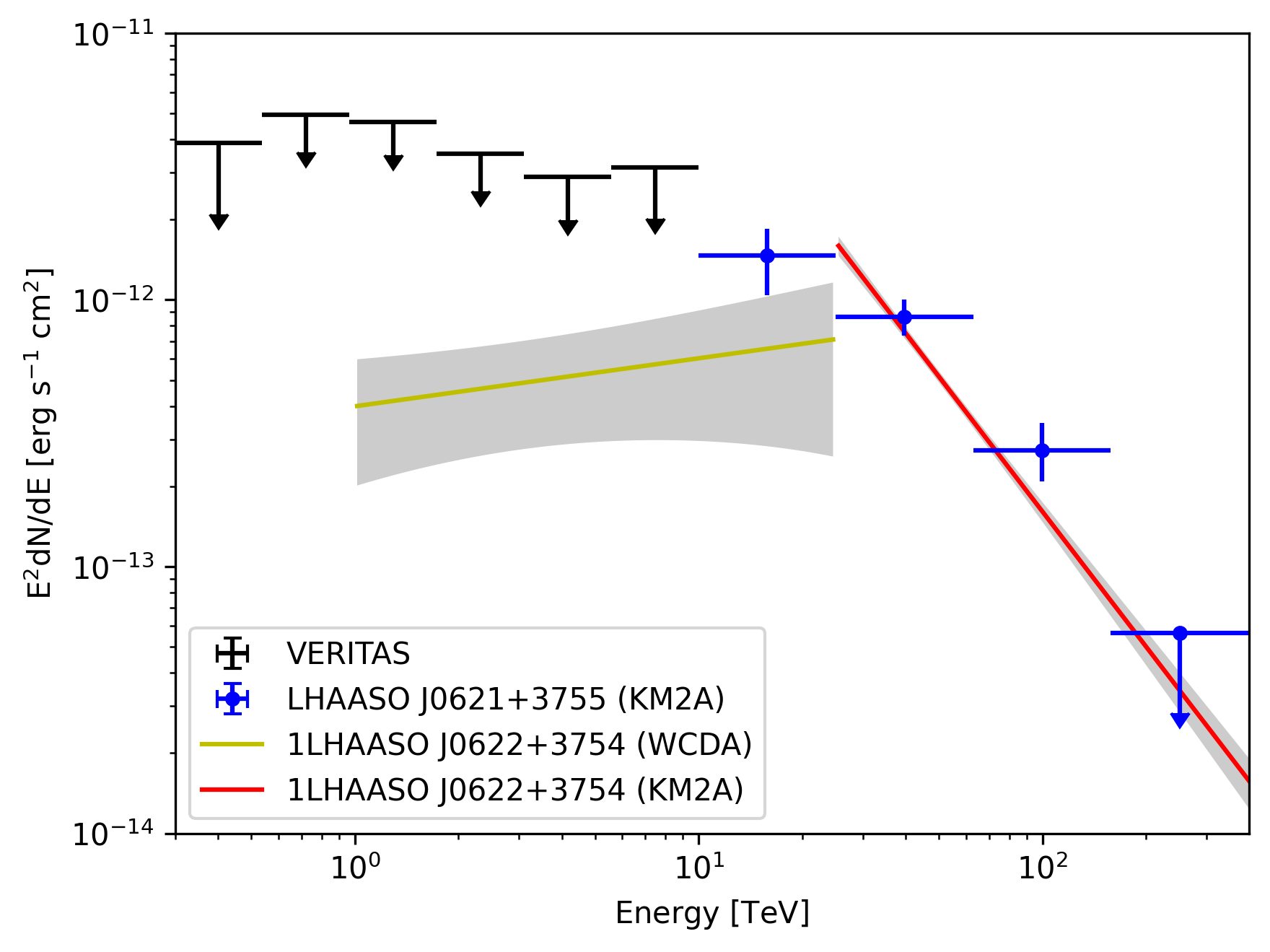}
    \caption{\textbf{\textit{Left}}: A significance map ($3.5\degree\times3.5\degree$) of the region around \lhaaso{} in 0.3--10 TeV. The source region (86\% containment region of \lhaasocat{}, a circular region with a radius of $1\degree$ centered at the centroid of \lhaaso{}) is overlaid with a black solid line. The pixel size is $0.05\degree\times0.05\degree$, and the image was smoothed using a Gaussian kernel ($\sigma=1$ pixel). \textbf{\textit{Right}}: TeV gamma-ray SED of \lhaaso{}. The 95\% VERITAS flux upper limits for a circular region with a radius of $1\degree$ from this work (black), LHAASO KM2A flux points and 95\% flux upper limit from \cite{lhaasoj0621} (blue), LHAASO WCDA and KM2A best-fit power-law spectral models (yellow and red solid lines, respectively) and statistical uncertainties (grey shaded area) for \lhaasocat{} are overlaid. The LHAASO flux points, upper limits, spectral models and uncertainties were scaled to the same region size as the VERITAS source region size for comparability.}
    \label{fig:veritas}
\end{figure}

Figure \ref{fig:veritas} \textit{left} panel shows a significance \footnote{$\sqrt{\textrm{TS}}$ where TS (test statistic) is the Cash statistic \citep{cash} for Poisson-distributed data and modeled background.} map around \lhaaso{} in 0.3--10 TeV. No significant gamma-ray emission was detected within the source region shown by the black solid line in the figure. The significance distributions drawn from the bins within both the entire FoV and the FoV excluding the source region are consistent with a standard normal distribution expected from statistical fluctuations. The total significance of excess counts in 0.3--10 TeV within the source region (a circle with a radius of $1\degree$) is 1$\sigma$. We derived the 95\% flux upper limits in six logarithmic bins in 0.3--10 TeV (bin edges 0.3, 0.5, 1.0, 1.7, 3.1, 5.6 and 10 TeV) as shown in Figure \ref{fig:veritas} \textit{right} panel.\footnote{To scale the LHAASO data for the VERITAS source region size, we calculated the corresponding containment fraction using the 2D Gaussian sigma of the LHAASO source and multiplied it by the LHAASO spectrum.}
%\SM{Our results are compared with available LHAASO measurements, see \ref{sec:discussion} for more discussion.}

\section{\xmm{} analysis} \label{sec:xmm}

\xmm{} is a space-based observatory carrying three coaligned X-ray telescopes and the European Photon Imaging Camera (EPIC) sensitive in the 0.1--15 keV band. The EPIC consists of three CCD cameras (MOS1, MOS2, PN), each providing a FoV $\sim0.5\degree$ in diameter with an angular resolution of $\sim6''$ (FWHM). We observed \lhaaso{} with \xmm{} in March--April 2023 (observation IDs 0923400101, 0923400601, and 0923401501) with a total exposure of 74 ks. The telescope was pointed at \psr{} for all three observations to capture the brightest part of the pulsar halo. We used the Science Analysis System (SAS)\footnote{Due to software errors from the extended source analysis commands, we used SAS v21 up to event cleaning and v20 afterward. For the pulsar analysis in \S \ref{subsec:psr_timing} and \S \ref{subsec:psr_spec}, SAS v21 was used.} for \xmm{} data analysis. We produced event files using the \texttt{emchain} (for MOS) and \texttt{epchain} (for PN) tasks and filtered out the good time intervals affected by soft proton flares using the \texttt{espfilt} task. After filtering, the cleaned event files have a net exposure of 48 ks. Most of observation 0923401501 was affected by strong soft proton flares, and hence, it was not used in this work. 

\begin{figure} 
    \includegraphics[width=\linewidth]{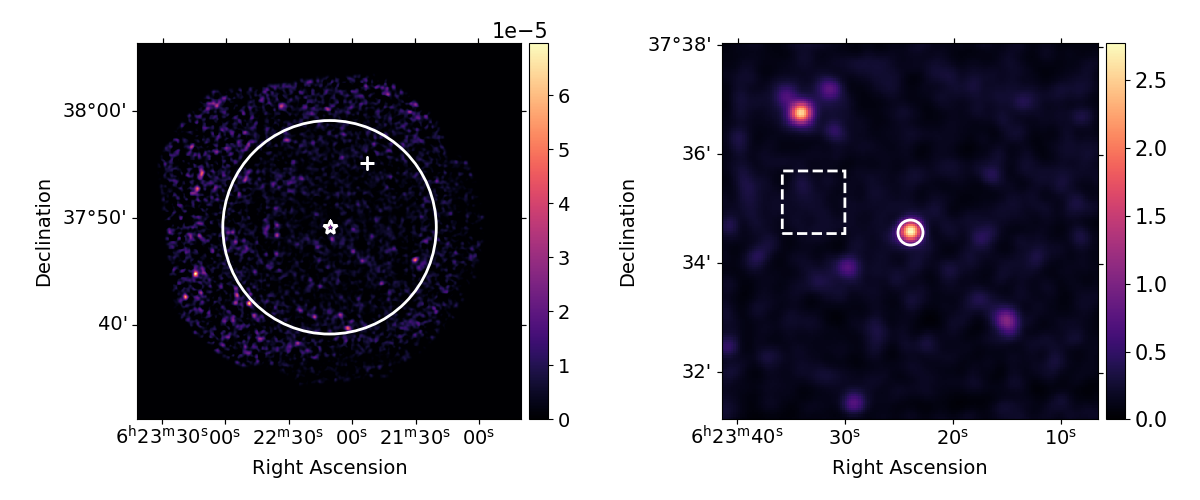}
    \caption{\textbf{\textit{Left}}: Mosaiced, background-subtracted, and exposure-corrected image of the \xmm{} FoV in 2--7 keV. The white circle (radius 10$'$ centered at \psr{}) marks the source region from which the flux upper limit for a putative halo was calculated. The location of \psr{} and the centroid of \lhaaso{} are marked with a white star and a white cross, respectively. \textbf{\textit{Right}}: Observation ID 923400101 MOS2 counts map in 0.2--3 keV around \psr{} ($0.2\degree\times0.2\degree$). The source and background regions for the pulsar spectrum are marked with solid circle and dashed box, respectively.}
\label{fig:xmm-fov}
\end{figure}

We used the \texttt{cheese} task on the cleaned event files to detect point sources and create Swiss-cheese masks to remove them. Among several point sources, we detected \psr{} for the first time in the X-ray band. We focus on the analysis of this central engine of the pulsar halo in \S \ref{subsec:psr_timing} and \S \ref{subsec:psr_spec} and leave the study of other point sources for future work. We generated images and vignetting-corrected exposure maps of the entire FoV using the \texttt{mos-spectra} and \texttt{pn-spectra} tasks. The particle background images were modeled using the filter wheel closed data and the corner chips data by the \texttt{mos\_back} and \texttt{pn\_back} tasks. We set the energy range of our analysis to 2--7 keV to avoid instrumental and solar charge exchange lines \citep{Snowden2004:xmm-swcx,Kuntz2008:xmm-particle-bkg}. We applied the Swiss-cheese masks to the background-subtracted FoV images, mosaiced the masked and background-subtracted images, and divided it by the mosaiced exposure map to create a flux map of the entire FoV. No hint of diffuse emission is present in the FoV as shown in Figure \ref{fig:xmm-fov} \textit{left} panel.

We extracted the source spectra using the \texttt{mos-spectra} and \texttt{pn-spectra} tasks from a circular region with a radius of $10'$ centered at the location of \psr{}, the largest region MOS2 and PN can cover. Since MOS1 can only cover a much smaller region due to the nonoperational CCD chips (CCD 3 and 6), we excluded MOS1 from the spectral analysis. The particle background spectra were modeled using the \texttt{mos\_back} and \texttt{pn\_back} tasks. The Swiss-cheese masks from the \texttt{cheese} task were applied to both the source and background spectra. Thanks to the location of \lhaaso{} off the Galactic Plane and far outside of the Galactic Bulge ($l=175.76\degree$, $b=10.95\degree$), we did not expect significant galactic background emission. We modeled the cosmic X-ray background (CXB) using an absorbed power-law with an index $\Gamma=1.41$ and normalization of 11.6 photons cm$^{-2}$ s$^{-1}$ sr$^{-1}$ keV$^{-1}$ at 1 keV \citep{cxb}, and adopt a hydrogen column density $N_{H}=3.14\times10^{21}$ cm$^{-2}$ (Galactic column density towards \psr{}).\footnote{\url{https://www.swift.ac.uk/analysis/nhtot/}} We used \texttt{tbabs} model in Xspec \citep{xspec} for the absorption model with the \texttt{wilm} abundance table \citep{WilmAbund} for all the X-ray analyses presented in this work. The particle and CXB background components dominate the source spectra in the energy range of our analysis (2--7 keV), leaving no room for putative emission associated with \lhaaso{}. We calculated a $2\sigma$ flux upper limit in 2--7 keV of $5.2\times10^{-14}$ erg cm$^{-2}$ s$^{-1}$, assuming a fully absorbed power-law spectrum with $\Gamma=2$.

\subsection{Pulsar timing analysis} \label{subsec:psr_timing}
We utilized the X-ray timing analysis software {\tt Stingray} and {\tt HENDRICS} \citep{matteo_bachetti_2022_6394742, hendrics} to generate lightcurves and power density spectra (PDSs) of \psr{}. We applied a barycenter correction to photon arrival times for each observation using the {\tt SAS} tool {\tt barycen} before extracting the events from a $r = 24''$ circular region centered at the coordinates of \psr{}.

\begin{figure}
    \centering
    \includegraphics[width=0.395\linewidth]{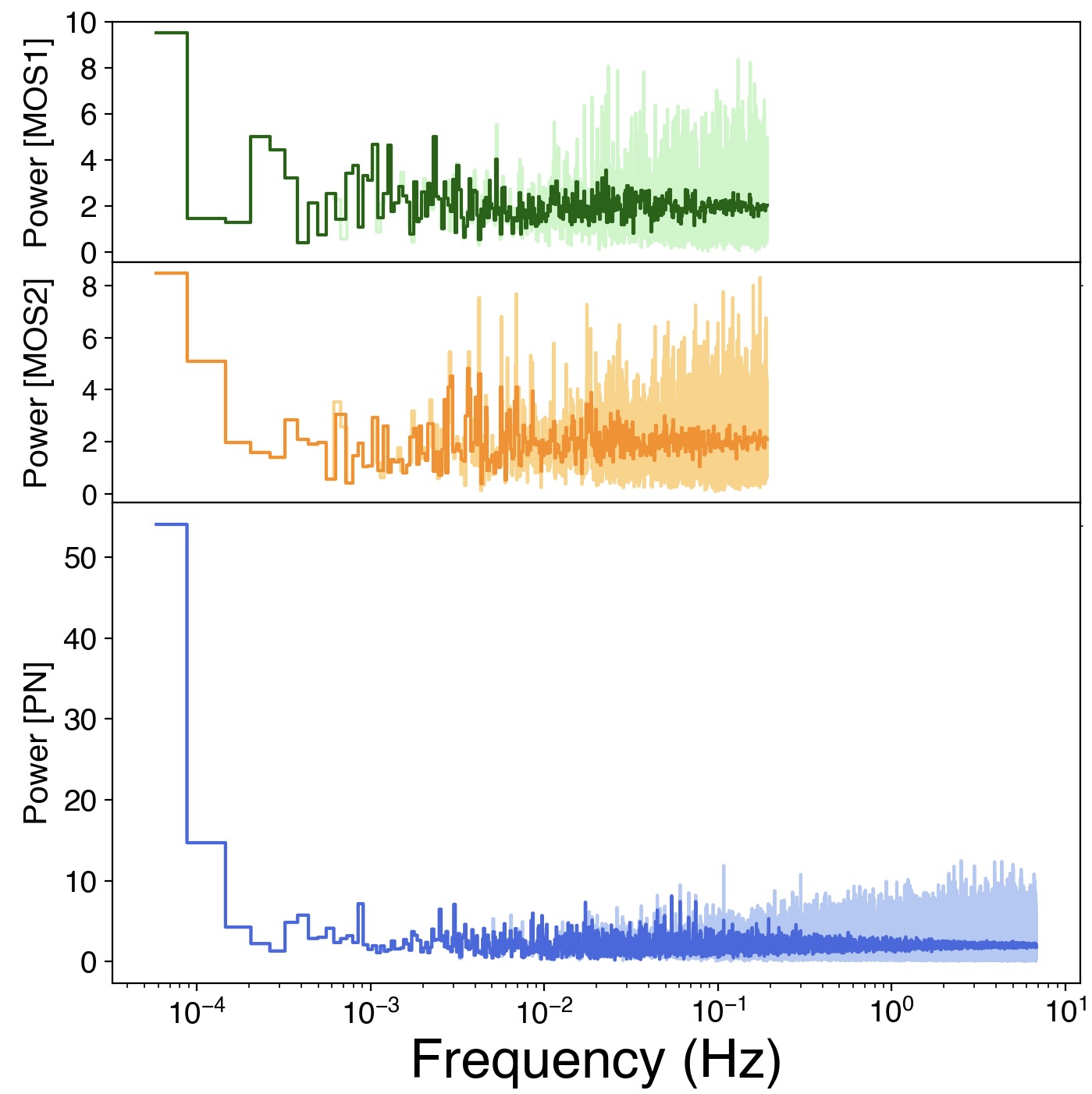}
    \includegraphics[width=0.595\linewidth]{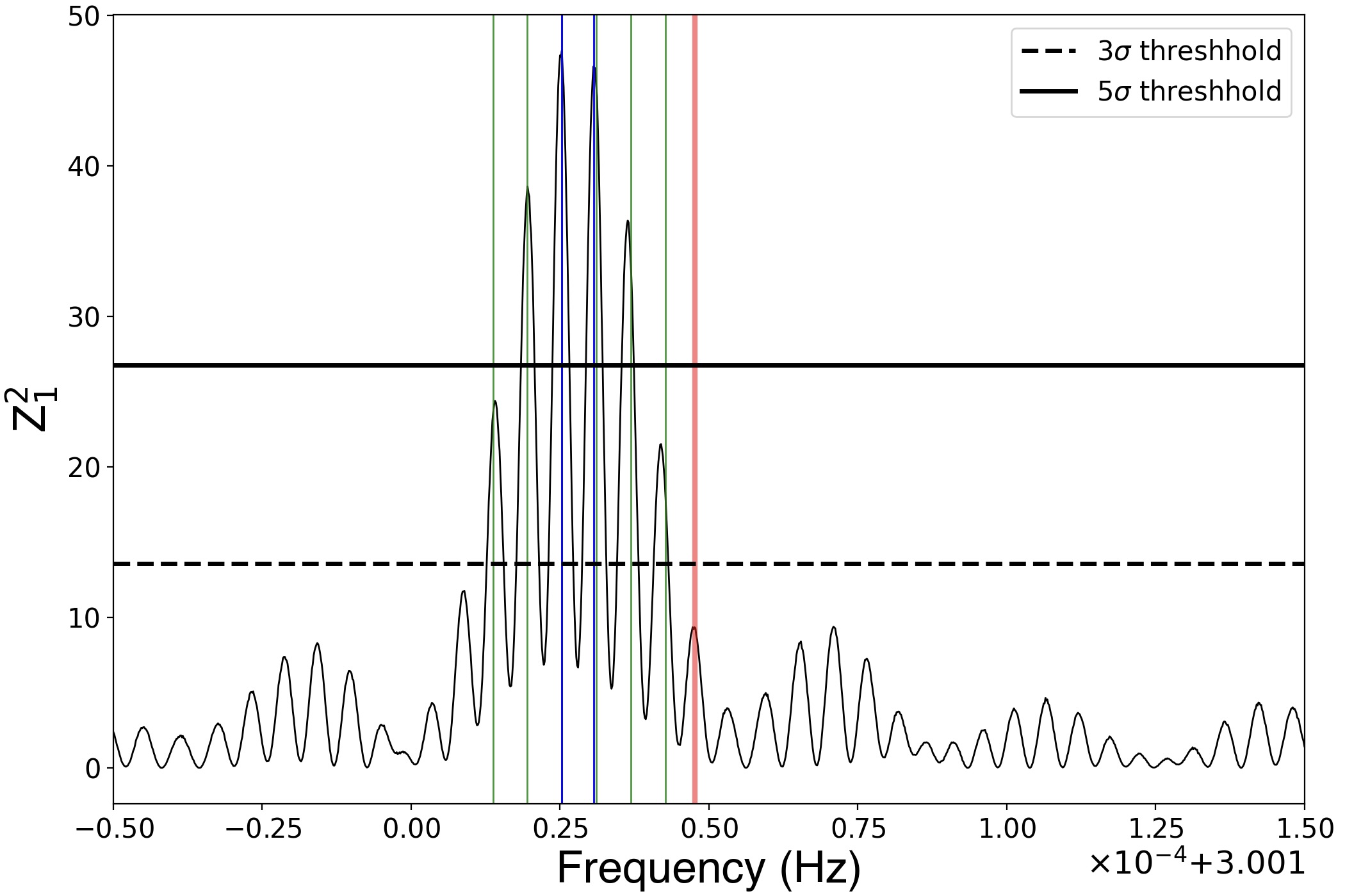}
    \caption{\textbf{\textit{Left}}: The PDS from the MOS1, MOS2, and PN lightcurves (ordered from top to bottom). Both the unbinned (ligher colors) and log-rebinned (darker colors) PDSs have been presented. The PDSs are cut off at $f=0.192$ Hz for MOS1/2 and $f=6.81$ Hz for PN, adhering to the timing resolution of each instrument. \textbf{\textit{Right}}: The $Z^2_1$ periodogram for the combined PN lightcurve. An oversampling factor of 32 was used. The red line is the expected frequency calculated from the gamma-ray ephemeris. The blue lines are the most significant peak (at 3.001025 Hz) and the second most significant peak (at 3.001030 Hz). The green lines represent the aliasing of the most significant peak due to the orbital period of \xmm{}. }
    \label{PDS + Z^2_Test}
\end{figure}

We present the PDS from each of the cameras (\emph{left} panel of Figure \ref{PDS + Z^2_Test}). The MOS1, MOS2, and PN PDSs are each the product of averaging three individual PDSs made from lightcurve segments of length 17060 s. The PDSs are cut off at a maximum frequency of 0.192 Hz for MOS1/2 and 6.81 Hz for PN, determined by the camera's timing resolution ($f_{max} = 1/(2\Delta t)$). All three averaged PDSs lack distinct features and appear to be dominated by red and white noise. 

We proceeded with searching for a periodic signal near the expected spin period of \psr{} from previous observations. Using the $P_{0}$ and $\dot{P}_{0}$ given by the gamma-ray ephemeris \citep{3pc}, we calculated the expected spin period of \psr{} at the time of the observation to be $\sim 0.333217$ s. This corresponds to a frequency of $\sim 3.00105$ Hz, which is well above the maximum resolvable frequency for the MOS1/2 cameras. Thus, we utilized only data from the PN camera and extracted one lightcurve across the two observations to construct a $Z_1^2$ (Z-squared statistics with one degree of freedom; \cite{Buccheri1983:z2test}) periodogram using {\tt Stingray} (\emph{right} panel of Figure \ref{PDS + Z^2_Test}). We searched between 3.00095 and 3.00115 Hz with an oversampling factor of 32 and found the strongest peak to be centered at $f = 3.001025$ Hz ($P_{1} = 0.3332195$ s). However, we also detected numerous weaker but still significant peaks at the alias frequencies $f_{peak}\pm n(f_{\mathrm{xmm}})$, where $n$ admits integer values, and $f_{\mathrm{xmm}}=5.78\times10^{-6}$ Hz is the \xmm{} orbital frequency \citep{Jansen2001:xmm-orbit}. In particular, we found another almost equally significant peak at $f = 3.001030$ Hz $\approx f_{peak}+f_{xmm}$ ($P_{1} = 0.3332189$ s). We also conducted an alternative epoch-folding search within the same frequency range, and the same peaks and aliasing behavior were again observed.

\begin{figure}
    \centering
    \includegraphics[width=0.495\linewidth]{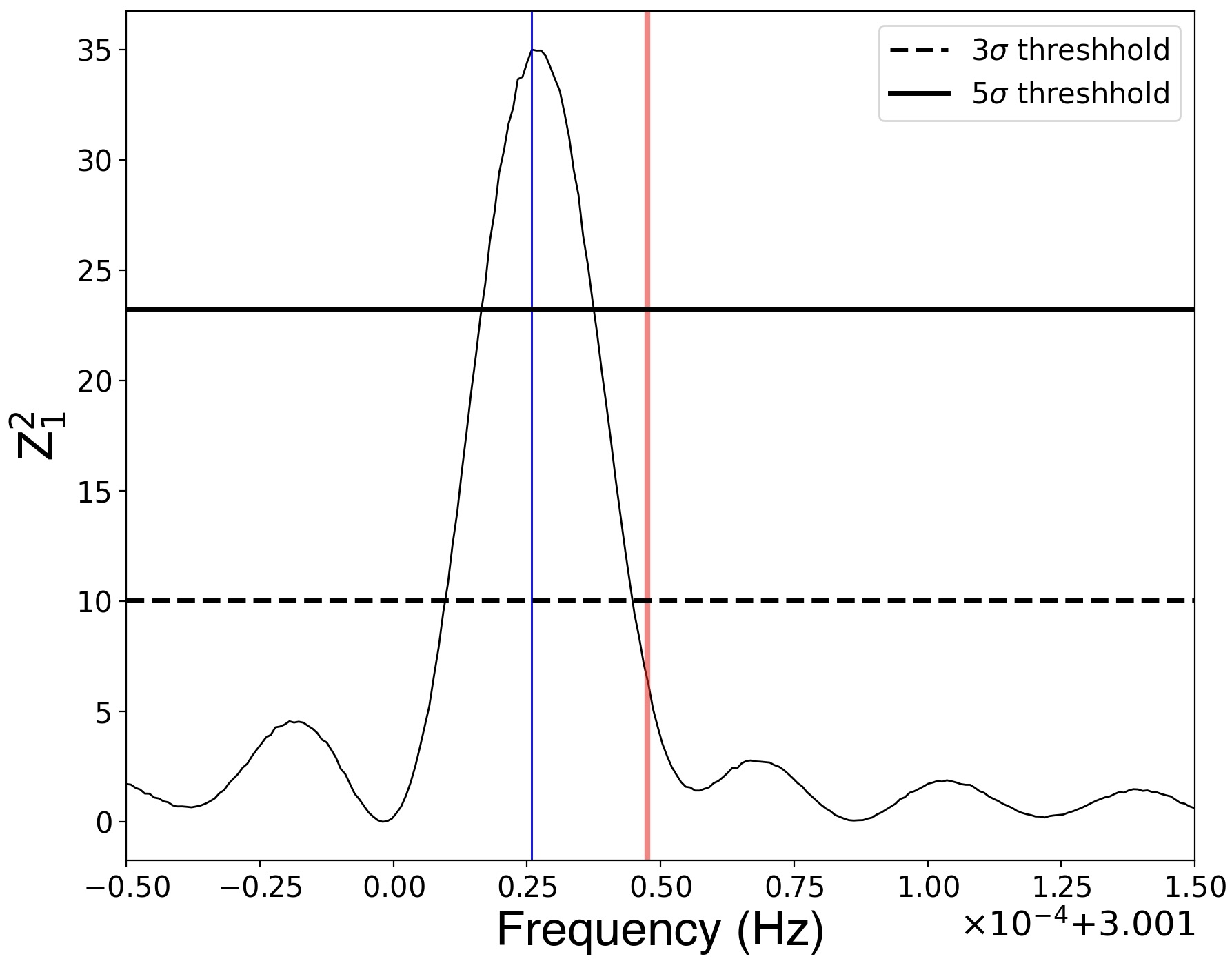}
    \includegraphics[width=0.495\linewidth]{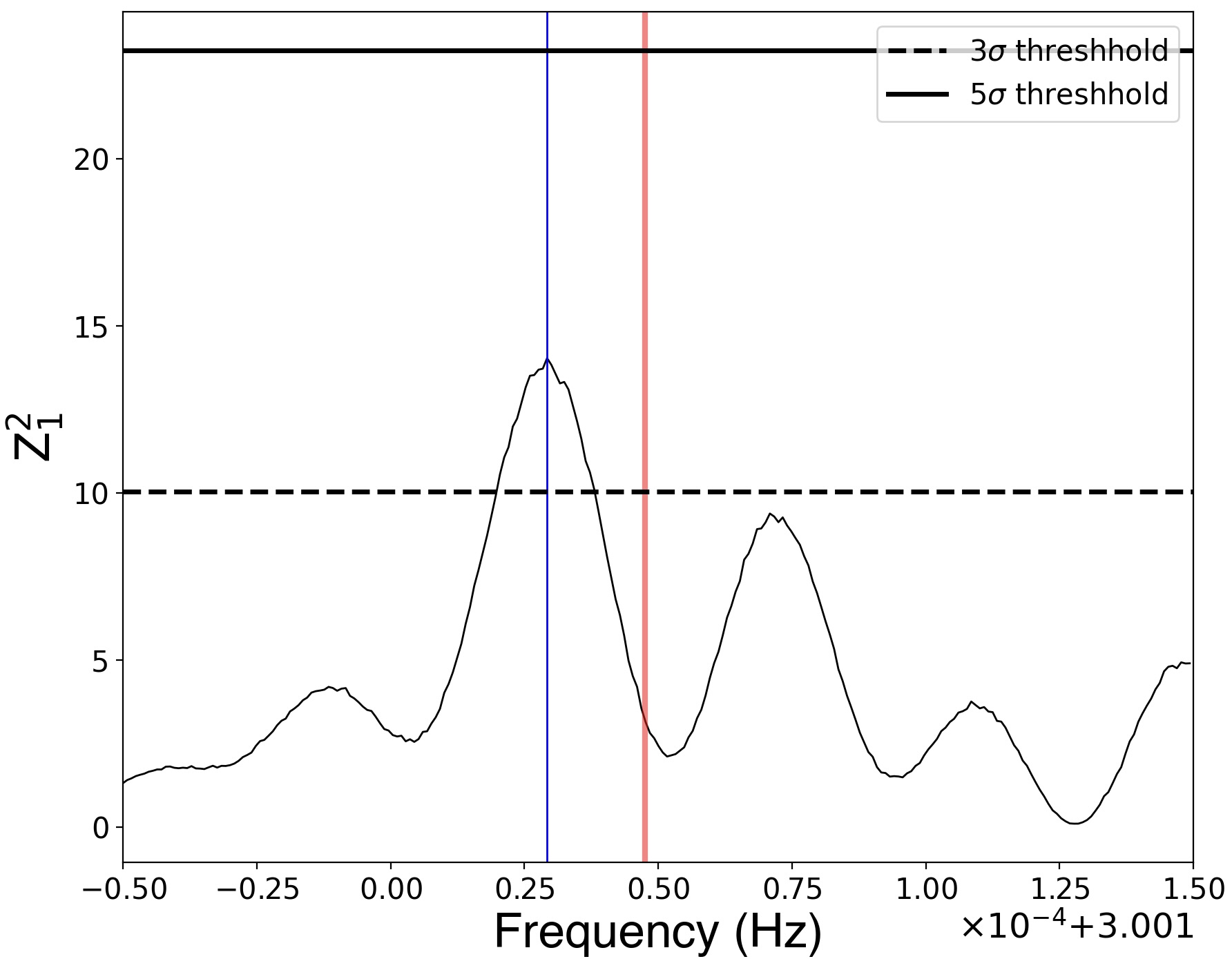}
    \caption{\textbf{\textit{Left}}: The $Z^2_1$ periodogram for the PN lightcurve of the 0923400101 observation. The blue line is the most significant peak (at 3.001026 Hz). \textbf{\textit{Right}}: The $Z^2_1$ periodogram for the PN lightcurve of the 0923400601 observation. The blue line is the most significant peak (at 3.001029 Hz). As in Figure \ref{PDS + Z^2_Test}, the red line in both panels is the expected frequency calculated from the gamma-ray ephemeris.}
    \label{Obs_Z^2_Test}
\end{figure}

The strong aliasing is an artifact of jointly analyzing the two observations where there exists a 1.5-day gap between observations. In an attempt to remedy this, we extracted individual lightcurves for the 0923400101 and 0923400601 observations and repeated the $Z_1^2$ search for them separately (Figure \ref{Obs_Z^2_Test}). For the 0923400101 observation, we detect a frequency of $f = 3.001026$ Hz ($P_{1} = 0.3332194$ s). For the 0923400601, we detect a frequency of $f = 3.001029$ Hz ($P_{1} = 0.3332190$ s) though at a lower significance than any of the previous frequencies. 

\begin{figure}
    \centering
    \includegraphics[width=0.495\linewidth]{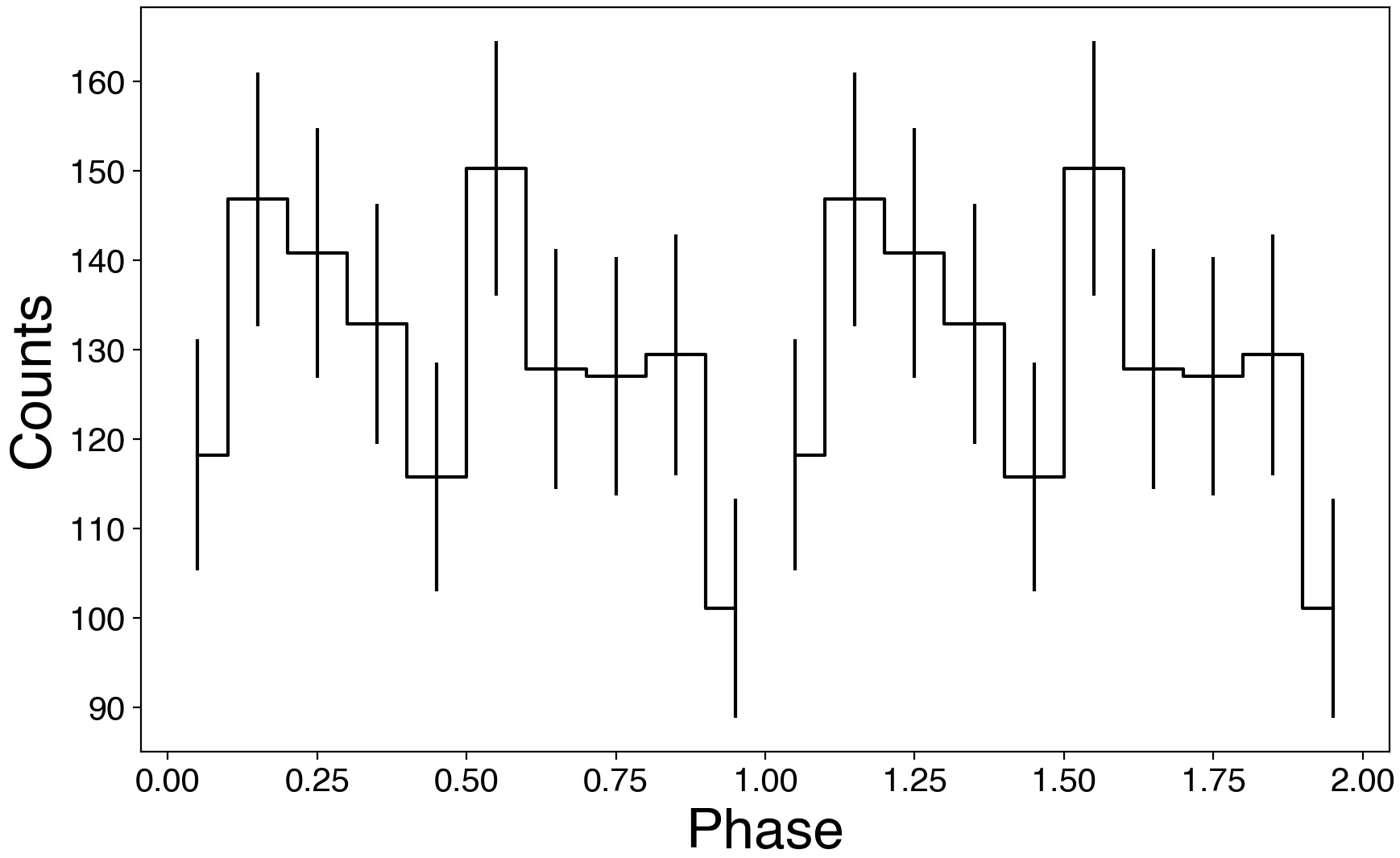}
    \includegraphics[width=0.495\linewidth]{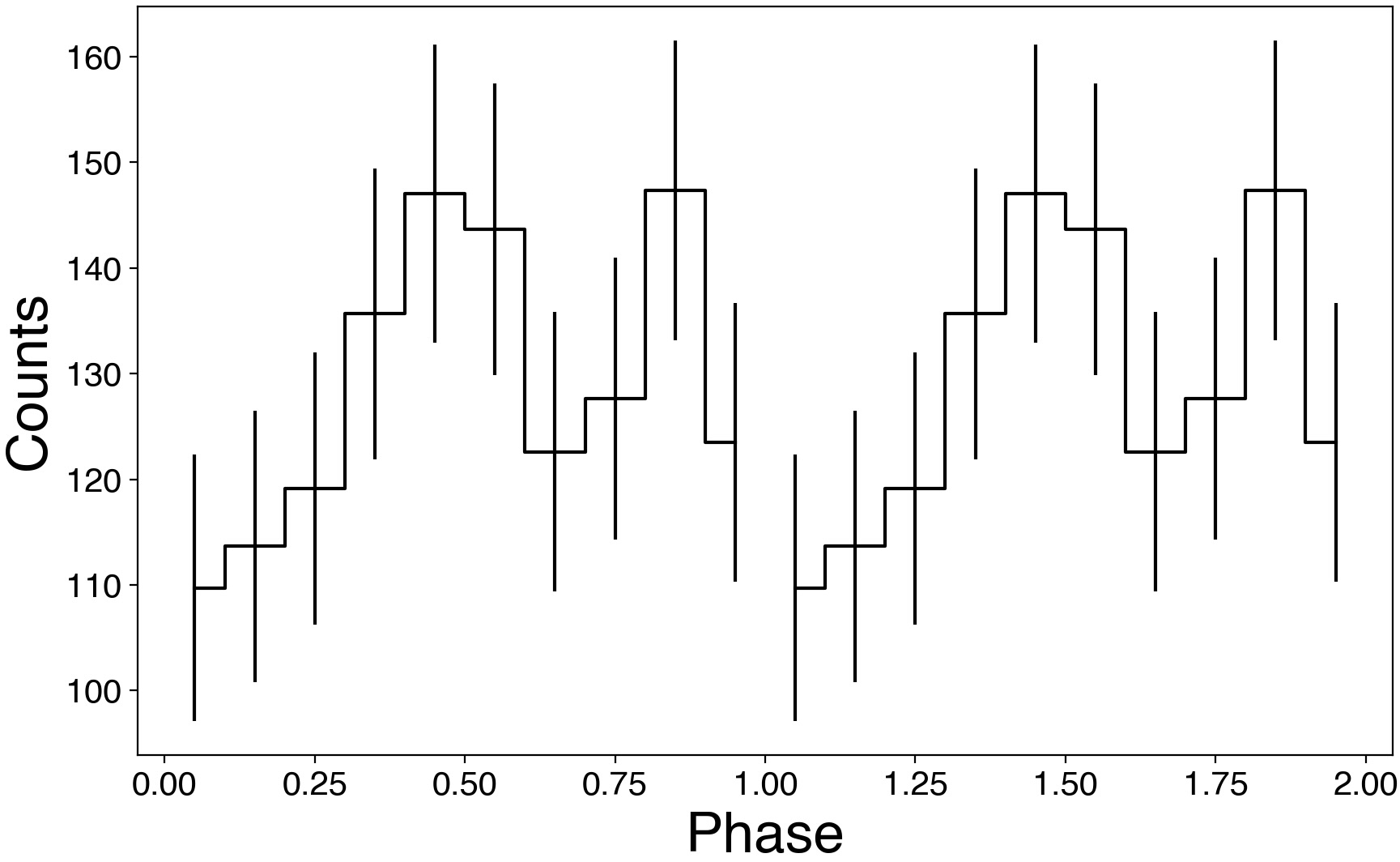}
    \caption{\textbf{\textit{Left}}: The background-subtracted PN lightcurve folded over $P=0.332195$ into 10 phase bins. \textbf{\textit{Right}}: The background-subtracted PN lightcurve folded over $P=0.332189$ into 10 phase bins. The folding is done over these two nearly identical periods as the heavy aliasing seen in Figure \ref{PDS + Z^2_Test} caused both periods to be detected with equal likelihood. In both cases, the large error bars preclude us from pursuing further analysis.}
    \label{folded_lc}
\end{figure}

We folded the background-subtracted lightcurve extracted from the PN camera over the two most significant periods found from the combined $Z^2_1$ periodogram into 10 phase bins (Figure \ref{folded_lc}). Qualitatively, both folded lightcurves exhibit some degree of sinusoidal-like pulse shape. However, the error bars associated with each phase bin are large, a consequence of the low full-frame timing resolution of the PN camera ($\Delta t=73.4$ ms) relative to the folding period. For $P_1 = 0.332195$ s, we computed a pulsed fraction of $19.6\pm 10.5$\% whereas for $P_1 = 0.332189$ s, we computed a pulsed fraction of $14.6\pm 10.3$\%. These large error bars preclude us from reliably fitting the folded lightcurve with a sinusoid, and we did not pursue further phase-resolved analysis as we were unable to clearly demarcate the on and off phases.

\subsection{Pulsar spectral analysis} \label{subsec:psr_spec}
To analyze the phase-averaged spectra from the processed \xmm{} data, we extractd the source counts from the same $24''$ circular region used for the timing analysis and the background from a $120''\times120''$ square away from the source as shown in Figure \ref{fig:xmm-fov} \textit{right} panel. We extractd a total of six EPIC spectra (PN, MOS1 and MOS2 from the two observations) and binned them individually by imposing $>3\sigma$ signal-to-noise ratio in each energy bin. We analyzed these spectra in Xspec using the 0.2-3.0 keV data, where the source is detected above the background. To perform this analysis, we fit various models to the data and each spectral model presented is multiplied by {\tt const*tbabs}. The {\tt const} component accounts for systematic variations in the flux normalization between the different X-ray spectra. {\tt tbabs} considers the interstellar medium (ISM) absorption.  

\begin{figure}
    \centering
    \includegraphics[width=0.495\linewidth]{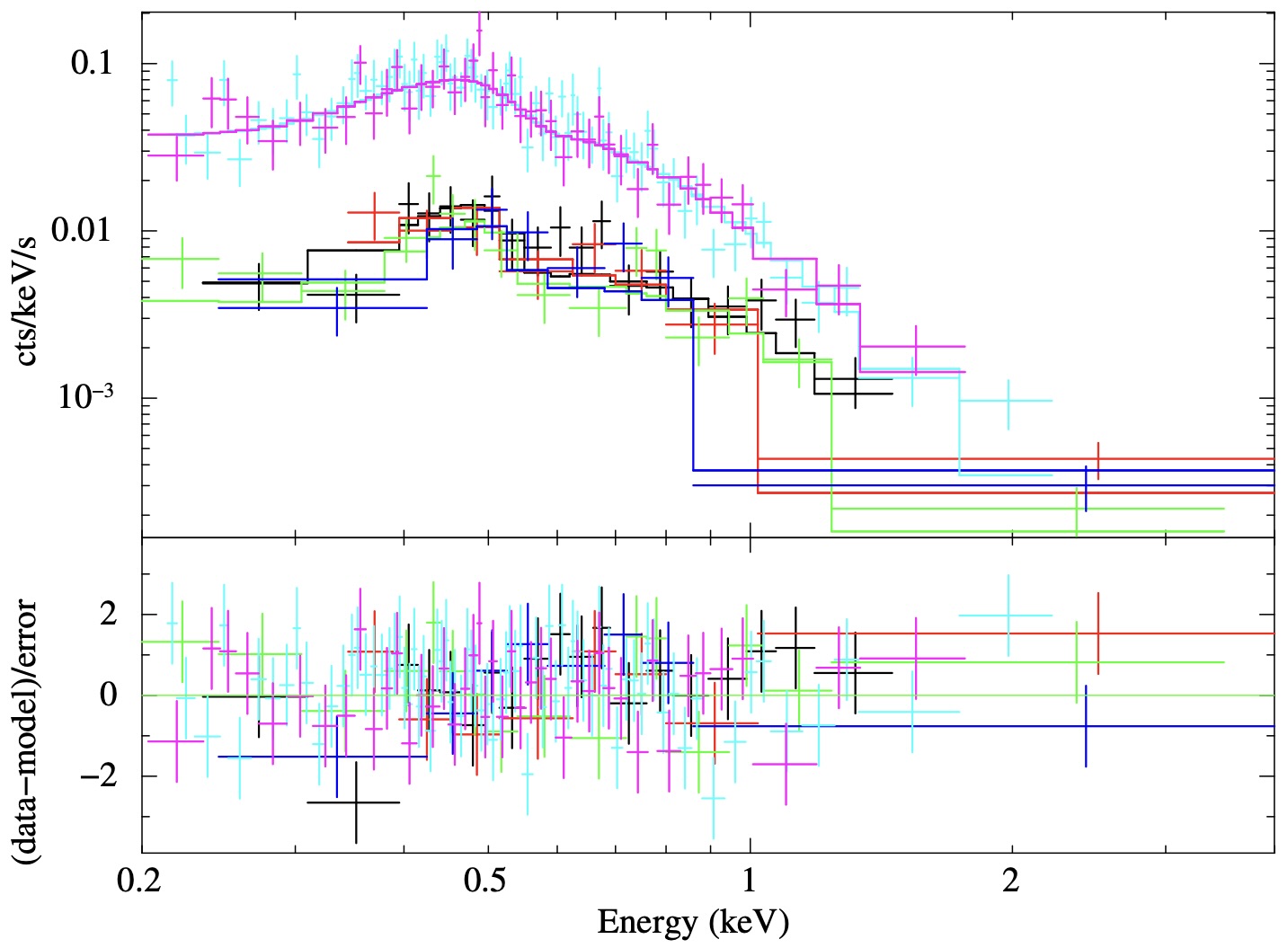}
    \includegraphics[width=0.495\linewidth]{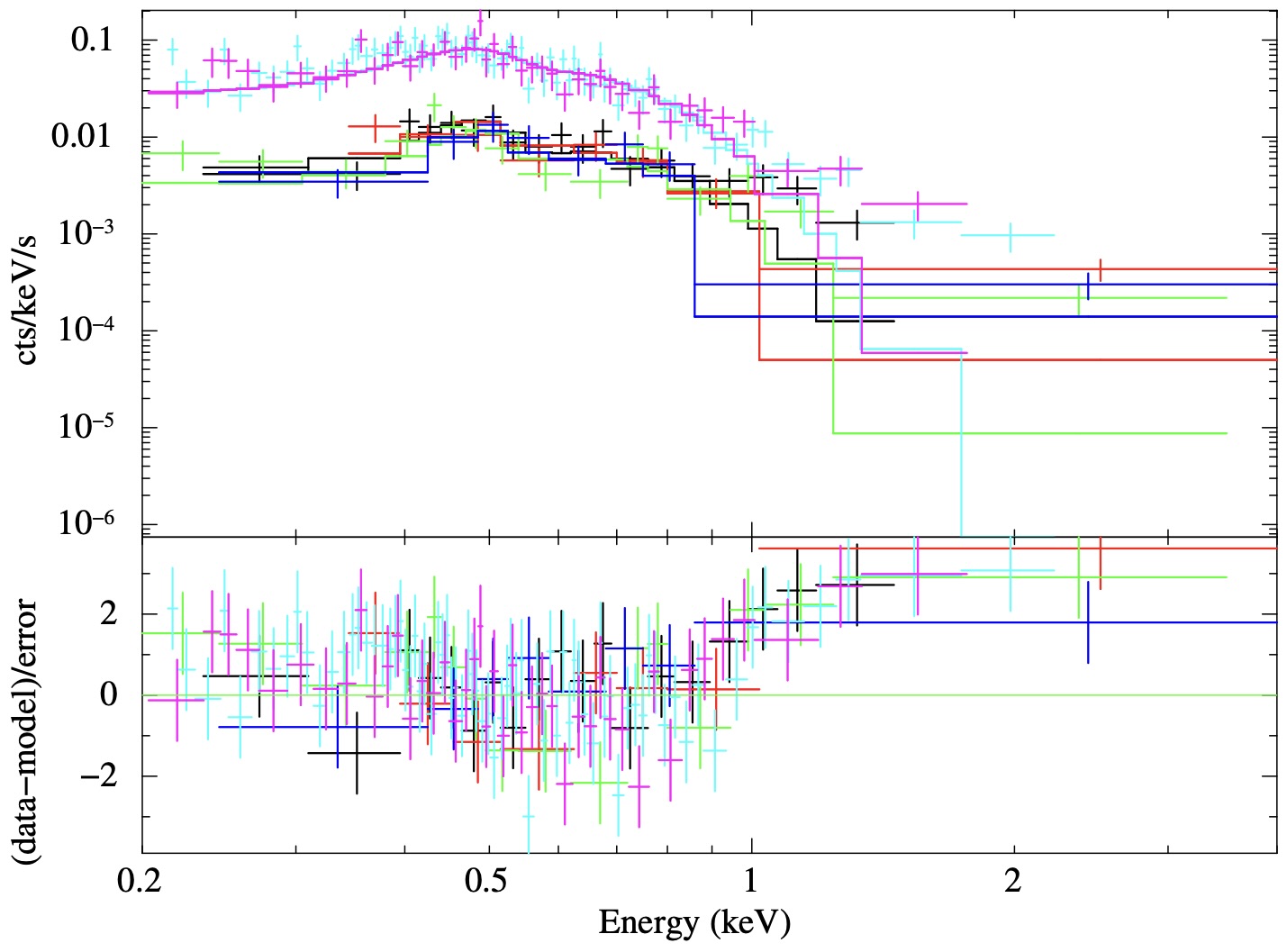}
    \caption{\textbf{\textit{Left}}: The phase-averaged \xmm\ spectra of \psr{} fit with an absorbed {\tt pow} model. \textbf{\textit{Right}}: The same spectra fit with an absorbed {\tt bbodyrad} model.  In both panels, different colors are for different spectra (three instruments $\times$ two observations = six spectra). }
    \label{Spectral_Fits_PL_BB}
\end{figure}

\begin{figure}
    \centering
    \includegraphics[width=0.495\linewidth]{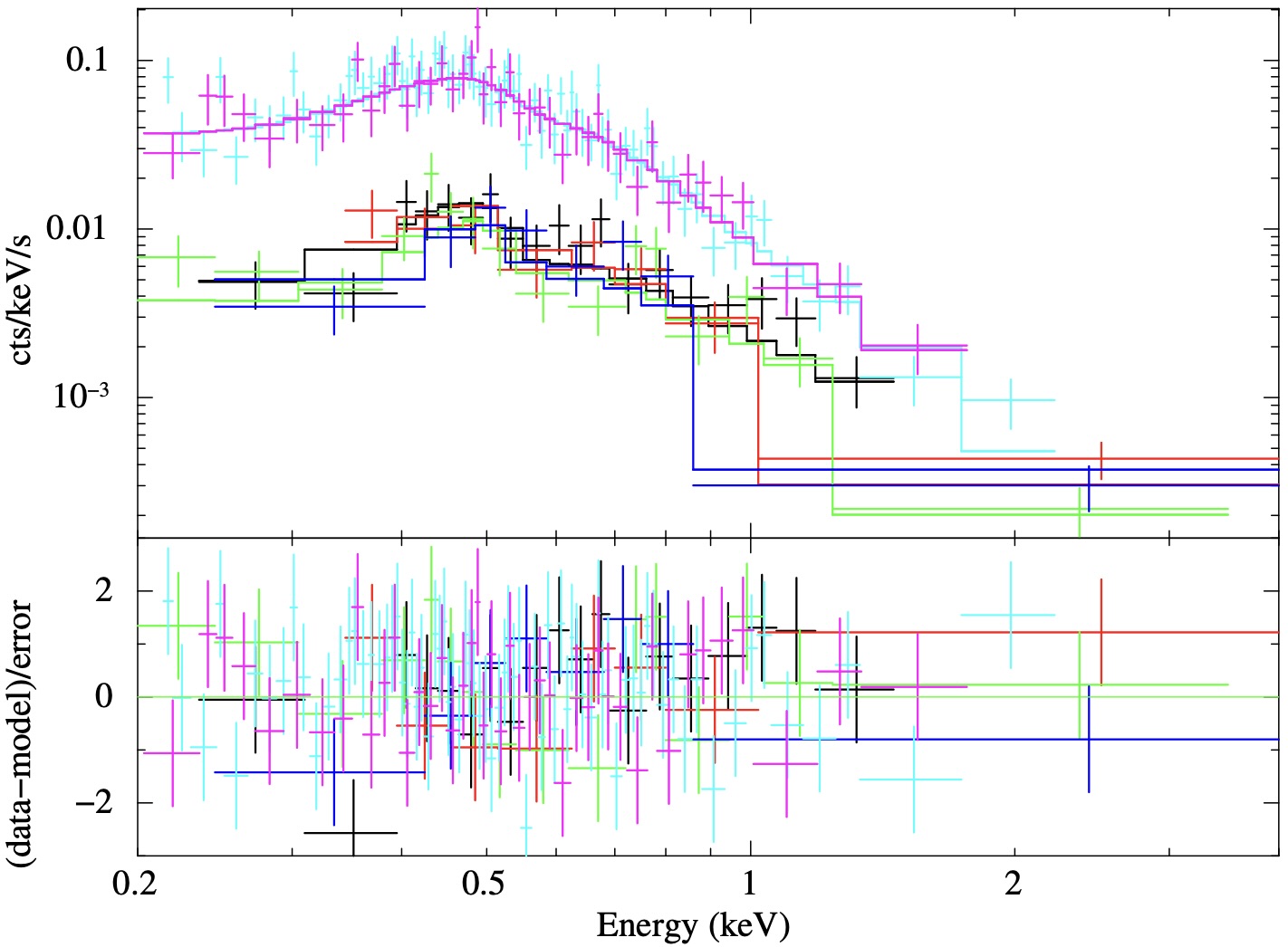}
    \includegraphics[width=0.495\linewidth]{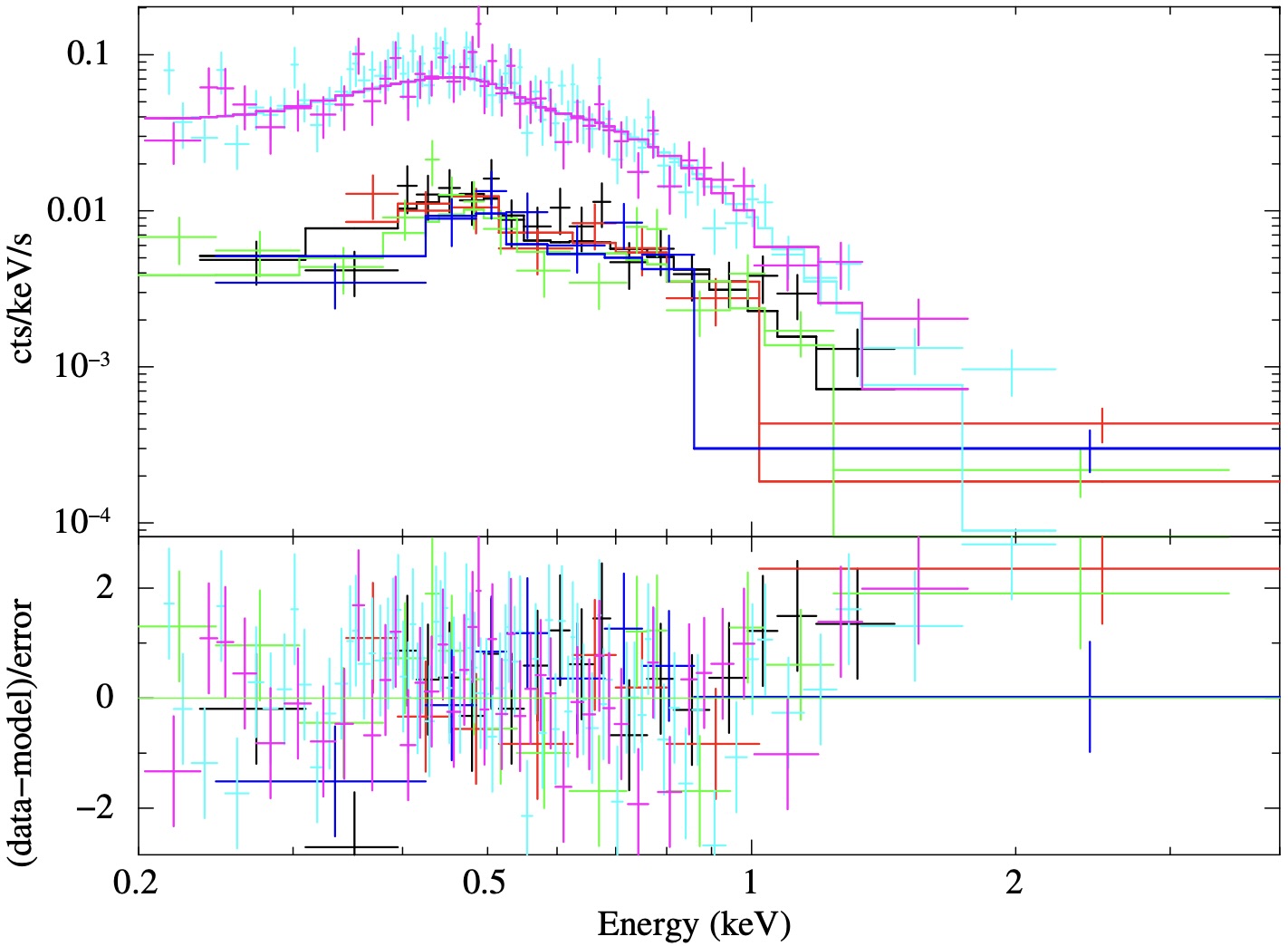}
    \caption{\textbf{\textit{Left}}: The phase-averaged \xmm\ spectra of \psr{} fit with an absorbed 2*{\tt bbodyrad} model. \textbf{\textit{Right}}: The same spectra fit with an absorbed {\tt nsmax}(H) model. The $B$-field is fixed to $B = 4\times10^{12}$G while the gravitational redshift is fixed to $z_g = 0.30$. In both panels, different colors are for different spectra (three instruments $\times$ two observations = six spectra). }
    \label{Spectral_Fits_2BB_NSMAX}
\end{figure}

We first performed a phenomenological fit of the data using a single power-law (PL) model (Table \ref{tab_phenom}). Although the fit is acceptable (Figure \ref{Spectral_Fits_PL_BB}) with $\chi^2/\textrm{dof}=\chi^2_\nu = 0.945$ (reduced $\chi^2$; Table \ref{tab_phenom}), the very soft photon index of $\Gamma = 5.81^{+0.26}_{-0.23}$ indicates a thermal origin of the pulsar's X-ray emission. Fitting a single blackbody (BB) model yielded a poor fit (Figure \ref{Spectral_Fits_PL_BB}) with $\chi^2_\nu = 1.66$ (Table \ref{tab_phenom}). Though a combination BB+PL model showed a markedly improved $\chi^2_\nu$ value of 0.945 (Table \ref{tab_phenom}) compared to the single BB model, the PL component photon index of $\Gamma = 4.21^{+0.61}_{-0.65}$ is still too soft to represent nonthermal emission from the pulsar. We found the most viable variation of the BB model to be a BB+BB model, which fits the data well (Figure \ref{Spectral_Fits_2BB_NSMAX}) and shows a similarly improved $\chi^2_\nu$ value of 0.919 (Table \ref{tab_phenom}). This best-fit model yielded an unabsorbed flux of $3.9^{+1.6}_{-1.1}\times10^{-13}$ erg cm$^{-2}$ s$^{-1}$ composed of a cooler component of $kT_1=78.8^{+5.5}_{-5.2}$ eV as well as a hotter component of $kT_2=227^{+38}_{-29}$ eV (Table \ref{tab_phenom}). Assuming the canonical neutron star (NS) radius $R=10$ km for the cooler BB component, the model flux normalization implies a source distance of $2.82^{+1.04}_{-0.80}$ kpc. The flux normalization ratio between the two BB components is $\sim 1600$, roughly consistent with the fractional polar cap area $(2 \pi R/c P_{\rm spin} \sim 1700$, where $c$ is the speed of light and $P_{\textrm{spin}}$ is the NS spin period; e.g., \cite{Ruderman1975:polar-cap}). Thus, it is plausible that the two BB components represent hot and cold thermal emission from polar caps and the rest of the NS surface, respectively. 
%Both a single blackbody (BB) model as well as a BB+BB model yielded a poor fit with $\chi^2_\nu = 1.64$ and $\chi^2_\nu = 1.51$ respectively (Table \ref{tab_phenom}). Though a combination BB+PL model fit the data well, the addition of the BB componet is statistically not very significant, yielding an F-test statistic of $p=0.0120$. Thus, we rule out any spectral models associated with blackbody components.  
%The simplest thermal model is that of blackbody radiation and we attempt to fit the data using different combinations of the blackbody radiation and power law models  
%However. all such models result in a poor $\chi^2$ fit statistic and generate a best-fit $N_H$ value which is highly unphysical, leading us to rule out the blackbody radiation model for fitting the data.
\begin{deluxetable*}{lcccc}[!ht] \label{tab_phenom}
\tablecaption{Power law and blackbody radiation model fits to the \xmm{} spectra.}
\tablecolumns{5}
\tablehead{
\colhead{Parameter}   
& 
\colhead{{\tt pow}}
& 
\colhead{{\tt bbodyrad}}
& 
\colhead{{\tt pow} + {\tt bbodyrad}}
& 
\colhead{2*{\tt bbodyrad}}
}
\startdata  
$N_H (10^{21} \rm{cm}^{-2})^a$ & $2.39^{+0.24}_{-0.23}$ & $2.39^{*}$ & $1.83^{+0.32}_{-0.29}$ & $1.61^{+0.28}_{-0.25}$\\
$\Gamma^b$ & $5.81^{+0.26}_{-0.23}$ & $\ldots$ & $4.21^{+0.61}_{-0.65}$ & $\ldots$\\
$K^{\text{\tt pow}}(10^{-5})^c$ & $1.89^{+0.18}_{-0.17}$ & $\ldots$ & $1.22^{+0.30}_{-0.28}$ & $\ldots$\\
$kT_1$ (eV)$^d$ & $\ldots$ & $82.0\pm{+1.5}$ & $73.7^{+6.2}_{-6.4}$  & $78.8^{+5.5}_{-5.2}$ \\
$K^{\text{\tt bbodyrad}}_1{}^e$ & $\ldots$ & $1.94^{+0.31}_{-0.26}\times10^{3}$ & $1.66^{+1.95}_{-0.83}\times10^{3}$ & $1.26^{+1.18}_{-0.59}\times10^{3}$\\
$kT_2$ (eV)$^d$ & $\ldots$ & $\ldots$ & $\ldots$ & $227^{+38}_{-29}$ \\
$K^{\text{\tt bbodyrad}}_2{\ }^e$ & $\ldots$ & $\ldots$ & $\ldots$ & $0.74^{+1.02}_{-0.45}$\\
$\chi_{\nu}^2$ (dof) & 0.945 (167) & 1.66 (168) & 0.906 (165) & 0.919 (165) \\
\enddata
All errors shown are $1\sigma$ confidence intervals. Parameters not applicable to the model are marked ``$\ldots$".\\
$^a$ The ISM hydrogen column density is associated with {\tt tbabs}. \\
$^b$ The photon index associated with the {\tt pow} model. \\
$^c$ Normalization parameter associated with the {\tt pow} model, defined as photons keV$^{-1}$ cm$^{-2}$ s$^{-1}$ at 1 keV. \\
$^d$ The blackbody temperature associated with the {\tt bbodyrad} model. \\
$^e$ Normalization parameter associated with the {\tt bbodyrad} model, defined as $(R_{km}/D_{10})^2$, where $R_{km}$ is the source radius (in km) and $D_{10}$ is the distance (in 10 kpc) to the object.\\
$^{*}$ The $N_H$ has been fixed to $2.39\times10^{21}$ cm$^{-2}$ (value from the {\tt pow} fit) as fitting the parameter would result in an unreasonably low value. 
\end{deluxetable*}

To model the thermal X-ray emission from the NS surface more realistically, we applied the NS atmosphere model \texttt{nsmax} available in Xspec. \texttt{nsmax} fully employs atomic structure calculation, partially-ionized plasma, opacity tables in different polarization modes, non-ideal equation of state and radiative transfer effects in the high magnetic field regime ($B > 10^9$ G) \citep{Mori2007, Ho2008}. The input parameters are $B$ (surface magnetic field strength), $z$ (gravitational redshift), $Z$ (chemical composition), $kT_{\rm eff}$ (effective temperature), and flux normalization (which is proportional to the emission area divided by $d^2$, where $d$ is the distance to the NS).  
%A more sophisticated model for fitting thermal emissions from neutron stars would be to use ionized atmosphere models. 
%We use the {\tt nsmax} model in XSPEC (citation) which provides the most accurate atmosphere model. The {\tt nsmax} model requires the surface $B$-field and the chemical makeup of the atmospheric plasma. 
%These are the B-field grids available in the \texttt{nsmax} model and closest to the dipole B-field. 
%the $P_{0}$ and $\dot{P}_{0}$ given by the gamma-ray empheris (citation), we calculate the $B$-field of \psr{} to be $\sim 3.5\times10^{12}G$ and assume that any changes in this value since that observation are negligible. 
We considered all available chemical compositions in {\tt nsmax} (H, C, O and Ne). Although an optically thick hydrogen layer should be present due to the fast gravitational sedimentation on the NS surface, diffusive nuclear burning can leave heavier element atmospheres \citep{Chang2004, Chang2010}. Non-hydrogen atmosphere compositions have been identified from X-ray observations of a handful of isolated NS \citep{Mori2006, Mori2007b, Ho2009}. Hydrogen atmosphere models exhibit harder X-ray spectra for the same effective temperature because the (energy-dependent) free-free absorption process is dominant. In contrast, heavier element atmosphere models are characterized by softer X-ray spectra and multiple absorption features because of the larger number of bound-bound and bound-free transitions in the X-ray band.

\begin{deluxetable*}{lcccccc}[!ht] \label{tab_nsmax}
\tablecaption{{\tt nsmax} model fits to the \xmm{} spectra.}
\tablecolumns{7}
\tablehead{
\colhead{Parameter}   
& 
\colhead{{\tt nsmax}(H)}
& 
\colhead{{\tt nsmax}(C)}
& 
\colhead{{\tt nsmax}(O)}
& 
\colhead{{\tt nsmax}(Ne)}
& 
\colhead{{\tt nsmax}(H)+{\tt pow}}
& 
\colhead{2*{\tt nsmax}(H)}
}
\startdata  
$N_H (10^{21} \rm{cm}^{-2})^a$ & $1.31^{+0.25}_{-0.16}$ & $0.39\pm0.17$ & $1.75\pm0.13$ & $0.499^{+0.063}_{-0.062}$ & $2.13^{+0.31}_{-0.30}$ & $2.06^{+0.33}_{-0.28}$ \\
$B (10^{12} \text{G})^b$$^*$ & $4.0$ & $10.0$ & $10.0$ & $10.0$ & $4.0$ & $4.0$\\
$z_g$$^c$$^*$ & $0.30$ & $0.30$ & $0.30$ & $0.30$ & $0.30$ & $0.30$ \\
$kT_1$ (eV)$^d$ & $56.7^{+4.4}_{-3.9}$ & $167^{+4}_{-3}$ & $177^{+2}_{-1}$  & $161^{+2}_{-3}$ & $40.2^{+4.7}_{-4.1}$ & $41.8^{+4.4}_{-3.8}$ \\
$K^{\text{\tt nsmax}}_1{}^e$ & $82^{+100}_{-33}$ & $0.142\pm0.035$ & $0.537\pm0.092$ & $0.219^{+0.006}_{-0.008}$ & $2.3^{+4.6}_{-1.5}\times10^{3}$ & $1.6^{+3.1}_{-1.0}\times10^{3}$\\
$kT_2$ (eV)$^d$ & $\ldots$ & $\ldots$ & $\ldots$ & $\ldots$ & $\ldots$ & $234^{+165}_{-72}$ \\
$K^{\text{\tt nsmax}}_2{}^e$ & $\ldots$ & $\ldots$ & $\ldots$ & $\ldots$ & $\ldots$ & $6.6^{+38}_{-5.9}\times10^{-3}$\\
$\Gamma^f$ & $\ldots$ & $\ldots$ & $\ldots$ & $\ldots$ & $3.0^{+1.0}_{-1.1}$ & $\ldots$\\
$K^{\text{\tt pow}}(10^{-5})^g$ & $\ldots$ & $\ldots$ & $\ldots$ & $\ldots$ & $0.62^{+0.38}_{-0.29}$ & $\ldots$\\
$\chi_{\nu}^2$ (dof) & 1.10 (167) & 2.39 (167) & 2.26 (167) & 1.76 (167) & 0.904 (165) & 0.905 (165) \\
\enddata
All errors shown are $1\sigma$ confidence intervals. Parameters not applicable to the model are marked ``$\ldots$".\\
$^a$ The ISM hydrogen column density is associated with {\tt tbabs} with the {\tt wilm} abundance table. \\
$^b$ The surface magnetic field strength with the {\tt nsmax} model. The model provides a discrete grid of $B$-field values, and only the H model has a finer grid. For the non-H elements, a $B$-field of $1\times10^{12}$G was tested and resulted in equally poor fits.\\
$^c$ The gravitational redshift associated with the {\tt nsmax} model. The fixed value corresponds to a redshift for a neutron star of $M = 1.4M_\odot$ and $R=10$ km. \\
$^d$ The atmosphere temperature associated with the {\tt nsmax} model. \\
$^e$ Normalization parameter associated with the {\tt nsmax} model, defined as $(R_{em}/d)^2$, where $R_{em}$ is the radius (in km) of the emission region, and $d$ is the distance (in kpc) to the object. \\
$^f$ The photon index associated with the {\tt pow} model. \\
$^g$ Normalization parameter associated with the {\tt pow} model, defined as photons keV$^{-1}$ cm$^{-2}$ s$^{-1}$ at 1 keV. \\
$^*$ These parameters are fixed. 
\end{deluxetable*}

To perform each fit, we assumed that $z = 0.3$, corresponding to the canonical NS mass ($M = 1.4 M_{\odot}$) and radius ($R =10$ km). The $B$-field lower limit derived from the period and spin-down rate measured by long-term \fermi\ observations \citep{3pc} is $B \sim 2.9\times10^{12}$ G. We adopted the lowest available $B$ value in the \texttt{nsmax} models which lie above the aforementioned limit ($B = 4\times10^{12}$ for H, $B=1\times10^{13}$ G for C, O and Ne).
Only the hydrogen model produced an acceptable fit to the \xmm\ EPIC spectra with $\chi^2_\nu = 1.10$ (Figure \ref{Spectral_Fits_2BB_NSMAX}). Non-hydrogen atmosphere models did not fit the \xmm\ spectra well largely because of the lack of the predicted absorption lines and edges (Table \ref{tab_nsmax}). As evident from the spectral shape (Figure \ref{Spectral_Fits_2BB_NSMAX}), the \textit{XMM-Newton} spectra are featureless. The hydrogen atmosphere model fit yielded an unabsorbed flux of $3.1^{+1.5}_{-0.6}\times10^{-13}$ erg cm$^{-2}$ s$^{-1}$ and a best-fit effective temperature of $kT_{\rm eff} = 56.7^{+4.4}_{-3.9}$ eV. Assuming the canonical NS radius of $R=10$ km, the model flux normalization implies a source distance of $1.1^{+0.3}_{-0.6}$ kpc. %\textcolor{red}{The surface temperature is consistent with the standard NS cooling curves for the spin-down age of *** kyr \citep{}.} 
Since a slight excess is observed in the residuals above $\sim1$ keV (Figure \ref{Spectral_Fits_2BB_NSMAX}), we also attempted the fitting with a {\tt nsmax}(H)+{\tt pow} model and a 2*{\tt nsmax}(H) model. For the {\tt nsmax}(H)+{\tt pow} (2*{\tt nsmax}(H)) model, a statistical improvement in the fit was observed with $\chi^2_\nu = 0.904$ ($0.905$) (Table \ref{tab_nsmax}) and the F-test probability of $p = 3.3\times10^{-8}$ ($1.4\times10^{-8}$) compared to the {\tt nsmax}(H) model. However, the flux normalization of the  {\tt nsmax} model implies a source distance of $0.209$ kpc and $0.248$ kpc for the {\tt nsmax}(H)+{\tt pow} (2*{\tt nsmax}) models, respectively. These distances are incompatible with other estimates (e.g., 1.6 kpc obtained from the phenomenological correlation between pulsar gamma-ray luminosity and spin-down luminosity \citep{fermipsr}) and too close for \psr{}, whose TeV halo appears significantly more compact than the Geminga halo at 250 pc. Furthermore, in the case of the 2*{\tt nsmax} model, the ratio of the normalization parameters $(K_1^{\textrm{\texttt{nsmax}}},\ K_2^{\textrm{\texttt{nsmax}}})$ implies an unphysically small hotspot with area of $\sim10^{-6}$ times the surface area of the NS. Therefore, we conclude that a single magnetized hydrogen atmosphere model accounts for the \xmm\ spectra most plausibly. 

In summary, the double-blackbody model fits the \xmm\ EPIC spectra best statistically. The model can be interpreted as hot and cold emission components from the polar cap and the rest of the NS surface. Although the magnetized hydrogen atmosphere model at $B = 4\times10^{12}$ G represents the most physically plausible solution, it leaves an X-ray excess above $E\sim 1$ keV (Figure \ref{Spectral_Fits_2BB_NSMAX}), whose nature could be investigated better in a future X-ray observation with improved statistics. 

\section{Pulsar halo modeling} \label{sec:modeling}

%\subsection{Introduction}
Like the prototypical gamma-ray pulsar halos of Geminga and Monogem \citep{Lopez-Coto:2022igd,Liu:2022hqf,Fang:2022fof,amato_halo_review}, the TeV extended emission of \lhaaso{} is believed to be produced by electrons diffusing away from the pulsar \psr{} and upscattering ambient photons, producing IC emission \citep{lhaasoj0621}.
The spatial properties of the emission observed by LHAASO are well explained, assuming that, in the vicinity of the pulsar, the diffusion coefficient of energetic electrons is different from the average values inferred for galactic CRs. Specifically, the emission morphology suggests that diffusion is suppressed by at least one order of magnitude with respect to the average Galactic value, 
although an alternative interpretation in terms of ballistic propagation has been proposed \citep{Recchia:2021kty,Bao:2021hey}.

\psr{} is a middle-aged (208 kyr) pulsar similar to Geminga and Monogem (Table \ref{tab:pulsars}). The extent of the TeV emission of \lhaaso{} ($\theta_d\sim0.9\degree$, \cite{lhaasoj0621}) is smaller than that of Geminga ($\theta_d\sim5.5\degree$, \cite{HAWC_Halos_and_Positron_Flux}), which may indicate the distance to \lhaaso{} is much larger than that to Geminga ($250^{+120}_{-62}$ pc, \cite{geminga-distance}). In this case, a significant part of the X-ray counterpart of \lhaaso{} may be accessible to the limited FoV of X-ray telescopes. For our modeling, we adopt the ``pseudo-distance" of 1.6 kpc obtained from the phenomenological correlation between pulsar gamma-ray luminosity and spin-down luminosity \citep{fermipsr}.

%but its larger distance of 1.6~kpc implies that the gamma-ray emission, and possibly its X-ray counterpart, are more concentrated around the pulsar, as demonstrated by the extension observed in TeV by LHAASO \citep{lhaasoj0621}, and thus potentially more accessible to X-ray observations with smaller field of view. 
%\jw{The distance of 1.6 kpc is a ``pseudo-distance" from the gamma-ray luminosity vs. distance correlation, so it's not even a measurement. I suggest changing the order of things in this sentence: the LHAASO source size is smaller than Geminga which might be because of a larger distance (pseudo-distance 1.6 kpc vs. Geminga 250 pc). If it is indeed farther away, angular size of X-ray halo would be also smaller. Maybe also mention the physical size of the LHAASO and Geminga halo normalized to the distance?} \SM{I see your point, I agree with changing the order of the sentence. I don't think it's necessary to give the physical size normalized to distance here, since this depends on energy.}
%
In addition to the multi-TeV gamma rays observed by LHAASO, observation at lower gamma-ray energies provided by VERITAS and X-ray observation with \xmm{} are crucial to constrain the properties of the electron population emitted by the pulsar and the environment in which they propagate, such as the strength of the magnetic field in which pairs are expected to produce synchrotron radiation. 
This has been extensively explored for the Geminga pulsar halo in \cite{Manconi:2024wlq}, where GeV and TeV gamma rays have been used together with novel and robust X-ray upper limits to constrain the ambient magnetic field. 
In this paper, we use the model extensively introduced in \cite{DiMauro:2019yvh,DiMauro:2019hwn,DiMauro:2020jbz} for the IC emission and expanded in \cite{Manconi:2024wlq} to compute the multiwavelength spectral energy distribution (SED) down to X-ray energies. The rationale and ingredients of our modeling are summarized below. We refer the reader to the aforementioned publications for more details on the model's assumptions. 

%\subsection{Benchmark model}
Our model assumes that electrons are injected by the pulsar continuously at a rate proportional to the pulsar's spin down $\dot{E}(t) = \dot{E}_0 ( 1+ t/\tau_0)^{-\frac{n+1}{n-1}}$, with characteristic decay time fixed to $\tau_0=12$~kyr and dipole braking index of $n=3$. 
%\jw{So the electron injection rate is $\eta\dot{E}$, but this sentence sounds like it's $\dot{E}$. Suggestion: at a rate proportional to the pulsar spin-down luminosity, $\eta\dot{E}$, and provide the definition of $\eta$ and $\dot{E}$ here} \SM{agree with the suggestion; but I would refer to previous papers (Eqs 9-12 from \cite{DiMauro:2019yvh})for the definition of $\eta$ for being concise.} \jw{By the definition of $\eta$ I meant the last sentence of this paragraph, but if there's more, yes I agree we can cite the paper!}
Accelerated electrons have an energy spectrum that follows a power-law model with an exponential cutoff, $Q(E)=Q_0 E^{-\gamma} \exp(-E/E_c)$, where $E_c$ is the cutoff energy and $\gamma$ is the slope of the injection spectrum. An efficiency $\eta$ of conversion of the spin-down luminosity into electron and positron pairs is used to determine the normalization of the energy spectrum $Q_0$; see Equations 9--12 from \cite{DiMauro:2019yvh} for more details.

Electrons are propagated using a transport equation that includes diffusion and radiative cooling by synchrotron emission in the ambient magnetic field and IC emission on the local interstellar radiation field, modeled following the local model of \cite{Vernetto:2016alq}. As discussed in \cite{Manconi:2024wlq}, we approximate the magnetic field to be a uniform, random ambient field of value $B$ extending at least to the scale of the observed TeV emission.
Diffusion around the pulsar is described by the diffusion coefficient  $D(E)=D_0 (E/1 \textrm{ GeV})^\delta$, where $D_0$ is a reference value at 1 GeV, and  $\delta=0.33$ as in \cite{DiMauro:2019hwn,Recchia:2021kty}.
A two-zone diffusion model is also explored, in which suppression of the diffusion is confined to a region up to a radius of $r_b$ \citep{Tang:2018wyr,DiMauro:2019hwn,Osipov:2020lty}, while typical Galactic $\delta$ and $D_0$ values are assumed outside this region \citep{Kappl:2015bqa}.

The particle density at different positions around the pulsar is then integrated along the line of sight to obtain the synchrotron and IC emissions; see Equations 1--6 in \cite{Manconi:2024wlq}.  
To properly compare with the data and the upper limits provided by different experiments at very different angular scales,  both emissions are integrated within the angular distances of $10'$ and $1\degree$ from the central pulsar for \xmm{} and VERITAS, respectively. 
%\FD{, \sout{(ten arcminutes, and 1 degree, for XMM-Newton and VERITAS, respectively). }}

We built a theoretical benchmark model 
compatible with the LHAASO-KM2A SED and surface brightness profile \citep{lhaasoj0621} as well as the VERITAS and \xmm{} upper bounds from this work. The model parameters are the ones that describe the TeV gamma-ray emission as in  \citet{Recchia:2021kty,Bao:2021hey,Fang:2021qon}, and the magnetic field strength that determines the synchrotron emission.
Specifically, the electron injection spectral index is set to $\gamma=1.4$, with $E_c=200$ TeV. 
The surface brightness profile measured by LHAASO-KM2A \citep{lhaasoj0621} is well described by $D_0=2\times10^{25}$ cm$^2$ s$^{-1}$ and $\eta=0.2$. 
We note that the diffusive extension of the TeV halo emission approximately scales as  $\theta_d \sim \sqrt{D(E) t_E}/d$ \citep{lhaasoj0621}, where $t_E$ is the cooling time for pairs with energy $E$. A change in the uncertain source distance $d$ would, therefore, directly affect the observed TeV halo extension and, thus, the best-fit value of the parameter $D_0$ describing the LHAASO-KM2A surface brightness. See \cite{DiMauro:2019hwn} for a precise estimate of the effect of pulsar's distance and age on halo extension.

\begin{figure}
\centering
\includegraphics[width=0.6\textwidth]{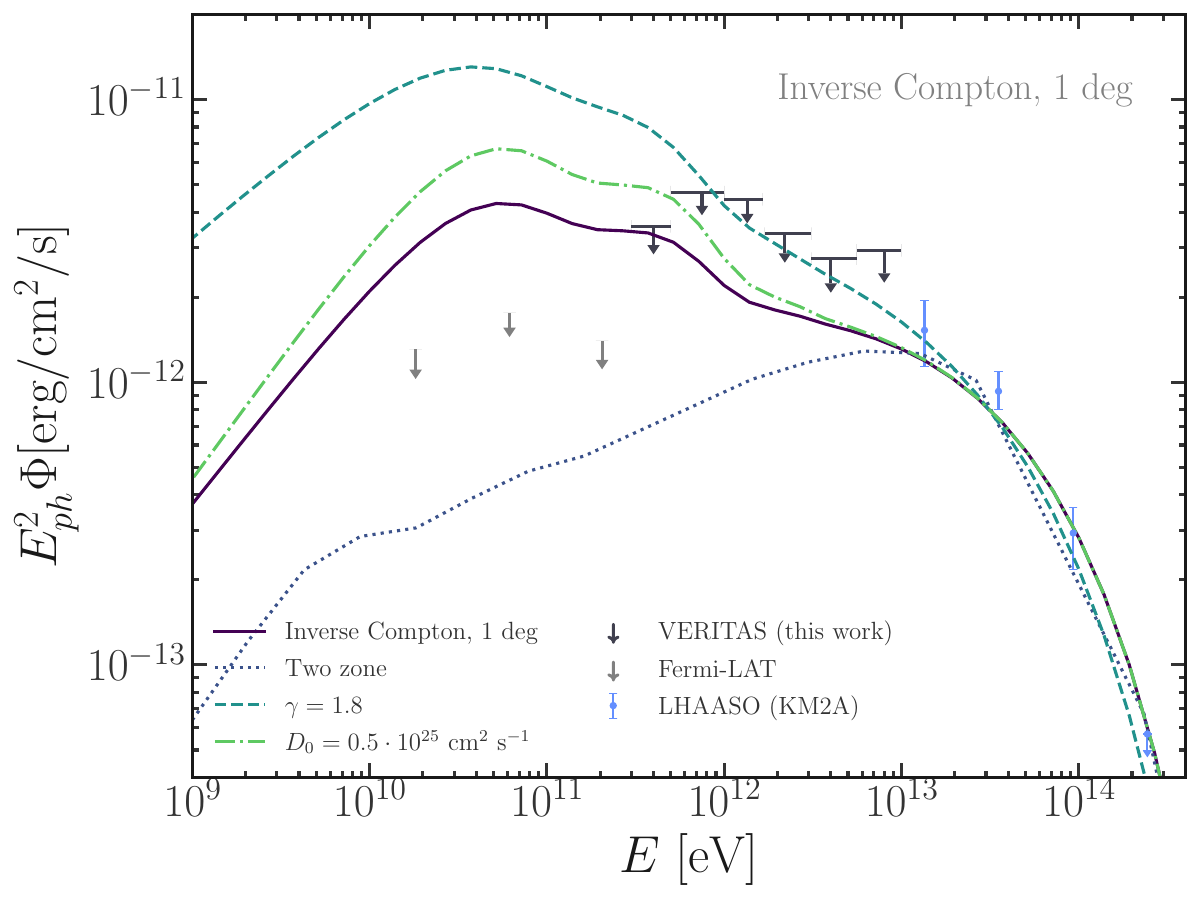}
\caption{IC SED models of \lhaaso{} integrated over a $1\degree$ region around \psr{}. The LHAASO-KM2A and \fermi{} flux points (blue points and gray arrows, \cite{lhaasoj0621}) and VERITAS flux upper limits (black arrows, this work) are overlaid. The solid black line represents the benchmark model (see text for details). The effects of changing the diffusion model (blue dotted line), the injection spectral index (blue dashed line), and the diffusion coefficient (green dot-dashed line) are illustrated to showcase the constraining power of VERITAS and \fermi{} upper limits.}
\label{fig:ic_veritas}
\end{figure}

%\subsection{Inverse Compton emission: impact of VERITAS upper limits}
Figure~\ref{fig:ic_veritas} shows the benchmark SED model for the IC emission integrated over a $1\degree$ region around the  PSR J0622+3749. 
In addition, the SED models with a different electron injection spectral index ($\gamma=1.8$; blue dashed line), diffusion coefficient ($D_0=0.5\times10^{25}$ cm$^2$ s$^{-1}$; green dot-dashed line), and transport model (two-zone diffusion with $r_b=30$~pc; blue dotted line) are overlaid for comparison. $\eta$ was varied for different models to match the observed TeV spectrum. 
The models are compared with the dedicated LHAASO-KM2A observation of \lhaaso{} \citep{lhaasoj0621}, the VERITAS observation presented in this work, and the \fermi{} upper limits from \cite{lhaasoj0621}.
As investigated in detail in \cite{Manconi:2024wlq} and \cite{DiMauro:2019hwn}, the electron injection spectral index and transport properties, namely the diffusion coefficient and the one or two-zone propagation, strongly influence the normalization and shape of the SED in the GeV to TeV energy range. Moreover, the spectrum at energies above 100 GeV is influenced by the spectral cutoff and temporal profile of electron injection, here fixed to a dipole braking with $n=3$.
The VERITAS upper limits from this work (in particular, below 1 TeV) refute the one-zone models with a softer spectral index of $\gamma=1.8$ and a diffusion coefficient roughly two times lower than the benchmark model. Moreover, the \fermi{} upper limits from \cite{lhaasoj0621} suggest that a two-zone diffusion model with $r_b=30$~pc more accurately describes the emission in the GeV energies.  This demonstrates that observations at a few TeV and GeV energies are crucial to constrain the properties of the electron population in pulsar halos. 
%, see discussion in \cite{Manconi:2024wlq} and \cite{DiMauro:2019hwn} for more details. 
We do not consider the best-fit spectrum of \lhaasocat{} in the first LHAASO catalog \citep{lhaasocat} since the results from the generic approach for catalog generation may not be optimal for use in our modeling. However, we verified that the pulsar halo model could be tuned to describe the steep spectrum at a few TeV measured by LHAASO-WCDA, for example, by adopting the two-zone transport model (blue dotted line). This is because the larger propagation lengths of lower-energy electrons result in a hard spectrum within the spectral integration region. A systematic exploration of the parameter space of the pulsar halo model, including, e.g., correlated spin-down, injection, and diffusion properties, is left for future investigation. 

\begin{figure}
\centering
\includegraphics[width=0.49\textwidth]{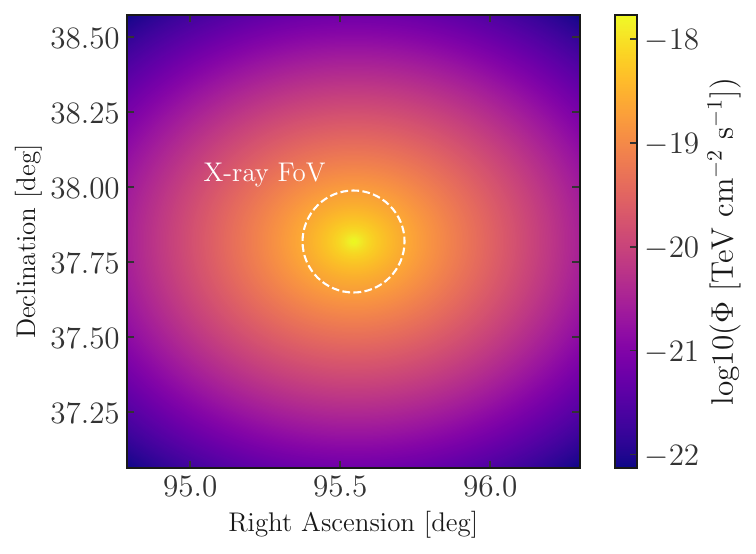}
\includegraphics[width=0.49\textwidth]{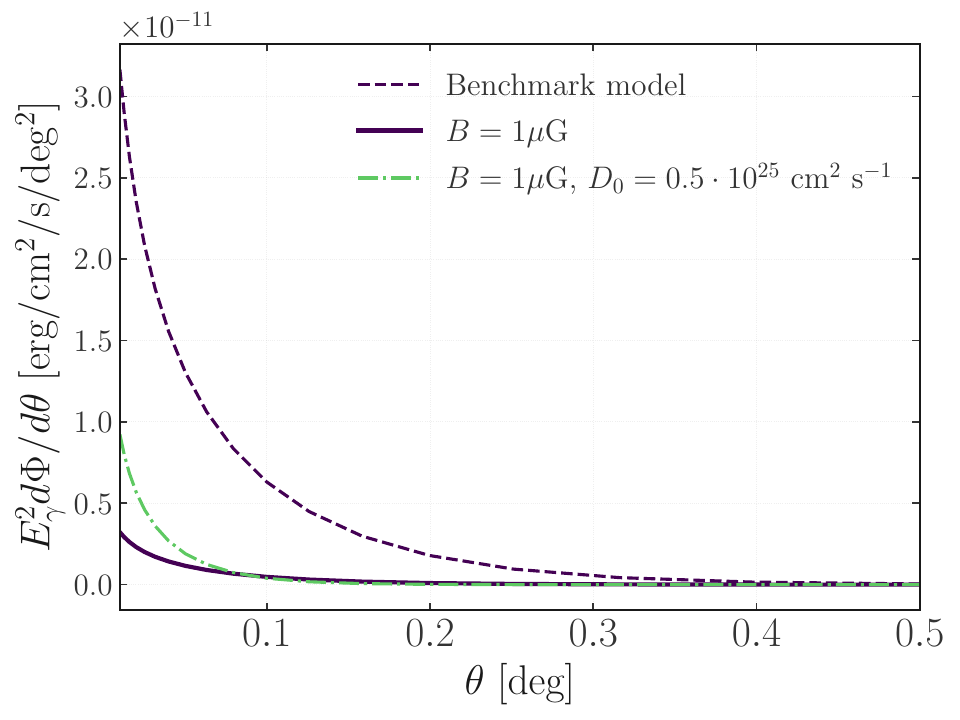}
\caption{\textit{Left}: The $1.5\degree\times1.5\degree$ model sky map of the synchrotron emission of \lhaaso{} in 2--7 keV for the benchmark model. The color map is on a logarithmic scale. The \xmm{} integration region is overlaid with a white dashed line. \textit{Right}: Surface brightness profile of the synchrotron halo in 2--7 keV for the benchmark model ($B=3\ \mu$G, dashed line), reduced magnetic field (solid line), and reduced diffusion coefficient (dot-dashed line); see the text for the details.}
\label{fig:sync}
\end{figure}

%\subsection{Synchrotron emission: impact of XMM-Newton upper limits}
Once the pulsar halo model was tuned to the gamma-ray data, we proceeded with the prediction for the synchrotron emission. 
We note that, within our model, the emission from the pulsar or a potential PWN was not included. Moreover, we used a one-zone model to compute the synchrotron emission at keV energies---as we verified, the effect of the second zone is negligible for the highest-energy electrons emitting TeV gamma rays by IC scattering and X-rays by synchrotron radiation.
Figure~\ref{fig:sync} \textit{left} panel shows the synchrotron flux map predicted by the benchmark model for a $1.5\degree\times1.5\degree$ region around the pulsar in 2--7 keV. The FoV of the \xmm{} observation is overlaid with a dashed white line. The flux map is in logarithmic scale to highlight the predicted extension of the pulsar halo. 
The Figure \ref{fig:sync} \textit{right} panel shows the surface brightness profile of the synchrotron emission out to $0.5\degree$ from the pulsar integrated over the energy range of 2--7 keV predicted by the benchmark model with a magnetic field strength of 3 $\mu$G (dashed line), the same model with a reduced magnetic field strength of 1 $\mu$G (solid line), and with a reduced magnetic field and diffusion coefficient (green dot-dashed line). As discussed in detail for Geminga in \cite{Manconi:2024wlq}, the magnetic field and the diffusion coefficient govern the transport of energetic particles around the pulsar, and hence represent the main parameters that determine the synchrotron emission from pulsar halos observable with typical FoV of X-ray instruments. Specifically, the synchrotron X-ray flux is the number of electrons times the synchrotron intensity of the individual electrons. While the low diffusion coefficient concentrates the X-ray-emitting electrons within a small angular scale and increases the number of electrons, the low magnetic field decreases the intensity of their synchrotron emission, balancing out the effect of slow diffusion.

\begin{figure}
\centering
\includegraphics[width=0.6\textwidth]{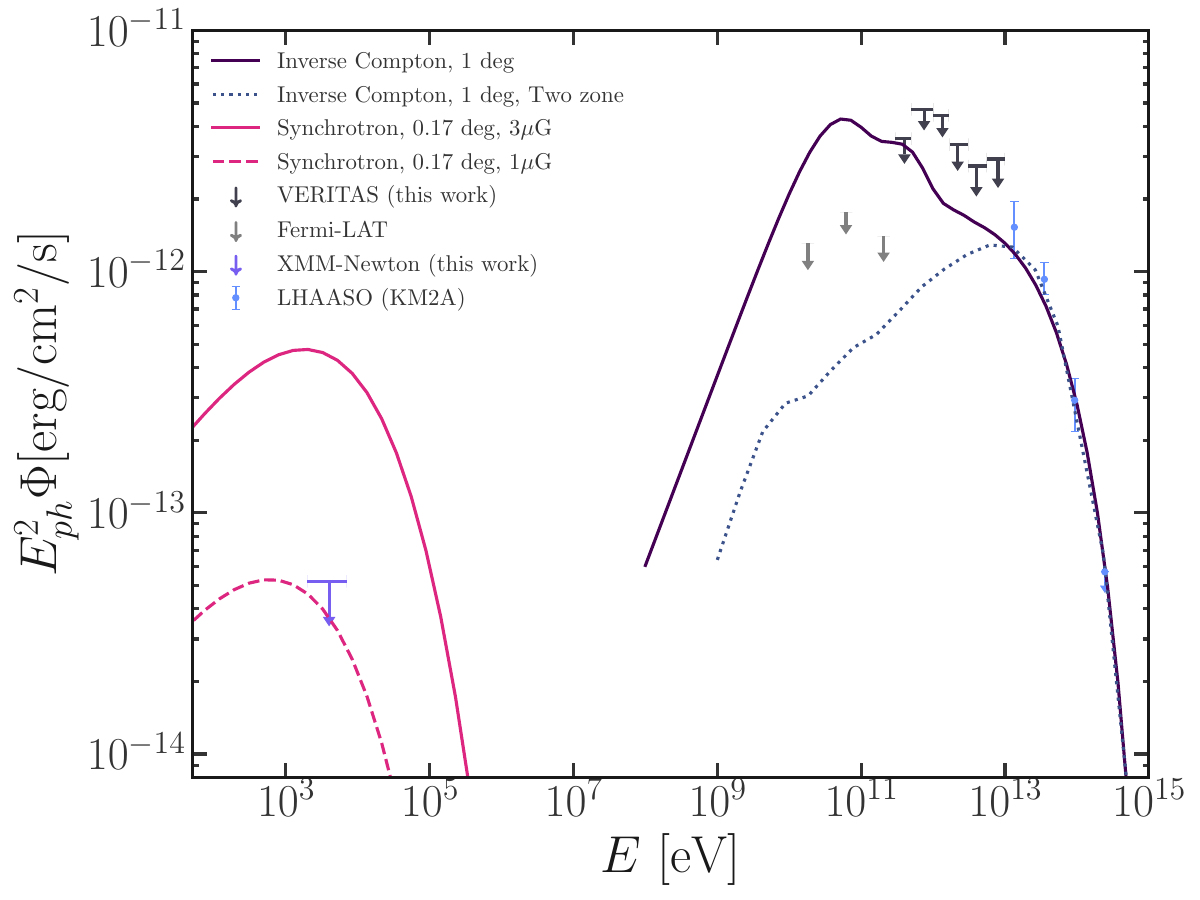}
\caption{The multiwavelength SED model for \lhaaso{} as a pulsar halo. The gamma-ray data from LHAASO-KM2A and \fermi{} (blue and gray points, \cite{lhaasoj0621}) and VERITAS (black, this work) and the X-ray data from \xmm{} (magenta, this work) are overlaid. The pulsar halo model is tuned to be compatible with the IC emission in the TeV range (solid line, obtained by integrating the model within 1 degree from the pulsar). The synchrotron spectrum is integrated over a $10'$ region around \psr{} and reported for two values of the magnetic field, 3 $\mu$G and 1 $\mu$G (magenta dashed and dotted lines, respectively).}
\label{fig:interpr}
\end{figure}

To compare our models with the X-ray data, the synchrotron emission is integrated over the region from which our X-ray upper limits were derived ($10'=0.17\degree$ from the pulsar).
In Figure~\ref{fig:interpr}, we present the predicted multiwavelength SED along with the X-ray and gamma-ray data.  
The purple line represents the modeled IC emission integrated over a $1\degree$ region around the pulsar as in Figure \ref{fig:ic_veritas}. This model is compatible with the LHAASO-KM2A (blue) and VERITAS (black) data. Moreover, introducing the two-zone model with the low-diffusion region radius of $r_b=30$~pc guarantees compatibility with the \fermi{} upper limits at GeV energies.
The solid ($B=3$ $\mu$G) and dashed ($B=1$ $\mu$G) lines in magenta represent the model synchrotron emission integrated over the $0.17\degree$ region around the pulsar. 
Our \xmm{} observation constrains the magnetic field strength to be below $1\ \mu$G, a result in line with what was found for the Geminga pulsar halo \citep{Manconi:2024wlq}. 

\section{Discussion} \label{sec:discussion}
%\SM{@Jooyun: we discussed with Fiorenza and would like to propose to: anticipate the current section 5 (Discussion) and leave the current section 4 on the modeling after. In this way, the discussion can focus on the observational part, and we have a full section to illustrate the model and its results, without mixing. Everything is then summarized in the conclusions!}
%\jw{Sounds good!}

%\subsection{Interpretation of the pulsar halo modeling result}

In this section, we discuss the implications of our VERITAS and \xmm{} observations for the physical properties of \lhaaso{} as a pulsar halo and the implications of our \xmm{} observation for the physical properties of \psr{}. 

\subsection{\lhaaso{} as a pulsar halo}

\psr{} possesses properties similar to the Geminga pulsar, as shown in Table \ref{tab:pulsars}. In addition, the diffusion coefficient and magnetic field of \lhaaso{} found by our pulsar halo modeling are consistent with those of Geminga (e.g., \citet{HAWC_Halos_and_Positron_Flux,Manconi:2024wlq}). However, the IACT and X-ray observations of \lhaaso{} significantly differ from those of Geminga. H.E.S.S. detected the Geminga halo with a radius of at least $3\degree$ in 0.5--40 TeV and a flux normalization of $(2.8\pm0.7)\times10^{-12}$ cm$^{-2}$ s$^{-1}$ TeV$^{-1}$ at 1 TeV within a radius of $1\degree$ around the pulsar \citep{geminga-hess}. Our VERITAS observation yielded a null detection of a halo within a radius of $1\degree$ around the \lhaaso{} centroid. Let us assume the physical size of the two sources are identical in all energies, and the difference in their angular sizes is due to their different distances from Earth. In this case, the ratio of the halo extensions measured by HAWC and LHAASO, $\theta_d($Geminga$) / \theta_d($\lhaaso{}$)=5.5\degree / 0.91\degree=6$ \citep{HAWC_Halos_and_Positron_Flux,lhaasoj0621}, indicates the expected extension of \lhaaso{} in the VERITAS energy range is $\sim0.5\degree$, i.e., six times smaller than the extension of Geminga measured by H.E.S.S. Since the expected extension of $0.5\degree$ is well within our source region, it is unlikely that background over-subtraction causes the null detection. A more plausible explanation for the null detection is that the halo flux is below the VERITAS sensitivity limit. The 50-hr point-source sensitivity of VERITAS is comparable to the flux of \lhaaso{} measured by WCDA \citep{veritas-sensitivity}. Given the much larger extension ($0.5\degree$--$1\degree$) of \lhaaso{}, VERITAS may not be able to achieve detection of this source even with a few times longer exposure time. On the other hand, the spectrum measured by WCDA is inconsistent with both measurements by KM2A presented in \cite{lhaasoj0621} and \cite{lhaasocat} in the energy range common to the two instruments as shown in Figure \ref{fig:veritas}. The WCDA spectrum forms an extremely sharp peak at a few tens of TeV, for which it is difficult to provide a physical explanation. The inconsistency between the WCDA and KM2A spectra and a sharp peak in the overlapping energy range between the WCDA and KM2A are seen in other LHAASO catalog sources, such as LHAASO J2108+5157. This discrepancy may arise from the generic nature of the data analysis for catalog generation in \citet{lhaasocat}. A dedicated WCDA analysis will enable more detailed spectral measurement of \lhaaso{} in the TeV range. Our VERITAS observation provides an alternative constraint on the TeV flux of \lhaaso{} consistent with the KM2A spectrum and physically plausible as shown in \S \ref{sec:modeling}. 

In the X-ray band, thanks to its proximity to Earth ($250^{+120}_{-62}$ pc, \citet{geminga-distance}), complicated sub-structures of the PWN were resolved around the Geminga pulsar. The sub-structures include a bow shock ($\sim8''$), compact central nebula ($\sim2'$), diffuse axisymmetric wings ($\sim3'$) and a tail ($\sim45''$) \citep{geminga-pwn2,geminga-pwn1,geminga-pwn3,Mori2014}. These structures, if they existed in \lhaaso{}, would be too small to be resolved at a distance 6 times larger than that of Geminga (i.e., 1.5 kpc). This is consistent with our \xmm{} observation, where \psr{} was detected as a point source. The contribution from a putative PWN to the point source emission is likely minimal as no significant nonthermal (power-law) component was found from our spectral analysis (\S \ref{subsec:psr_spec}). No extended X-ray halo emission was detected from Geminga \citep{Manconi:2024wlq} or \lhaaso{}. Even if \lhaaso{} is farther away from Earth than Geminga, the X-ray halo is likely much larger than the FoV covered by typical pointing X-ray telescopes as shown in the \textit{left} panel of Figure \ref{fig:sync}. Still, pointing the telescope at the central pulsar will allow the most stringent estimate of flux upper limits from the brightest part of a halo. All-sky survey telescopes such as the eROSITA and Einstein Probe have the advantage of observing these extended halos. However, the sensitivity of their spectral measurement is limited by the shallow exposure (a few 100 s, \citet{erosita1,erosita2}).

Independent from the WCDA spectrum, our VERITAS flux upper limits constrain the electron injection spectral index to $\gamma=1.4$ as shown in Figure \ref{fig:ic_veritas}, a spectrum much harder than that of the Geminga halo found by \cite{HAWC_Halos_and_Positron_Flux,DiMauro:2019yvh} and other works $(\gamma\simgt2)$. The electrons emitting TeV gamma rays have energies of tens to hundreds of TeV \citep{HAWC_Halos_and_Positron_Flux} and a cooling time of only a few kyr. Considering such a short cooling time, a diffusion coefficient 100 times smaller than the Galactic average is necessary to confine such high-energy electrons within the observed source size $\sim30$ pc. On the other hand, electrons emitting gamma rays with energies of tens to hundreds of GeV have a cooling time comparable to the age of the pulsar $\sim100$ kyrs, and hence, the number of low-energy electrons around the pulsar is much greater in case of stronger suppression of diffusion. This leads to an observed GeV flux that is much higher than what is expected from the injection spectrum, as seen from the Geminga or Monogem halo whose IC spectra are most likely peaked at or below GeV energies (e.g., \cite{DiMauro:2019yvh}). This is not the case for , as the \fermi{} upper limits tightly constrain the extent of the region in which such diffusion suppression can be present. The combined effect of a hard injection spectrum and a narrow region of suppressed diffusion creates a peak in the broadband gamma-ray spectrum at 1--10 TeV. Our findings, along with the magnetic field strength $\lesssim1$ $\mu$G obtained from our \xmm{} observation, provide ingredients for studying the formation of magnetohydrodynamic turbulence around a pulsar -- a possible origin of suppressed diffusion.

\subsection{X-ray emission from the pulsar}

We detected an X-ray pulsation around the expected spin period of $P\approx 333.2$ ms. The pulsation was detected only below $\sim2$ keV (with $\sim5\sigma$ significance), although the pulsed fraction was found to be low ($\sim17$\%) with large uncertainties. This is likely because of the limited timing resolution of the EPIC cameras operating in the Full Window mode ($\Delta t = 73.4$~ms), in addition to intrinsically small X-ray pulsed fractions observed from thermally-emitting isolated NS, which are typically $\simlt 20$\% \citep{Bogdanov2024}. Given the X-ray detection of the pulsar, a future \xmm\ observation with the fast timing mode will allow us to characterize the X-ray light curve and align with the gamma-ray ephemeris data better. 

Our \xmm\ spectral analysis confirmed the soft X-ray emission appearing only below $\sim2$ keV. 
We found that the \xmm\ spectra of the pulsar are predominantly thermal as the power-law model fit resulted in an extremely soft photon index ($\Gamma \sim 6$). This contrasts with the Geminga pulsar, where both thermal and nonthermal components have been found by \xmm\ and \nustar\ observations \citep{Halpern1997}. Compared to the nonthermal X-ray emission of the Geminga pulsar extending to $\sim20$ keV \citep{Mori2014}, synchrotron emission from the magnetosphere of \psr\ seems absent or significantly weaker. Applying various phenomenological models composed of BB and PL components, we found an absorbed double blackbody model produced an acceptable fit. In this case, the X-ray spectra are thermal from a large portion of the NS surface ($kT_{\rm c} = 79\pm5$ eV) and hot polar cap ($kT_{\rm h} = 227_{-29}^{+38}$ eV). Similarly, thermal X-ray spectra of the Geminga pulsar are most consistent with a double BB model with lower temperatures $kT_{\rm c} = 44.0\pm0.8$ eV and $kT_{\rm h} = 195\pm14$ eV  \citep{Mori2014}. 
%{\color{red}  KM: There is a relation between $\dot{E}$ and $L_X$ like $L_X \sim 10^{-3} \dot{E}$. Check if the PL model luminosity upper limit gives the X-ray conversion factor of $\sim10^{-3}$ or less. Or more rubustly, the $F_X/F_{\rm GeV}$ flux ratio could be mentioned as it does not depend on the (unknown) source distance. } 
To model the thermal X-ray emission more realistically, we applied the \texttt{nsmax} model with different surface elements and magnetic field strengths. We found that the hydrogen atmosphere model fit the \xmm\ spectra best with $kT = 57\pm4$ eV. Other heavier element atmosphere models failed to reproduce the \xmm\ spectra largely because the expected absorption lines and edges are absent in the featureless X-ray spectra. In contrast, magnetized hydrogen atmosphere models were ruled out for the Geminga pulsar because they yielded emission radii $\simgt 400$ km, too large for an isolated NS, and overpredicted the UV flux \citep{Mori2014}.

Intriguingly, the two middle-aged pulsars, \psr\ and Geminga, exhibit distinct X-ray spectra. Firstly, nonthermal X-ray emission is present in Geminga but not in \psr{}. The BB temperature of the cooler component, supposedly originating from a large fraction of the NS surface, is hotter for \psr. This aligns with our expectations because \psr\ is younger ($\tau = 208$ kyr) than the Geminga pulsar ($\tau = 342$ kyr), assuming their characteristic ages are good approximations of their true ages. The hydrogen atmosphere is preferred for \psr, whereas it was ruled out for the Geminga pulsar. The hydrogen layer remains optically thick on the surface \psr. Although the diffusive nuclear burning should be more effective for the higher surface temperature of \psr, accretion from the ISM may compensate for this effect. In addition, other factors such as the surface magnetic field and underlying chemical composition in the envelope can influence the rate of diffusive nuclear burning \citep{Chang2004, Chang2010}. The \xmm\ detection and characterization of the thermal X-ray emission from \psr\ will be valuable for investigating the evolution of the surface composition and temperature of middle-aged pulsars.

\section{Summary and conclusion} 
\label{sec:summary}

We observed a pulsar halo candidate \lhaaso{} with VERITAS in the TeV gamma-ray band and \xmm{} in the X-ray band. 
To address the challenge of extended source analysis using IACTs with limited FoV, as part of this work, we developed a sophisticated technique to accurately measure the telescope acceptance and estimate the background of the entire FoV. The code implementing this technique is publicly available and can be useful for elucidating the mysterious nature of many extended Galactic PeVatrons discovered by LHAASO and HAWC. 

Our VERITAS and \xmm{} observations resulted in null detection of emission associated with \lhaaso{} within radii of $1\degree$ and $10'$, respectively. We modeled the multiwavelength SED and gamma-ray surface brightness profile of \lhaaso{} as synchrotron and IC emissions from a pulsar halo accounting for electron injection, cooling, and diffusion over the characteristic age of \psr{} (208 kyr) at an assumed distance of 1.6 kpc. Our VERITAS flux upper limits in the 0.3--10 TeV band constrain the electron injection spectral index to $\gamma\sim1.4$ and the diffusion coefficient to $2\times10^{25}$ cm$^2$ s$^{-1}$. This diffusion coefficient is similar to that of Geminga, and lower than the Galactic average by two orders of magnitude.
%The electron injection spectrum is much harder than that of Geminga ($\gamma\simgt2$).
Moreover, our VERITAS flux upper limits indicate a break in the gamma-ray spectrum at 1--10 TeV. Utilizing additional flux upper limits from the \fermi{}, this spectral break constrains the diffusion suppression region to $\sim30$ pc around the pulsar. Our \xmm{} flux upper limit in the 2--7 keV band ($5.2\times10^{-14}$ erg cm$^{2}$ s$^{-1}$) constrains the magnetic field strength to $\lesssim1$ $\mu$G. Our findings of the electron injection spectrum, diffusion coefficient, diffusion suppression length, and magnetic field strength can be used to study the formation of magnetohydrodynamic turbulence around the pulsar as the origin of suppressed diffusion. Even though \lhaaso{} is likely at a much larger distance than Geminga, the large extent and low surface brightness of the halo are major challenges of observing this source with the current-generation IACTs and pointing X-ray telescopes. The Cherenkov Telescope Array Observatory (CTAO), the next-generation high-sensitivity and wide-field IACT, and X-ray survey telescopes with large FoV, such as the eROSITA and the Einstein Probe, are expected to play a significant role in studying pulsar halos. 

In addition, we detected \psr{} in the X-ray band for the first time with our \xmm{} observation. The X-ray pulsation is detected at the gamma-ray pulse period, but the limited timing resolution and exposure of our observation hinder phase-resolved spectroscopy. Future observation with a higher timing resolution and longer exposure will enable a more detailed study of \psr{}. Nevertheless, we were able to perform phase-averaged spectral analysis and found that \psr{} likely has a hydrogen atmosphere.

%% Also note that the akcnowlodgment environment does not support long amounts of text. If you have a lot of people and institutions to acknowledge, do not use this command. Instead, create a new \section{Acknowledgments}.
\begin{acknowledgments}
This research is supported by grants from the U.S. Department of Energy Office of Science, the U.S. National Science Foundation and the Smithsonian Institution, by NSERC in Canada, and by the Helmholtz Association in Germany. This research used resources provided by the Open Science Grid, which is supported by the National Science Foundation and the U.S. Department of Energy's Office of Science, and resources of the National Energy Research Scientific Computing Center (NERSC), a U.S. Department of Energy Office of Science User Facility operated under Contract No. DE-AC02-05CH11231. We acknowledge the excellent work of the technical support staff at the Fred Lawrence Whipple Observatory and at the collaborating institutions in the construction and operation of the instrument.

Jooyun Woo and Kaya Mori are partially supported by NASA and ESA through the \xmm\ AO-22 observation program (XMMNC22). Samar Safi-Harb acknowledges support from the Natural Sciences and Engineering Research Council of Canada (NSERC) through the Canada Research Chairs and the Discovery Grants programs, and from the Canadian Institute for Theoretical Astrophysics and the Canadian Space Agency. Silvia Manconi acknowledges the European Union's Horizon Europe research and innovation program for support under the Marie Sklodowska-Curie Action HE MSCA PF–2021,  grant agreement No.10106280, project \textit{VerSi}.
The work of Fiorenza Donato is supported by the 
Research grant {\sl The Dark Universe: A Synergic Multimessenger Approach}, No.
2017X7X85K funded by the {\sc Miur}. Fiorenza Donato and Mattia Di Mauro acknowledge support from the research grant {\sl TAsP (Theoretical Astroparticle Physics)} funded by Istituto Nazionale di Fisica Nucleare (INFN). 
\end{acknowledgments}


\vspace{5mm}
\facilities{VERITAS,  \xmm{} (EPIC)}


\software{Eventdisplay \citep{Eventdisplay,Eventdisplay_v490p2}, Gammapy \citep{gammapy,gammapy_v1.1}, SAS, Xspec \citep{xspec}, Fermipy \citep{fermipy}, Astropy \citep{astropy:2013,astropy:2018,astropy:2022}, Numpy \citep{numpy}, Matplotlib \citep{matplotlib}}



\appendix

\section{3D acceptance map and field-of-view technique in Gammapy} \label{app:gammapy}

Observing nearby pulsar halos with IACTs enables high-resolution ($<0.1\degree$) morphology studies, a unique advantage for studying CR transport mechanisms. However, such observations bear challenges in background estimation due to the IACTs' limited FoV (diameter of $3.5\degree$ for VERITAS) and nearby halos' source extensions over $1\degree$. We developed a technique to accurately estimate the telescope acceptance map of the entire FoV. The acceptance map reflects the spatial and energy dependence (``3D acceptance") of gamma-like CR background, the dominant background component for IACT observations originating from different observing conditions and CR shower properties. The acceptance map can be scaled utilizing a minimal available source-free region in the FoV to generate a background for the entire FoV. This technique is useful for any IACT data analyses of extended sources. The code was developed based on Gammapy v1.1 and is available at \url{https://github.com/VERITAS-Observatory/gammapy-fov.git}.

\subsection{Generating a 3D acceptance map} \label{appsec:acceptance}

A 3D acceptance map has two spatial axes (detector X and detector Y) and one energy axis. In this map, each spatial and energy bin ($0.1^{\circ}\times0.1^{\circ}$ six logarithmic bins in 0.3--10 TeV for this work) contains an acceptance value. The size of this map is $3.5^{\circ}\times3.5^{\circ}$; that is, the map covers the range of detector X coordinates [$-1.75^{\circ}$, $1.75^{\circ}$] and detector Y coordinates [$-1.75^{\circ}$, $1.75^{\circ}$]. A single 3D acceptance map is used for all the observing runs of \lhaaso{} (``on runs"). The steps for acceptance map generation are summarized below and elaborated upon in the following text.

\begin{enumerate}
    \item Select ``off runs," observations with either no gamma-ray source or a point-like gamma-ray source within the FoV, taken under the observing conditions similar to those of the on runs. 
    \item For each off run, excise any gamma-ray sources and stars in the FoV and patch up the excised regions with reflected regions within the same FoV.
    \item Create a stacked count map by adding up the count maps of the off runs.
    \item Create a stacked exposure map (in units of TeV s sr) by adding up the exposure map (in units of s) of the off runs and multiplying it by the volume of each spatial and energy bin (in the unit of TeV sr).
    \item Calculate the acceptance (gamma-like event rate) in each spatial and energy bin (in units of TeV$^{-1}$ s$^{-1}$ sr$^{-1}$) by dividing the stacked counts map by the stacked exposure map.
\end{enumerate}

\begin{figure}[t!]
\centering
\includegraphics[width=\textwidth]{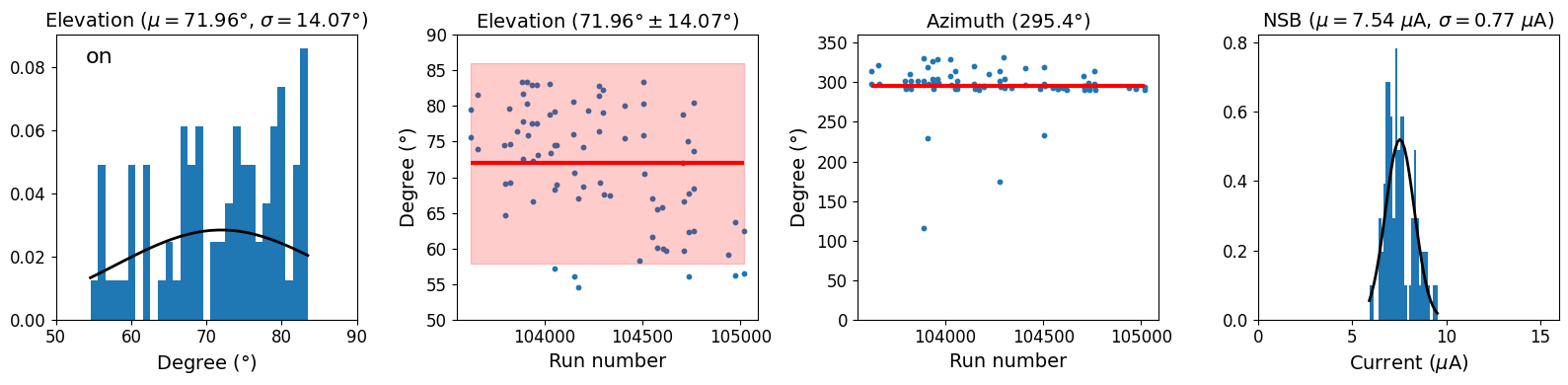}
\includegraphics[width=\textwidth]{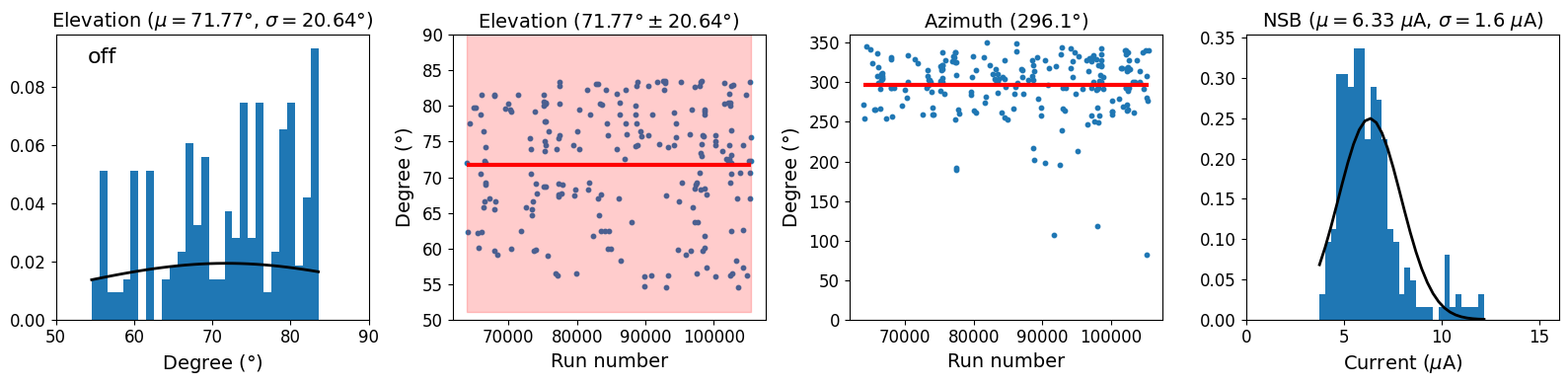}
\caption{Observing conditions of the on runs (top row, labeled ``on" ) and off runs for the 3D acceptance map (bottom row, labeled ``off"). On each row, from the left, the \textbf{first plot} is a histogram of elevations (blue bar) and an exposure-weighted Gaussian fit to the histogram (black curve). The mean (mu) and standard deviation (std) of the fit are provided in the figure title. The \textbf{second plot} is a scatter plot of elevations (blue dot). The mean and (mean$\pm1\sigma$) are marked as a red line and a red-shaded region. The \textbf{third plot} is a scatter plot of azimuths (blue dot) and the exposure-weighted mean of azimuths (red line). The \textbf{last plot} is a histogram of a night sky background (NSB) averaged over a run duration (blue bar). An exposure-weighted Gaussian fit to the histogram is overlaid as a black curve, and the mean ($\mu$) and the standard deviation ($\sigma$) of the fit are provided in the figure title.}
\label{fig:condition1}
\end{figure}

Selecting the off runs that closely match the on runs' observing conditions is crucial for accurate telescope acceptance estimation. We create a pool of extragalactic (galactic latitude $b>10\degree$) observing runs with a duration of at least 10 minutes under good sky condition. The pool only includes the runs taken after August 2012, before which the hardware conditions were different from those of the \lhaaso{} observations \citep{Kieda2011:pmt-upgrade,Otte2011:pmt-upgrade}. Elevation and azimuth of an observing run are among the most important factors for the spatial dependence of telescope acceptance. For each on run, we narrow down the pool to the observing runs taken at the azimuth within $45\degree$ of the on-run azimuth. Out of this reduced pool, off runs taken at the elevation closest to the on-run elevation are selected until the total duration of the off runs reaches two times the on-run duration. Selected off runs are used only once for a particular on run, and are not used again for the rest of the analysis. The observing conditions of the on runs and the off runs selected for 3D acceptance map generation are consistent with each other, as shown in Figure \ref{fig:condition1}. The discrepancy in the current distributions likely originates from the proximity to the Galactic Plane and the presence of a bright point-like source and stars in the FoV. The off runs have much higher galactic latitudes on average ($b>40\degree$) than \lhaaso{} ($b=10.95\degree$), and hence the lower average current. On the other hand, the presence of bright stars in the off runs may create the high-current tail.

\begin{figure}[t!]
\centering
\includegraphics[width=\textwidth]{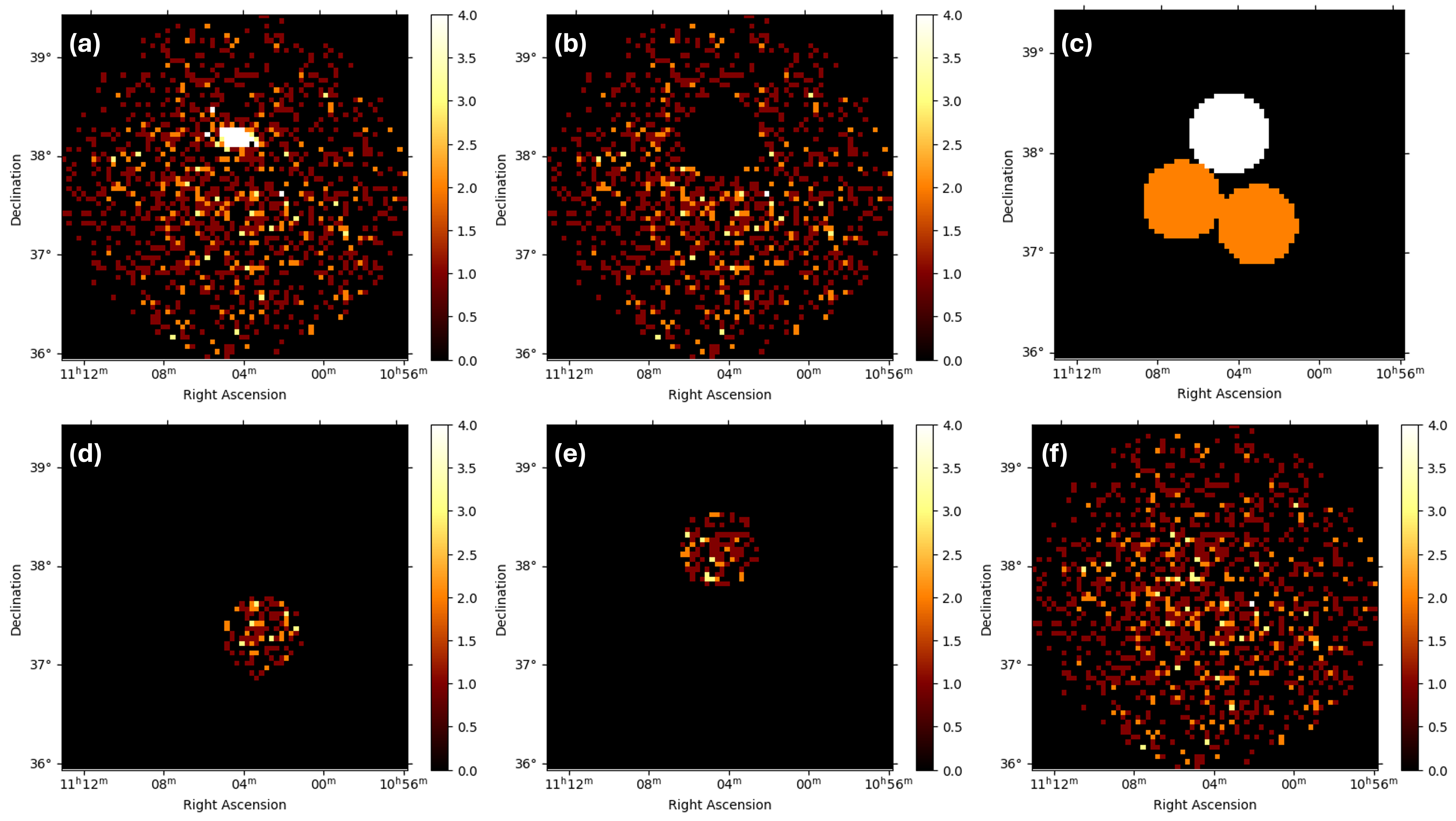}
\caption{Demonstration of excising a point-like gamma-ray source and filling the excised region with the events from a reflected region. All the sky maps are $3.5^{\circ}\times3.5^{\circ}$, and the bin size is $0.05^{\circ}\times0.05^{\circ}$. (c) is a mask map (black is zero, white and orange are one), and all the other maps are count maps whose color bars show the number of counts in each bin. For the counts map, all the gamma-like events were plotted (lowest energy 0.1 TeV, highest energy 9.8 TeV). (a) The bright gamma-ray source in the FoV is Markarian 421. The observation was taken at the $0.5^{\circ}$ south wobble. (b) Events from a circle with radius $0.4^{\circ}$ centered at the position of Markarian 421 are excised. (c) The white circle is the excised region, and the orange circles are the reflected regions. All three regions are $0.5^{\circ}$ offset from the center of the FoV. (d) Events from one of the reflected regions are plotted. (e) The events plotted in (d) are copied, and their coordinates are rotated around the center of the FoV such that the events are located at the excised region. (f) The events from (b) and (e) are added to create a blank sky map.}
\label{fig:excise}
\end{figure}

Point-like gamma-ray sources or stars in the FoV of the off runs are excised by a circle with $0.4^{\circ}$ radius. The excised region is filled with the events from a region with the same shape, size, and offset from the center of the FoV (``reflected region"). A Gammapy class \texttt{ReflectedRegionsFinder} is used to find such a region. The coordinates of the events within the reflected region are rotated around the center of the FoV so that those events fill the excised region. The impact of this procedure in the accurate estimation of the acceptance spatial distribution is minimal as only $\sim10\%$ of the off runs contain a point-like gamma-ray source at one of four locations: $0.5\degree$ offset to the east, west, south, and north of the center of the FoV. Figure \ref{fig:excise} demonstrates this procedure using an observation of Markarian 421 taken at the $0.5^{\circ}$ south wobble. This procedure is currently necessary for a technical reason, but an alternative way of leaving the excised region empty could be made possible in principle. 

\begin{figure}[t!]
\centering
\includegraphics[width=\textwidth]{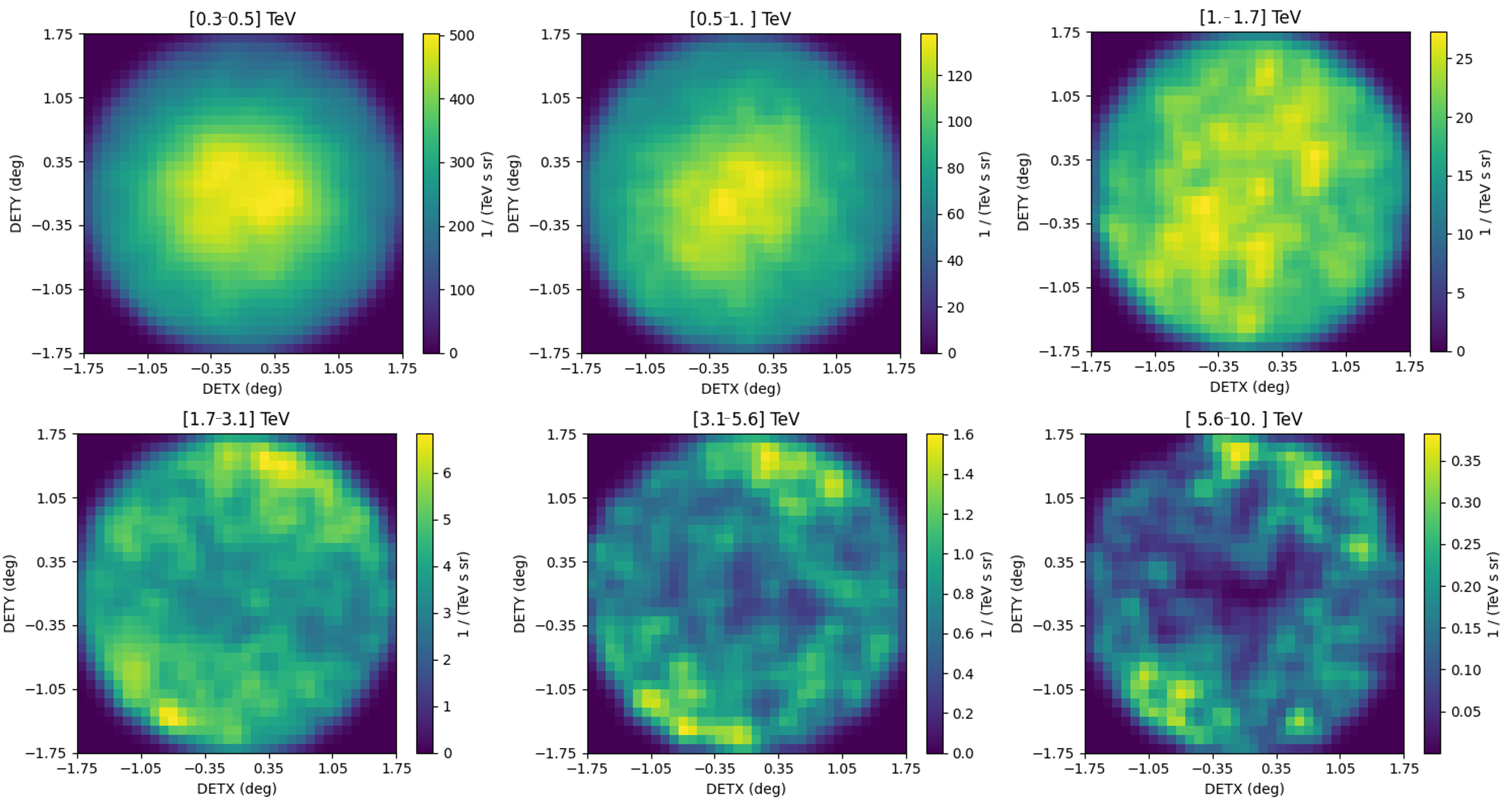}
\caption{3D acceptance map in the unit of TeV$^{-1}$ s$^{-1}$ sr$^{-1}$. Each image is a 2D (detector X and detector Y) acceptance map in each energy bin, as labeled in the figure titles.}
\label{fig:acceptance}
\end{figure}

Once the off runs are selected for all the on runs and sources are excised, the 3D acceptance map is calculated. After replacing zero acceptances with small definite values, the acceptance map is smoothed using a 2D Gaussian kernel with $\sigma=1$ bin $=0.1\degree$. Figure \ref{fig:acceptance} shows the resulting 3D acceptance map used for the analysis of the on runs. The azimuthal asymmetry of the acceptance is the outcome of the varying atmospheric depth along the line of sight (related to the elevation) and the Earth's magnetic field strength (related to the azimuth) throughout the FoV. The optical axis of the telescope, i.e., the center of the FoV, has the highest acceptance in the lower energies, as expected, but has the lowest acceptance in the higher energies due to the loss of more elongated, and hence truncated, shower images. 

\subsection{3D acceptance map validation and bias correction}

\begin{figure}[t!]
\centering
\includegraphics[width=\textwidth]{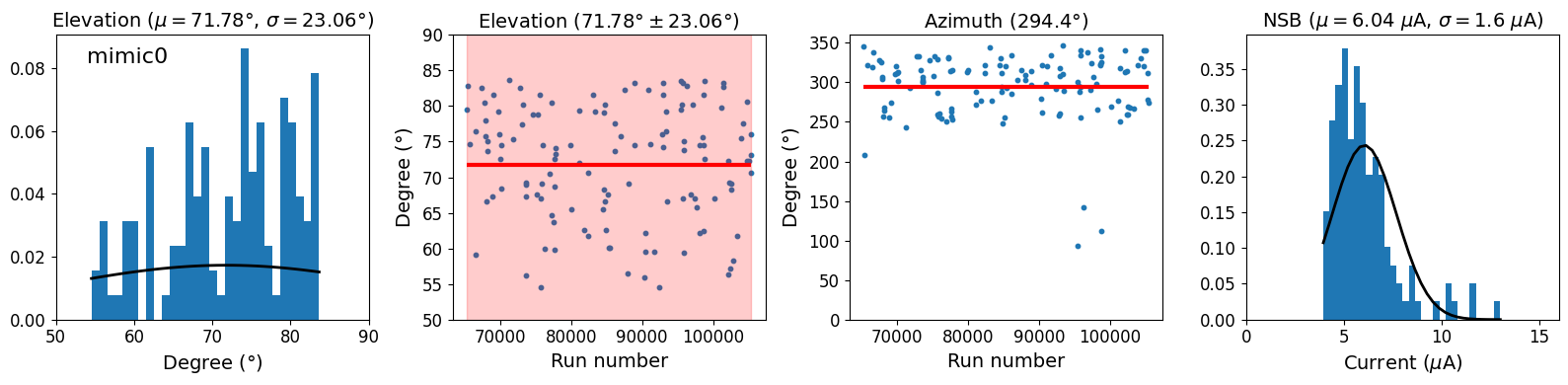}
\includegraphics[width=\textwidth]{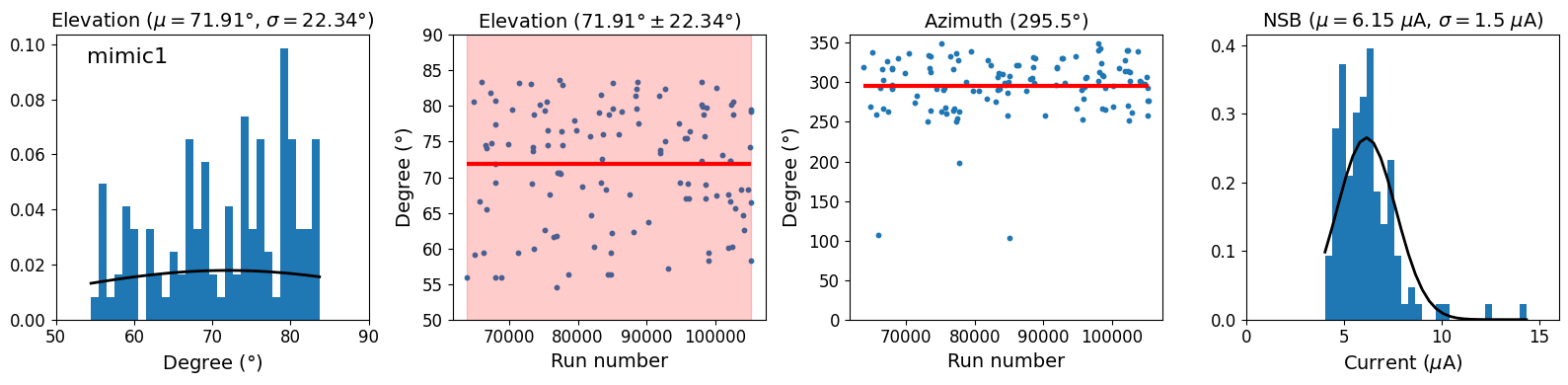}
\includegraphics[width=\textwidth]{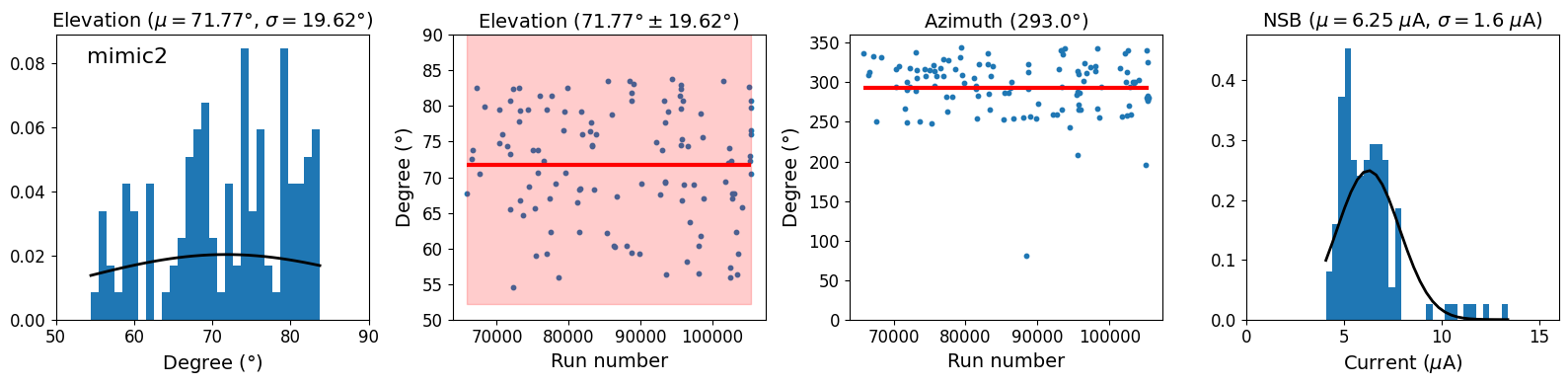}
\caption{The same plots as Figure \ref{fig:condition1} for the off runs for three mimic datasets (``mimic0--2").}
\label{fig:condition2}
\end{figure}

\begin{figure}[ht!]
\centering
\includegraphics[width=\textwidth]{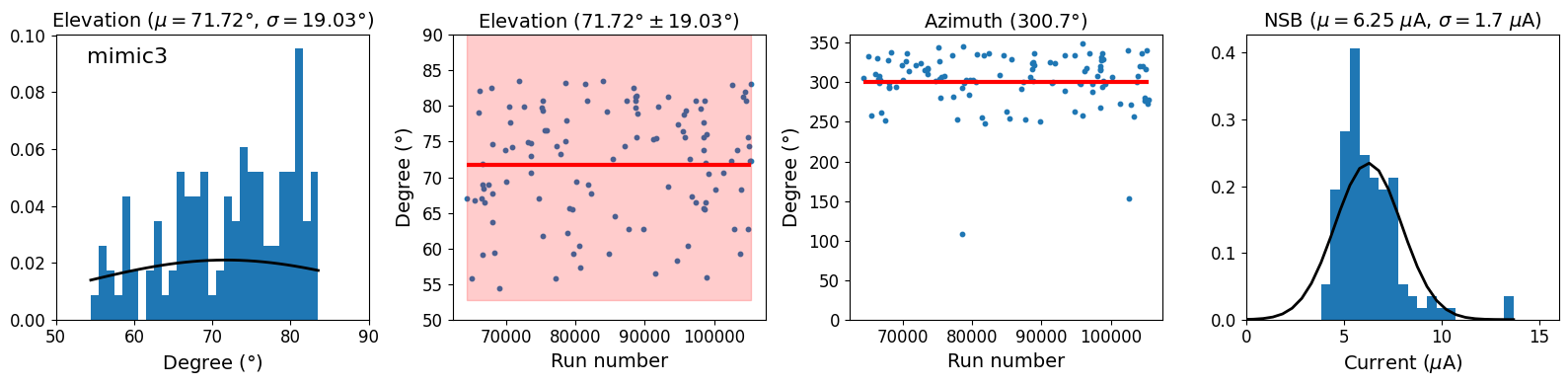}
\includegraphics[width=\textwidth]{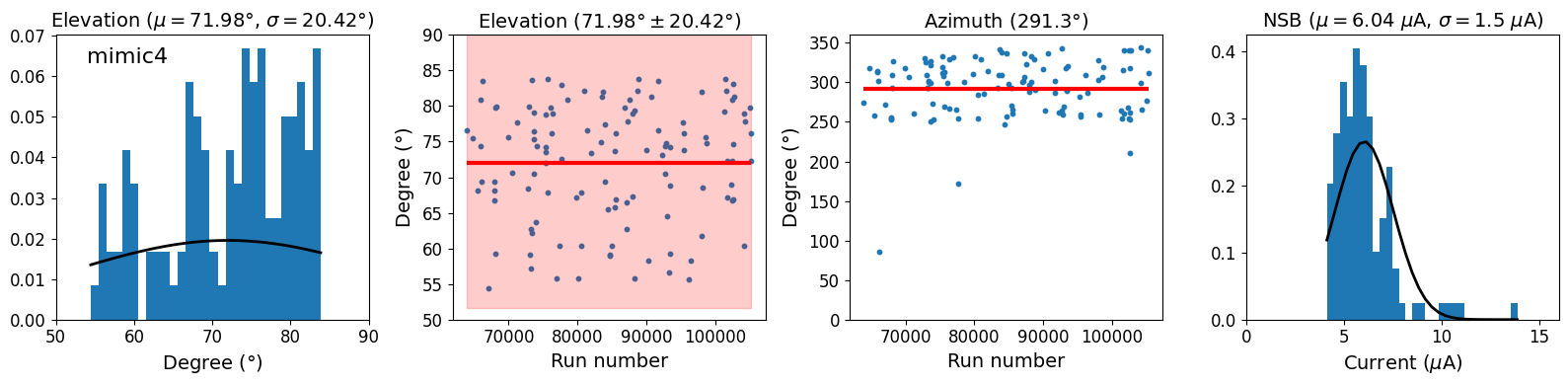}
\includegraphics[width=\textwidth]{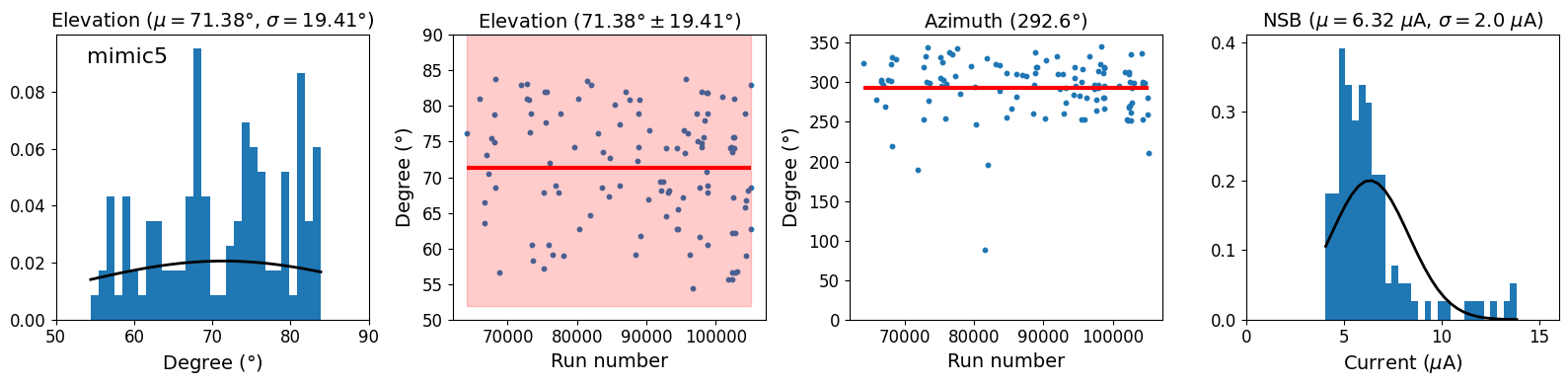}
\caption{The same plots as Figure \ref{fig:condition1} for the off runs for three mimic datasets (``mimic3--5").}
\label{fig:condition3}
\end{figure}

The 3D acceptance map is validated by comparing five mimic datasets with their backgrounds estimated using the 3D acceptance map and the FoV technique. A mimic dataset is constructed out of the remaining pool of off runs mentioned in \S \ref{appsec:acceptance} to mimic the on runs in terms of observing conditions and duration. The off runs are selected by the same criteria as those for the off runs used for acceptance map generation. Once an off run is selected for a mimic dataset, it is removed from the pool of off runs. Figures \ref{fig:condition2} and \ref{fig:condition3} show that the observing conditions of the five mimic datasets are consistent with each other and with the on runs (with the off runs for the current). Figure \ref{fig:condition3} also shows the observing conditions of the sixth mimic dataset used for validation of bias correction in the acceptance map, as described later in this section.

\begin{figure}[h!]
\centering
\includegraphics[width=0.31\textwidth]{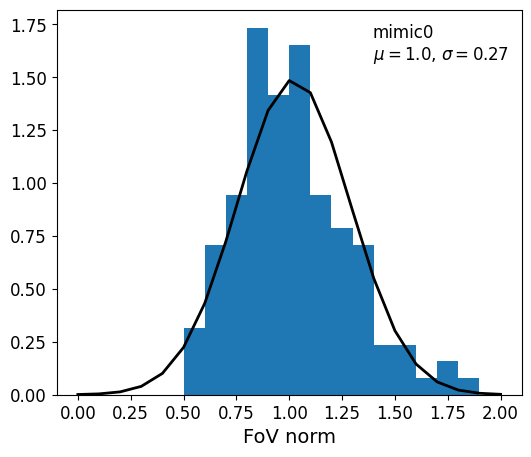}
\includegraphics[width=0.68\textwidth]{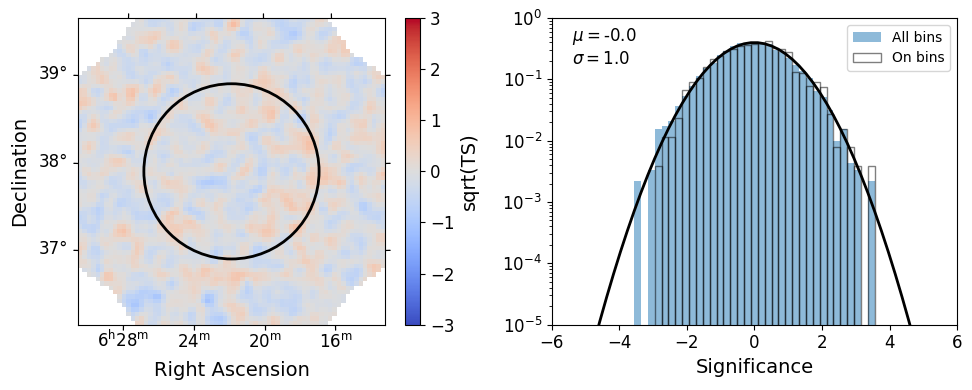}
\includegraphics[width=0.31\textwidth]{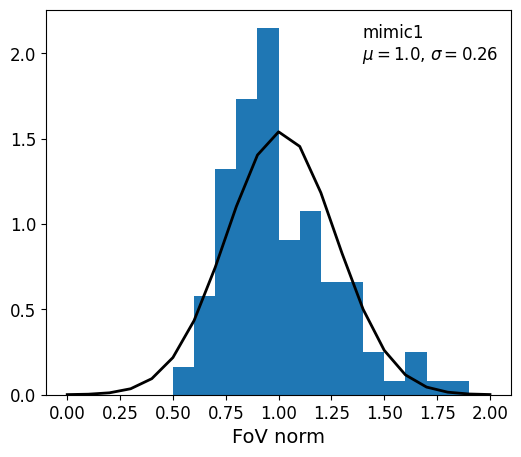}
\includegraphics[width=0.68\textwidth]{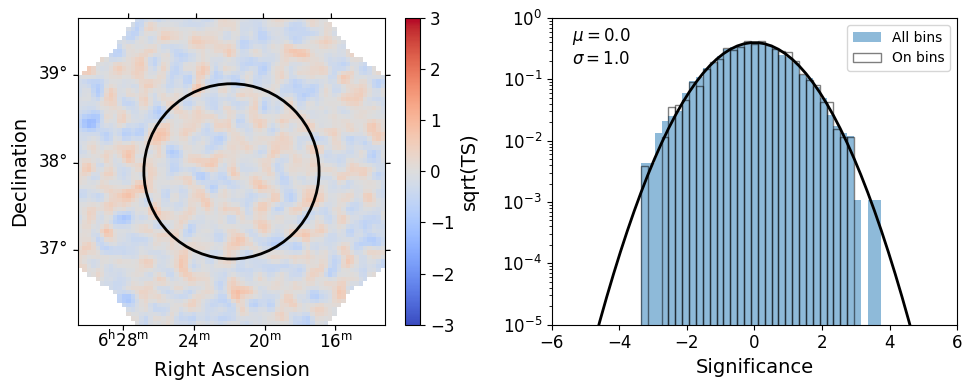}
\includegraphics[width=0.31\textwidth]{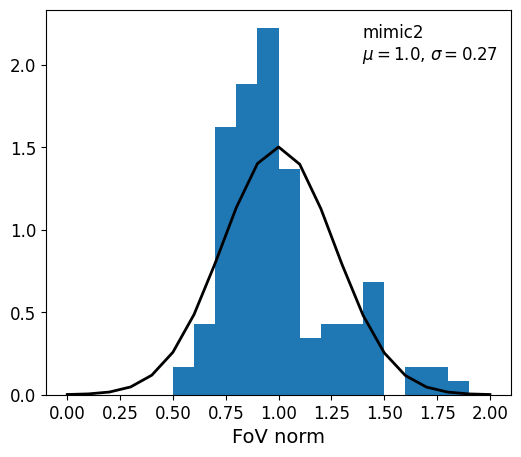}
\includegraphics[width=0.68\textwidth]{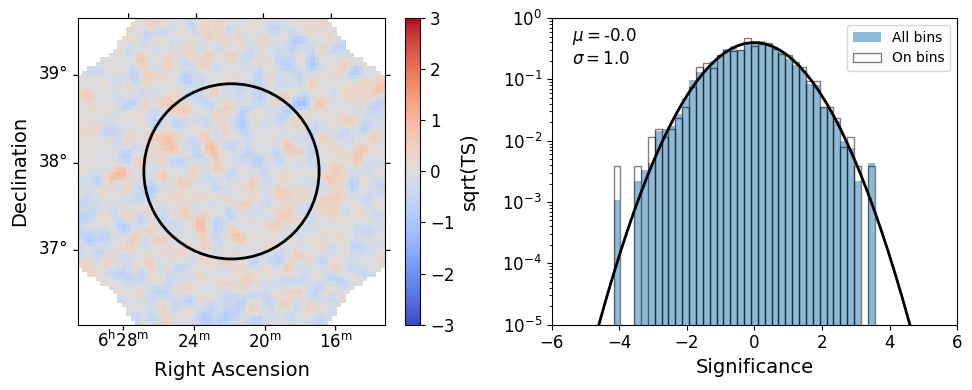}
\caption{Mimic datasets 0--2 on each row: the \textbf{first plot} is the distribution of FoV norms of each off run (blue bar) and a Gaussian fit to the distribution (black curve). The numbers on the top right corner are the mean $\pm$ standard deviation of the fit. The \textbf{second plot} is the significance map, and the \textbf{third plot} is the significance distribution from all (FoV) and on (source extraction region) regions as well as a Gaussian fit to the significances from the on region. The mean (mu) and standard deviation (std) of the fit are labeled on the left top corner of the plot.}
\label{fig:mimic1}
\end{figure}

\begin{figure}[t!]
\centering
\includegraphics[width=0.31\textwidth]{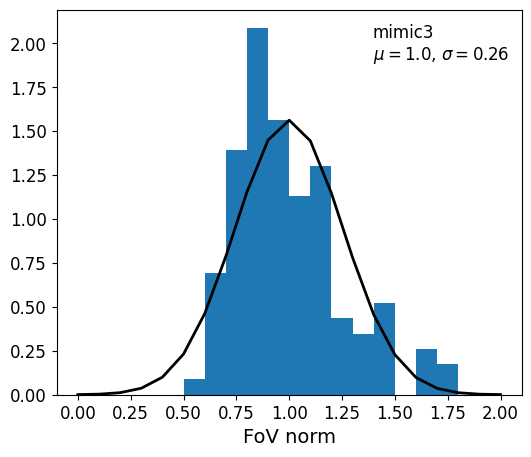}
\includegraphics[width=0.68\textwidth]{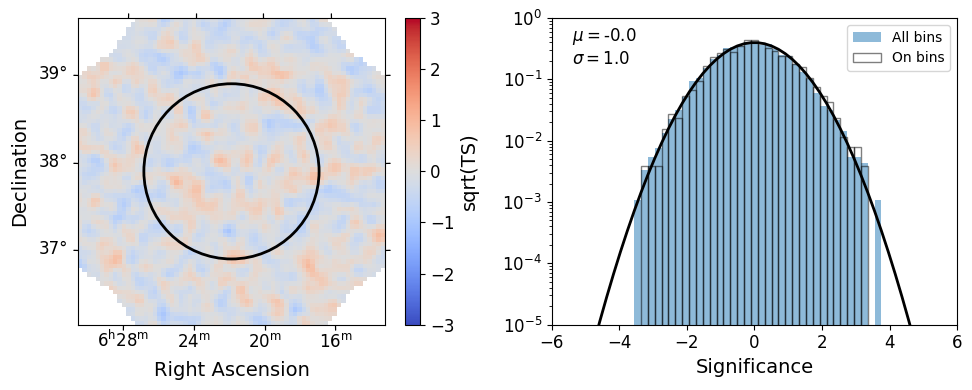}
\includegraphics[width=0.31\textwidth]{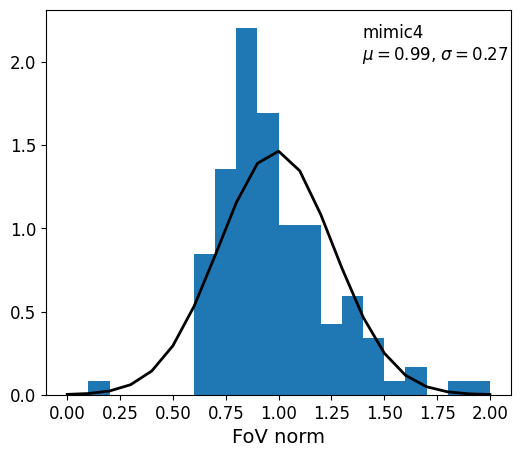}
\includegraphics[width=0.68\textwidth]{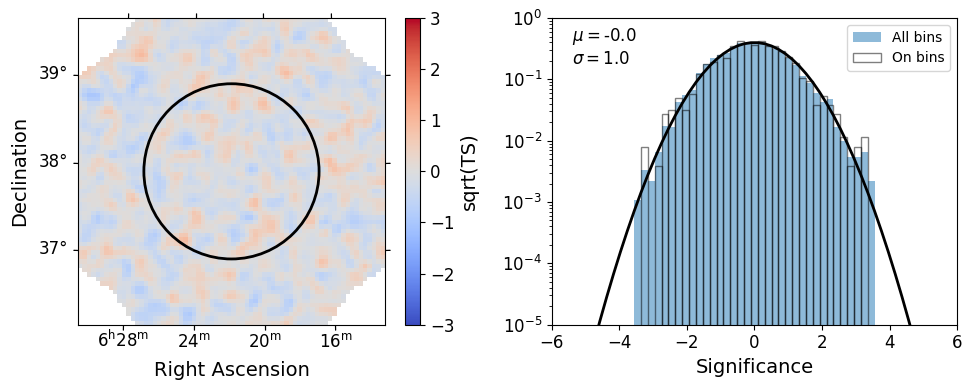}
\includegraphics[width=0.31\textwidth]{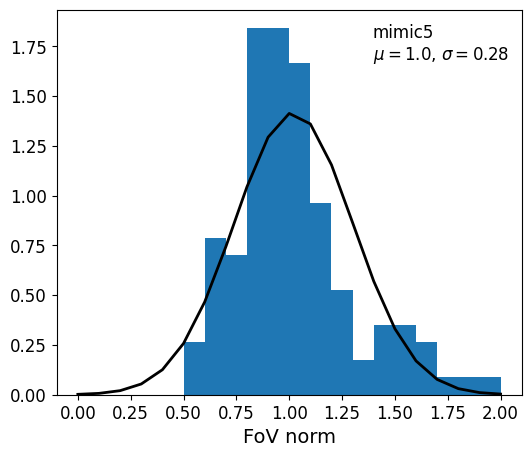}
\includegraphics[width=0.68\textwidth]{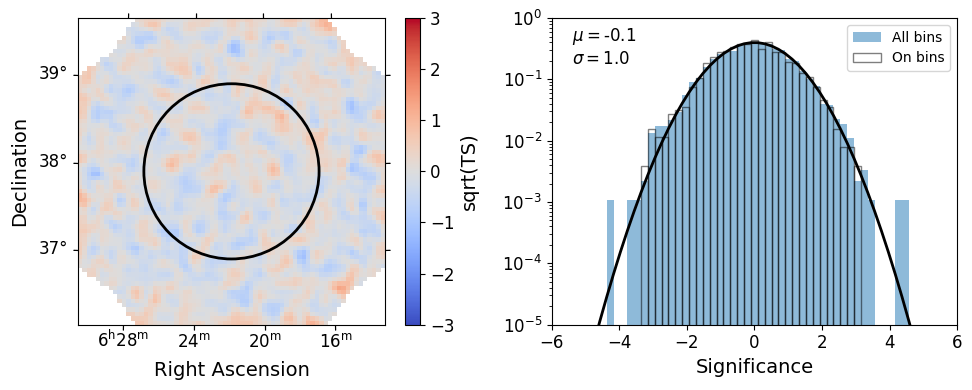}
\caption{The same plots as Figure \ref{fig:mimic1} for mimic datasets 3--5. Note that the significance map and distribution of mimic dataset 5 are after bias correction.}
\label{fig:mimic2}
\end{figure}

Any point-like gamma-ray sources or stars in the off runs are excised. The analysis of the mimic datasets is performed exactly as the on runs would be analyzed. Figures \ref{fig:mimic1} and \ref{fig:mimic2} show the analysis results of the mimic datasets. The distribution of the factors by which the acceptance map was scaled to estimate the background of the off runs (``FoV norm") is centered at 1 with a small spread, indicating the acceptance map is already a good approximation of the off runs. The sky maps of the mimic datasets show no noticeable bias, and the distribution of significance in each bin is consistent with that of statistical fluctuations. 

\begin{figure}[t!]
\centering
\includegraphics[width=\textwidth]{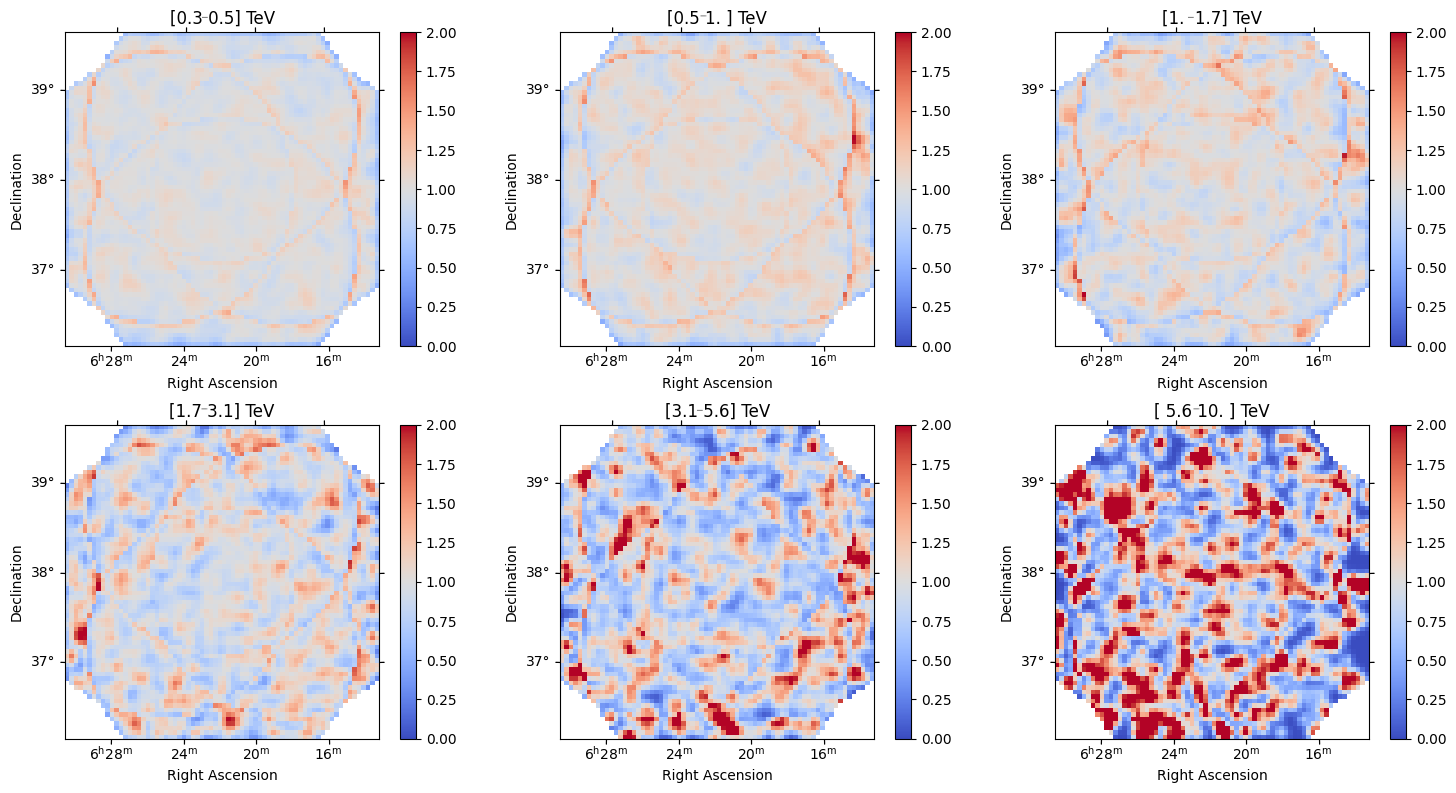}
\caption{Bias maps for the six energy bins as labeled in the figure titles.}
\label{fig:bias}
\end{figure}

Nevertheless, we calculate the bias in the acceptance map as a ratio of the observed background counts (stacked off runs) to the estimated background averaged over the five mimic datasets in each spatial and energy bin. The bias map is smoothed using a 2D Gaussian kernel with $\sigma=1$ bin. Figure \ref{fig:bias} shows the resulting 3D bias map. Biases are observed mainly along the FoV edge of each run where the exposure is small and the uncertainty on the counts is high. The bias map becomes patchy in the highest energy bins due to the low counts. 

\begin{figure}[t!]
\centering
\includegraphics[width=0.31\textwidth]{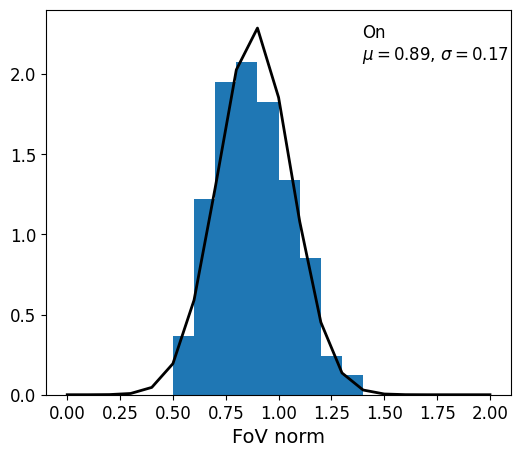}
\includegraphics[width=0.68\textwidth]{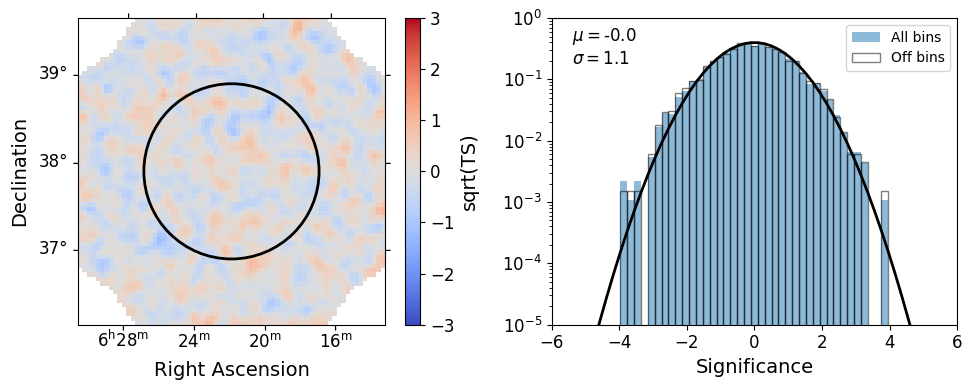}
\caption{The same \textbf{first} and \textbf{second} plots as Figure \ref{fig:mimic1} for on runs. The \textbf{third} plot is the significance distribution from all (Fov) and off (FoV minus source extraction region) regions as well as a Gaussian fit to the significances from the off region. Note that the significance map and distribution are after bias correction.}
\label{fig:on}
\end{figure}

We first validate the bias map by analyzing the last mimic dataset and applying bias correction to the estimated background. This last mimic dataset is analyzed in the same way as the previous five mimic datasets. Figure \ref{fig:mimic2} shows that the estimated background after bias correction is consistent with the observed counts of this last mimic dataset. Finally, the on runs are analyzed using the 3D acceptance map, 3D bias map, and the FoV technique as shown in Figure \ref{fig:on}. The significance distribution is consistent with the distribution of statistical fluctuations. The FoV norm is smaller than one due to the different proximity to the Galactic Plane between the on runs and the off runs used for the acceptance map generation.

\clearpage

\bibliography{main}{}

\begin{thebibliography}{}
\expandafter\ifx\csname natexlab\endcsname\relax\def\natexlab#1{#1}\fi
\providecommand{\url}[1]{\href{#1}{#1}}
\providecommand{\dodoi}[1]{doi:~\href{http://doi.org/#1}{\nolinkurl{#1}}}
\providecommand{\doeprint}[1]{\href{http://ascl.net/#1}{\nolinkurl{http://ascl.net/#1}}}
\providecommand{\doarXiv}[1]{\href{https://arxiv.org/abs/#1}{\nolinkurl{https://arxiv.org/abs/#1}}}

\bibitem[{{Abdo} {et~al.}(2009){Abdo}, {Allen}, {Aune}, {Berley}, {Chen}, {Christopher}, {DeYoung}, {Dingus}, {Ellsworth}, {Gonzalez}, {Goodman}, {Hays}, {Hoffman}, {H{\"u}ntemeyer}, {Kolterman}, {Linnemann}, {McEnery}, {Morgan}, {Mincer}, {Nemethy}, {Pretz}, {Ryan}, {Saz Parkinson}, {Shoup}, {Sinnis}, {Smith}, {Vasileiou}, {Walker}, {Williams}, \& {Yodh}}]{MilagroGalacticSources}
{Abdo}, A.~A., {Allen}, B.~T., {Aune}, T., {et~al.} 2009, \apjl, 700, L127, \dodoi{10.1088/0004-637X/700/2/L127}

\bibitem[{{Abeysekara} {et~al.}(2017){Abeysekara}, {Albert}, {Alfaro}, {Alvarez}, {{\'A}lvarez}, {Arceo}, {Arteaga-Vel{\'a}zquez}, {Avila Rojas}, {Ayala Solares}, {Barber}, {Bautista-Elivar}, {Becerril}, {Belmont-Moreno}, {BenZvi}, {Berley}, {Bernal}, {Braun}, {Brisbois}, {Caballero-Mora}, {Capistr{\'a}n}, {Carrami{\~n}ana}, {Casanova}, {Castillo}, {Cotti}, {Cotzomi}, {Couti{\~n}o de Le{\'o}n}, {De Le{\'o}n}, {De la Fuente}, {Dingus}, {DuVernois}, {D{\'\i}az-V{\'e}lez}, {Ellsworth}, {Engel}, {Enr{\'\i}quez-Rivera}, {Fiorino}, {Fraija}, {Garc{\'\i}a-Gonz{\'a}lez}, {Garfias}, {Gerhardt}, {Gonz{\'a}lez Mu{\~n}oz}, {Gonz{\'a}lez}, {Goodman}, {Hampel-Arias}, {Harding}, {Hern{\'a}ndez}, {Hern{\'a}ndez-Almada}, {Hinton}, {Hona}, {Hui}, {H{\"u}ntemeyer}, {Iriarte}, {Jardin-Blicq}, {Joshi}, {Kaufmann}, {Kieda}, {Lara}, {Lauer}, {Lee}, {Lennarz}, {Vargas}, {Linnemann}, {Longinotti}, {Luis Raya}, {Luna-Garc{\'\i}a}, {L{\'o}pez-Coto}, {Malone}, {Marinelli}, {Martinez}, {Martinez-Castellanos}, {Mart{\'\i}nez-Castro},
  {Mart{\'\i}nez-Huerta}, {Matthews}, {Miranda-Romagnoli}, {Moreno}, {Mostaf{\'a}}, {Nellen}, {Newbold}, {Nisa}, {Noriega-Papaqui}, {Pelayo}, {Pretz}, {P{\'e}rez-P{\'e}rez}, {Ren}, {Rho}, {Rivi{\`e}re}, {Rosa-Gonz{\'a}lez}, {Rosenberg}, {Ruiz-Velasco}, {Salazar}, {Salesa Greus}, {Sandoval}, {Schneider}, {Schoorlemmer}, {Sinnis}, {Smith}, {Springer}, {Surajbali}, {Taboada}, {Tibolla}, {Tollefson}, {Torres}, {Ukwatta}, {Vianello}, {Weisgarber}, {Westerhoff}, {Wisher}, {Wood}, {Yapici}, {Yodh}, {Younk}, {Zepeda}, {Zhou}, {Guo}, {Hahn}, {Li}, \& {Zhang}}]{HAWC_Halos_and_Positron_Flux}
{Abeysekara}, A.~U., {Albert}, A., {Alfaro}, R., {et~al.} 2017, Science, 358, 911, \dodoi{10.1126/science.aan4880}

\bibitem[{{Acharyya} {et~al.}(2024){Acharyya}, {Adams}, {Bangale}, {Bartkoske}, {Benbow}, {Buckley}, {Christiansen}, {Chromey}, {Duerr}, {Errando}, {Falcone}, {Feng}, {Foote}, {Fortson}, {Furniss}, {Hanlon}, {Hanna}, {Hervet}, {Hinrichs}, {Holder}, {Humensky}, {Jin}, {Kaaret}, {Kertzman}, {Kieda}, {Kleiner}, {Korzoun}, {Kumar}, {Lang}, {Lundy}, {Maier}, {McGrath}, {Millard}, {Millis}, {Mooney}, {Moriarty}, {Mukherjee}, {Ning}, {Ong}, {Park}, {Pohl}, {Pueschel}, {Quinn}, {Rabinowitz}, {Ragan}, {Ribeiro}, {Roache}, {Ryan}, {Sadeh}, {Saha}, {Sembroski}, {Shang}, {Splettstoesser}, {Talluri}, {Tucci}, {Valverde}, {Vassiliev}, {Weinstein}, {Williams}, {Wong}, {Woo}, {VERITAS Collaboration}, {Alfaro}, {Alvarez}, {Arteaga-Vel{\'a}zquez}, {Avila Rojas}, {Babu}, {Belmont-Moreno}, {Bernal}, {Caballero-Mora}, {Capistr{\'a}n}, {Carrami{\~n}ana}, {Casanova}, {Cotzomi}, {Couti{\~n}o de Le{\'o}n}, {de La Fuente}, {Depaoli}, {di Lalla}, {Diaz Hernandez}, {Duvernois}, {Espinoza}, {Fan}, {Fang}, {Fraija},
  {Garc{\'\i}a-Gonz{\'a}lez}, {Garfias}, {Gonz{\'a}lez}, {Goodman}, {Groetsch}, {Hern{\'a}ndez-Cadena}, {Hinton}, {Huang}, {Hueyotl-Zahuantitla}, {Iriarte}, {Kaufmann}, {Kieda}, {Lee}, {Le{\'o}n Vargas}, {Longinotti}, {Luis-Raya}, {Malone}, {Mart{\'\i}nez-Castro}, {Matthews}, {Miranda-Romagnoli}, {Morales-Soto}, {Moreno}, {Mostaf{\'a}}, {Nellen}, {P{\'e}rez-P{\'e}rez}, {Rho}, {Rosa-Gonz{\'a}lez}, {Salazar}, {Sandoval}, {Schneider}, {Serna-Franco}, {Son}, {Springer}, {Tibolla}, {Tollefson}, {Torres}, {Torres-Escobedo}, {Turner}, {Ure{\~n}a-Mena}, {Varela}, {Wang}, {Zhou}, {HAWC Collaboration}, {Eagle}, {Kumar}, \& {Fermi-Lat Collaboration}}]{matrix}
{Acharyya}, A., {Adams}, C.~B., {Bangale}, P., {et~al.} 2024, \apj, 974, 61, \dodoi{10.3847/1538-4357/ad698d}

\bibitem[{{Adams} {et~al.}(2022){Adams}, {Benbow}, {Brill}, {Buckley}, {Christiansen}, {Falcone}, {Feng}, {Finley}, {Foote}, {Fortson}, {Furniss}, {Giuri}, {Hanna}, {Hassan}, {Hervet}, {Holder}, {Hona}, {Humensky}, {Jin}, {Kaaret}, {Kleiner}, {Kumar}, {Lang}, {Lundy}, {Maier}, {Moriarty}, {Mukherjee}, {Nievas Rosillo}, {O'Brien}, {Park}, {Patel}, {Pfrang}, {Pohl}, {Prado}, {Pueschel}, {Quinn}, {Ragan}, {Reynolds}, {Ribeiro}, {Roache}, {Ryan}, {Santander}, {Weinstein}, {Williams}, \& {Williamson}}]{veritas-sensitivity}
{Adams}, C.~B., {Benbow}, W., {Brill}, A., {et~al.} 2022, \aap, 658, A83, \dodoi{10.1051/0004-6361/202142275}

\bibitem[{Aguasca-Cabot {et~al.}(2023)Aguasca-Cabot, Donath, Feijen, Gréaux, Giunti, Khélifi, Linhoff, Mender, Mohrmann, Nigro, Olivera-Nieto, Pintore, Regeard, Remy, Sinha, Streil, \& Terrier}]{gammapy_v1.1}
Aguasca-Cabot, A., Donath, A., Feijen, K., {et~al.} 2023, Gammapy: Python toolbox for gamma-ray astronomy, v1.1,  Zenodo, \dodoi{10.5281/zenodo.8033275}

\bibitem[{{Aharonian} {et~al.}(2021){Aharonian}, {An}, {Axikegu}, {Bai}, {Bao}, {Bastieri}, {Bi}, {Bi}, {Cai}, {Cai}, {Cao}, {Cao}, {Chang}, {Chang}, {Chang}, {Chen}, {Chen}, {Chen}, {Chen}, {Chen}, {Chen}, {Chen}, {Chen}, {Chen}, {Chen}, {Chen}, {Chen}, {Chen}, {Cheng}, {Cheng}, {Cui}, {Cui}, {Cui}, {Dai}, {Dai}, {Dai}, {Danzengluobu}, {Della Volpe}, {D'Ettorre Piazzoli}, {Dong}, {Fan}, {Fan}, {Fan}, {Fang}, {Fang}, {Feng}, {Feng}, {Feng}, {Feng}, {Gao}, {Gao}, {Gao}, {Gao}, {Ge}, {Geng}, {Gong}, {Gou}, {Gu}, {Guo}, {Guo}, {Guo}, {Guo}, {Han}, {He}, {He}, {He}, {He}, {He}, {He}, {Heller}, {Hor}, {Hou}, {Hou}, {Hu}, {Hu}, {Hu}, {Hu}, {Huang}, {Huang}, {Huang}, {Huang}, {Huang}, {Ji}, {Ji}, {Jia}, {Jiang}, {Jiang}, {Jin}, {Kuleshov}, {Levochkin}, {Li}, {Li}, {Li}, {Li}, {Li}, {Li}, {Li}, {Li}, {Li}, {Li}, {Li}, {Li}, {Li}, {Li}, {Li}, {Li}, {Li}, {Liang}, {Liang}, {Lin}, {Liu}, {Liu}, {Liu}, {Liu}, {Liu}, {Liu}, {Liu}, {Liu}, {Liu}, {Liu}, {Liu}, {Liu}, {Liu}, {Liu}, {Liu}, {Long}, {Lu}, {Lv}, {Ma}, {Ma},
  {Ma}, {Mao}, {Masood}, {Mitthumsiri}, {Montaruli}, {Nan}, {Pang}, {Pattarakijwanich}, {Pei}, {Qi}, {Ruffolo}, {Rulev}, {S{\'a}iz}, {Shao}, {Shchegolev}, {Sheng}, {Shi}, {Song}, {Stenkin}, {Stepanov}, {Sun}, {Sun}, {Sun}, {Tam}, {Tang}, {Tian}, {Wang}, {Wang}, {Wang}, {Wang}, {Wang}, {Wang}, {Wang}, {Wang}, {Wang}, {Wang}, {Wang}, {Wang}, {Wang}, {Wang}, {Wang}, {Wang}, {Wang}, {Wang}, {Wang}, {Wang}, {Wang}, {Wei}, {Wei}, {Wei}, {Wen}, {Wu}, {Wu}, {Wu}, {Wu}, {Wu}, {Xi}, {Xia}, {Xia}, {Xiang}, {Xiao}, {Xiao}, {Xin}, {Xin}, {Xing}, {Xu}, {Xu}, {Xue}, {Yan}, {Yang}, {Yang}, {Yang}, {Yang}, {Yang}, {Yang}, {Yang}, {Yao}, {Yao}, {Ye}, {Yin}, {Yin}, {You}, {You}, {Yu}, {Yuan}, {Zeng}, {Zeng}, {Zeng}, {Zeng}, {Zha}, {Zhai}, {Zhang}, {Zhang}, {Zhang}, {Zhang}, {Zhang}, {Zhang}, {Zhang}, {Zhang}, {Zhang}, {Zhang}, {Zhang}, {Zhang}, {Zhang}, {Zhang}, {Zhang}, {Zhang}, {Zhang}, {Zhang}, {Zhang}, {Zhao}, {Zhao}, {Zhao}, {Zhao}, {Zhao}, {Zheng}, {Zheng}, {Zhou}, {Zhou}, {Zhou}, {Zhou}, {Zhou}, {Zhou}, {Zhu}, {Zhu},
  {Zhu}, {Zhu}, {Zuo}, {LHAASO Collaboration}, \& {Huang}}]{lhaasoj0621}
{Aharonian}, F., {An}, Q., {Axikegu}, Bai, L.~X., {et~al.} 2021, \prl, 126, 241103, \dodoi{10.1103/PhysRevLett.126.241103}

\bibitem[{{Albert} {et~al.}(2020){Albert}, {Alfaro}, {Alvarez}, {Camacho}, {Arteaga-Vel{\'a}zquez}, {Arunbabu}, {Avila Rojas}, {Ayala Solares}, {Baghmanyan}, {Belmont-Moreno}, {BenZvi}, {Brisbois}, {Caballero-Mora}, {Capistr{\'a}n}, {Carrami{\~n}ana}, {Casanova}, {Cotti}, {Couti{\~n}o de Le{\'o}n}, {De la Fuente}, {Diaz Hernandez}, {Diaz-Cruz}, {Dingus}, {DuVernois}, {Durocher}, {D{\'\i}az-V{\'e}lez}, {Ellsworth}, {Engel}, {Espinoza}, {Fan}, {Fang}, {Alonso}, {Fleischhack}, {Fraija}, {Galv{\'a}n-G{\'a}mez}, {Garcia}, {Garc{\'\i}a-Gonz{\'a}lez}, {Garfias}, {Giacinti}, {Gonz{\'a}lez}, {Goodman}, {Harding}, {Hernandez}, {Hinton}, {Hona}, {Huang}, {Hueyotl-Zahuantitla}, {H{\"u}ntemeyer}, {Iriarte}, {Jardin-Blicq}, {Joshi}, {Kieda}, {Lara}, {Lee}, {Le{\'o}n Vargas}, {Linnemann}, {Longinotti}, {Luis-Raya}, {Lundeen}, {L{\'o}pez-Coto}, {Malone}, {Marandon}, {Martinez}, {Martinez-Castellanos}, {Mart{\'\i}nez-Castro}, {Matthews}, {Miranda-Romagnoli}, {Morales-Soto}, {Moreno}, {Mostaf{\'a}}, {Nayerhoda}, {Nellen},
  {Newbold}, {Nisa}, {Noriega-Papaqui}, {Olivera-Nieto}, {Omodei}, {Peisker}, {P{\'e}rez Araujo}, {P{\'e}rez-P{\'e}rez}, {Ren}, {Rho}, {Rivi{\`e}re}, {Rosa-Gonz{\'a}lez}, {Ruiz-Velasco}, {Salazar}, {Salesa Greus}, {Sandoval}, {Schneider}, {Schoorlemmer}, {Serna}, {Sinnis}, {Smith}, {Springer}, {Surajbali}, {Tollefson}, {Torres}, {Torres-Escobedo}, {Ukwatta}, {Ure{\~n}a-Mena}, {Weisgarber}, {Werner}, {Willox}, {Zepeda}, {Zhou}, {de Le{\'o}n}, {{\'A}lvarez}, \& {HAWC Collaboration}}]{3hwc}
{Albert}, A., {Alfaro}, R., {Alvarez}, C., {et~al.} 2020, \apj, 905, 76, \dodoi{10.3847/1538-4357/abc2d8}

\bibitem[{{Albert} {et~al.}(2023){Albert}, {Alfaro}, {Arteaga-Vel{\'a}zquez}, {Ayala Solares}, {Belmont-Moreno}, {Capistr{\'a}n}, {Carrami{\~n}ana}, {Casanova}, {Cotzomi}, {Couti{\~n}o De Le{\'o}n}, {De la Fuente}, {de Le{\'o}n}, {Diaz Hernandez}, {DuVernois}, {D{\'\i}az-V{\'e}lez}, {Espinoza}, {Fan}, {Fraija}, {Fang}, {Garc{\'\i}a-Gonz{\'a}lez}, {Garfias}, {Jardin-Blicq}, {Gonz{\'a}lez}, {Goodman}, {Harding}, {Hernandez}, {Huang}, {Hueyotl-Zahuantitla}, {H{\"u}ntemeyer}, {Iriarte}, {Joshi}, {Lara}, {Lee}, {Le{\'o}n Vargas}, {Linnemann}, {Longinotti}, {Luis-Raya}, {Malone}, {Martinez}, {Mart{\'\i}nez-Castro}, {Matthews}, {Morales-Soto}, {Moreno}, {Mostaf{\'a}}, {Nayerhoda}, {Nellen}, {Newbold}, {Nisa}, {P{\'e}rez Araujo}, {Son}, {P{\'e}rez-P{\'e}rez}, {Rho}, {Rosa-Gonz{\'a}lez}, {Sandoval}, {Schneider}, {Serna-Franco}, {Smith}, {Springer}, {Tollefson}, {Torres}, {Torres-Escobedo}, {Wang}, {Whitaker}, {Willox}, {Zhou}, \& {HAWC Collaboration}}]{j0359}
{Albert}, A., {Alfaro}, R., {Arteaga-Vel{\'a}zquez}, J.~C., {et~al.} 2023, \apjl, 944, L29, \dodoi{10.3847/2041-8213/acb5ee}

\bibitem[{{Amato} \& {Recchia}(2024)}]{amato_halo_review}
{Amato}, E., \& {Recchia}, S. 2024, Nuovo Cimento Rivista Serie, 47, 399, \dodoi{10.1007/s40766-024-00059-8}

\bibitem[{{Arnaud}(1996)}]{xspec}
{Arnaud}, K.~A. 1996, in Astronomical Society of the Pacific Conference Series, Vol. 101, Astronomical Data Analysis Software and Systems V, ed. G.~H. {Jacoby} \& J.~{Barnes}, 17

\bibitem[{{Astropy Collaboration} {et~al.}(2013){Astropy Collaboration}, {Robitaille}, {Tollerud}, {Greenfield}, {Droettboom}, {Bray}, {Aldcroft}, {Davis}, {Ginsburg}, {Price-Whelan}, {Kerzendorf}, {Conley}, {Crighton}, {Barbary}, {Muna}, {Ferguson}, {Grollier}, {Parikh}, {Nair}, {Unther}, {Deil}, {Woillez}, {Conseil}, {Kramer}, {Turner}, {Singer}, {Fox}, {Weaver}, {Zabalza}, {Edwards}, {Azalee Bostroem}, {Burke}, {Casey}, {Crawford}, {Dencheva}, {Ely}, {Jenness}, {Labrie}, {Lim}, {Pierfederici}, {Pontzen}, {Ptak}, {Refsdal}, {Servillat}, \& {Streicher}}]{astropy:2013}
{Astropy Collaboration}, {Robitaille}, T.~P., {Tollerud}, E.~J., {et~al.} 2013, \aap, 558, A33, \dodoi{10.1051/0004-6361/201322068}

\bibitem[{{Astropy Collaboration} {et~al.}(2018){Astropy Collaboration}, {Price-Whelan}, {Sip{\H{o}}cz}, {G{\"u}nther}, {Lim}, {Crawford}, {Conseil}, {Shupe}, {Craig}, {Dencheva}, {Ginsburg}, {Vand erPlas}, {Bradley}, {P{\'e}rez-Su{\'a}rez}, {de Val-Borro}, {Aldcroft}, {Cruz}, {Robitaille}, {Tollerud}, {Ardelean}, {Babej}, {Bach}, {Bachetti}, {Bakanov}, {Bamford}, {Barentsen}, {Barmby}, {Baumbach}, {Berry}, {Biscani}, {Boquien}, {Bostroem}, {Bouma}, {Brammer}, {Bray}, {Breytenbach}, {Buddelmeijer}, {Burke}, {Calderone}, {Cano Rodr{\'\i}guez}, {Cara}, {Cardoso}, {Cheedella}, {Copin}, {Corrales}, {Crichton}, {D'Avella}, {Deil}, {Depagne}, {Dietrich}, {Donath}, {Droettboom}, {Earl}, {Erben}, {Fabbro}, {Ferreira}, {Finethy}, {Fox}, {Garrison}, {Gibbons}, {Goldstein}, {Gommers}, {Greco}, {Greenfield}, {Groener}, {Grollier}, {Hagen}, {Hirst}, {Homeier}, {Horton}, {Hosseinzadeh}, {Hu}, {Hunkeler}, {Ivezi{\'c}}, {Jain}, {Jenness}, {Kanarek}, {Kendrew}, {Kern}, {Kerzendorf}, {Khvalko}, {King}, {Kirkby}, {Kulkarni},
  {Kumar}, {Lee}, {Lenz}, {Littlefair}, {Ma}, {Macleod}, {Mastropietro}, {McCully}, {Montagnac}, {Morris}, {Mueller}, {Mumford}, {Muna}, {Murphy}, {Nelson}, {Nguyen}, {Ninan}, {N{\"o}the}, {Ogaz}, {Oh}, {Parejko}, {Parley}, {Pascual}, {Patil}, {Patil}, {Plunkett}, {Prochaska}, {Rastogi}, {Reddy Janga}, {Sabater}, {Sakurikar}, {Seifert}, {Sherbert}, {Sherwood-Taylor}, {Shih}, {Sick}, {Silbiger}, {Singanamalla}, {Singer}, {Sladen}, {Sooley}, {Sornarajah}, {Streicher}, {Teuben}, {Thomas}, {Tremblay}, {Turner}, {Terr{\'o}n}, {van Kerkwijk}, {de la Vega}, {Watkins}, {Weaver}, {Whitmore}, {Woillez}, {Zabalza}, \& {Astropy Contributors}}]{astropy:2018}
{Astropy Collaboration}, {Price-Whelan}, A.~M., {Sip{\H{o}}cz}, B.~M., {et~al.} 2018, \aj, 156, 123, \dodoi{10.3847/1538-3881/aabc4f}

\bibitem[{{Astropy Collaboration} {et~al.}(2022){Astropy Collaboration}, {Price-Whelan}, {Lim}, {Earl}, {Starkman}, {Bradley}, {Shupe}, {Patil}, {Corrales}, {Brasseur}, {N{"o}the}, {Donath}, {Tollerud}, {Morris}, {Ginsburg}, {Vaher}, {Weaver}, {Tocknell}, {Jamieson}, {van Kerkwijk}, {Robitaille}, {Merry}, {Bachetti}, {G{"u}nther}, {Aldcroft}, {Alvarado-Montes}, {Archibald}, {B{'o}di}, {Bapat}, {Barentsen}, {Baz{'a}n}, {Biswas}, {Boquien}, {Burke}, {Cara}, {Cara}, {Conroy}, {Conseil}, {Craig}, {Cross}, {Cruz}, {D'Eugenio}, {Dencheva}, {Devillepoix}, {Dietrich}, {Eigenbrot}, {Erben}, {Ferreira}, {Foreman-Mackey}, {Fox}, {Freij}, {Garg}, {Geda}, {Glattly}, {Gondhalekar}, {Gordon}, {Grant}, {Greenfield}, {Groener}, {Guest}, {Gurovich}, {Handberg}, {Hart}, {Hatfield-Dodds}, {Homeier}, {Hosseinzadeh}, {Jenness}, {Jones}, {Joseph}, {Kalmbach}, {Karamehmetoglu}, {Ka{l}uszy{'n}ski}, {Kelley}, {Kern}, {Kerzendorf}, {Koch}, {Kulumani}, {Lee}, {Ly}, {Ma}, {MacBride}, {Maljaars}, {Muna}, {Murphy}, {Norman}, {O'Steen},
  {Oman}, {Pacifici}, {Pascual}, {Pascual-Granado}, {Patil}, {Perren}, {Pickering}, {Rastogi}, {Roulston}, {Ryan}, {Rykoff}, {Sabater}, {Sakurikar}, {Salgado}, {Sanghi}, {Saunders}, {Savchenko}, {Schwardt}, {Seifert-Eckert}, {Shih}, {Jain}, {Shukla}, {Sick}, {Simpson}, {Singanamalla}, {Singer}, {Singhal}, {Sinha}, {Sip{H{o}}cz}, {Spitler}, {Stansby}, {Streicher}, {{{S}}umak}, {Swinbank}, {Taranu}, {Tewary}, {Tremblay}, {Val-Borro}, {Van Kooten}, {Vasovi{'c}}, {Verma}, {de Miranda Cardoso}, {Williams}, {Wilson}, {Winkel}, {Wood-Vasey}, {Xue}, {Yoachim}, {Zhang}, {Zonca}, \& {Astropy Project Contributors}}]{astropy:2022}
{Astropy Collaboration}, {Price-Whelan}, A.~M., {Lim}, P.~L., {et~al.} 2022, \apj, 935, 167, \dodoi{10.3847/1538-4357/ac7c74}

\bibitem[{{Bachetti}(2015)}]{hendrics}
{Bachetti}, M. 2015, {MaLTPyNT: Quick look timing analysis for NuSTAR data}, Astrophysics Source Code Library, record ascl:1502.021

\bibitem[{Bachetti {et~al.}(2022)Bachetti, Huppenkothen, Khan, Mishra, Sharma, Stevens, Swinbank, Desai, Rashid, Ribeiro, Sipőcz, Vats, tappina, omargamal8, Davis, Rasquinha, Balm, Mumford, Mihir, Campana, parkma99, Garg, Tandon, Hota, Nick, Raj, Mishra, Smith, Mahlke, \& Sachidanand}]{matteo_bachetti_2022_6394742}
Bachetti, M., Huppenkothen, D., Khan, U., {et~al.} 2022, StingraySoftware/stingray: Version 1.0, v1.0,  Zenodo, \dodoi{10.5281/zenodo.6394742}

\bibitem[{Ballet {et~al.}(2024)Ballet, Bruel, Burnett, Lott, \& collaboration}]{4fgl-dr4}
Ballet, J., Bruel, P., Burnett, T.~H., Lott, B., \& collaboration, T. F.-L. 2024, Fermi Large Area Telescope Fourth Source Catalog Data Release 4 (4FGL-DR4).
\newblock \doarXiv{2307.12546}

\bibitem[{Bao {et~al.}(2022)Bao, Fang, Bi, \& Wang}]{Bao:2021hey}
Bao, L.-Z., Fang, K., Bi, X.-J., \& Wang, S.-H. 2022, Astrophys. J., 936, 183, \dodoi{10.3847/1538-4357/ac8b8a}

\bibitem[{{Berge} {et~al.}(2007){Berge}, {Funk}, \& {Hinton}}]{background}
{Berge}, D., {Funk}, S., \& {Hinton}, J. 2007, \aap, 466, 1219, \dodoi{10.1051/0004-6361:20066674}

\bibitem[{{Bogdanov} \& {Ho}(2024)}]{Bogdanov2024}
{Bogdanov}, S., \& {Ho}, W. C.~G. 2024, \apj, 969, 53, \dodoi{10.3847/1538-4357/ad452b}

\bibitem[{{Brisken} {et~al.}(2003){Brisken}, {Thorsett}, {Golden}, \& {Goss}}]{Brisken2003:monogem-distance}
{Brisken}, W.~F., {Thorsett}, S.~E., {Golden}, A., \& {Goss}, W.~M. 2003, \apjl, 593, L89, \dodoi{10.1086/378184}

\bibitem[{{Buccheri} {et~al.}(1983){Buccheri}, {Bennett}, {Bignami}, {Bloemen}, {Boriakoff}, {Caraveo}, {Hermsen}, {Kanbach}, {Manchester}, {Masnou}, {Mayer-Hasselwander}, {{\"O}zel}, {Paul}, {Sacco}, {Scarsi}, \& {Strong}}]{Buccheri1983:z2test}
{Buccheri}, R., {Bennett}, K., {Bignami}, G.~F., {et~al.} 1983, \aap, 128, 245

\bibitem[{{Cao} {et~al.}(2024{\natexlab{a}}){Cao}, {Aharonian}, {An}, {Axikegu}, {Bai}, {Bao}, {Bastieri}, {Bi}, {Bi}, {Cai}, {Cao}, {Cao}, {Cao}, {Chang}, {Chang}, {Chen}, {Chen}, {Chen}, {Chen}, {Chen}, {Chen}, {Chen}, {Chen}, {Chen}, {Chen}, {Chen}, {Chen}, {Cheng}, {Cheng}, {Cui}, {Cui}, {Cui}, {Cui}, {Dai}, {Dai}, {Dai}, {Danzengluobu}, {Dong}, {Duan}, {Fan}, {Fan}, {Fang}, {Fang}, {Feng}, {Feng}, {Feng}, {Feng}, {Feng}, {Gabici}, {Gao}, {Gao}, {Gao}, {Gao}, {Gao}, {Gao}, {Ge}, {Geng}, {Giacinti}, {Gong}, {Gou}, {Gu}, {Guo}, {Guo}, {Guo}, {Guo}, {Han}, {He}, {He}, {He}, {He}, {He}, {Hor}, {Hou}, {Hou}, {Hou}, {Hu}, {Hu}, {Hu}, {Huang}, {Huang}, {Huang}, {Huang}, {Huang}, {Huang}, {Huang}, {Ji}, {Jia}, {Jia}, {Jiang}, {Jiang}, {Jiang}, {Jin}, {Kang}, {Ke}, {Kuleshov}, {Kurinov}, {Li}, {Li}, {Li}, {Li}, {Li}, {Li}, {Li}, {Li}, {Li}, {Li}, {Li}, {Li}, {Li}, {Li}, {Li}, {Li}, {Li}, {Li}, {Li}, {Liang}, {Liang}, {Lin}, {Liu}, {Liu}, {Liu}, {Liu}, {Liu}, {Liu}, {Liu}, {Liu}, {Liu}, {Liu}, {Liu}, {Liu}, {Liu},
  {Liu}, {Lu}, {Luo}, {Lv}, {Ma}, {Ma}, {Ma}, {Mao}, {Min}, {Mitthumsiri}, {Mu}, {Nan}, {Neronov}, {Ou}, {Pang}, {Pattarakijwanich}, {Pei}, {Qi}, {Qi}, {Qiao}, {Qin}, {Ruffolo}, {S{\'a}iz}, {Semikoz}, {Shao}, {Shao}, {Shchegolev}, {Sheng}, {Shu}, {Song}, {Stenkin}, {Stepanov}, {Su}, {Sun}, {Sun}, {Sun}, {Tam}, {Tang}, {Tang}, {Tian}, {Wang}, {Wang}, {Wang}, {Wang}, {Wang}, {Wang}, {Wang}, {Wang}, {Wang}, {Wang}, {Wang}, {Wang}, {Wang}, {Wang}, {Wang}, {Wang}, {Wang}, {Wang}, {Wang}, {Wang}, {Wang}, {Wei}, {Wei}, {Wei}, {Wen}, {Wu}, {Wu}, {Wu}, {Wu}, {Wu}, {Xi}, {Xia}, {Xia}, {Xiang}, {Xiao}, {Xiao}, {Xin}, {Xin}, {Xing}, {Xiong}, {Xu}, {Xu}, {Xu}, {Xu}, {Xue}, {Yan}, {Yan}, {Yan}, {Yang}, {Yang}, {Yang}, {Yang}, {Yang}, {Yang}, {Yang}, {Yang}, {Yang}, {Yao}, {Yao}, {Ye}, {Yin}, {Yin}, {You}, {You}, {Yu}, {Yuan}, {Yue}, {Zeng}, {Zeng}, {Zeng}, {Zha}, {Zhang}, {Zhang}, {Zhang}, {Zhang}, {Zhang}, {Zhang}, {Zhang}, {Zhang}, {Zhang}, {Zhang}, {Zhang}, {Zhang}, {Zhang}, {Zhang}, {Zhang}, {Zhang}, {Zhang}, {Zhang},
  {Zhao}, {Zhao}, {Zhao}, {Zhao}, {Zhao}, {Zheng}, {Zheng}, {Zhou}, {Zhou}, {Zhou}, {Zhou}, {Zhou}, {Zhou}, {Zhou}, {Zhu}, {Zhu}, {Zhu}, {Zhu}, {Zou}, \& {Zuo}}]{j0248}
{Cao}, Z., {Aharonian}, F., {An}, Q., {et~al.} 2024{\natexlab{a}}, arXiv e-prints, arXiv:2410.04425, \dodoi{10.48550/arXiv.2410.04425}

\bibitem[{{Cao} {et~al.}(2024{\natexlab{b}}){Cao}, {Aharonian}, {An}, {Axikegu}, {Bai}, {Bao}, {Bastieri}, {Bi}, {Bi}, {Cai}, {Cao}, {Cao}, {Cao}, {Chang}, {Chang}, {Chen}, {Chen}, {Chen}, {Chen}, {Chen}, {Chen}, {Chen}, {Chen}, {Chen}, {Chen}, {Chen}, {Chen}, {Cheng}, {Cheng}, {Cui}, {Cui}, {Cui}, {Cui}, {Dai}, {Dai}, {Dai}, {Danzengluobu}, {Della Volpe}, {Dong}, {Duan}, {Fan}, {Fan}, {Fang}, {Fang}, {Feng}, {Feng}, {Feng}, {Feng}, {Feng}, {Gabici}, {Gao}, {Gao}, {Gao}, {Gao}, {Gao}, {Gao}, {Ge}, {Geng}, {Giacinti}, {Gong}, {Gou}, {Gu}, {Guo}, {Guo}, {Guo}, {Guo}, {Han}, {He}, {He}, {He}, {He}, {He}, {Heller}, {Hor}, {Hou}, {Hou}, {Hou}, {Hu}, {Hu}, {Hu}, {Huang}, {Huang}, {Huang}, {Huang}, {Huang}, {Huang}, {Huang}, {Ji}, {Jia}, {Jia}, {Jiang}, {Jiang}, {Jiang}, {Jin}, {Kang}, {Ke}, {Kuleshov}, {Kurinov}, {Li}, {Li}, {Li}, {Li}, {Li}, {Li}, {Li}, {Li}, {Li}, {Li}, {Li}, {Li}, {Li}, {Li}, {Li}, {Li}, {Li}, {Li}, {Li}, {Liang}, {Liang}, {Lin}, {Liu}, {Liu}, {Liu}, {Liu}, {Liu}, {Liu}, {Liu}, {Liu}, {Liu},
  {Liu}, {Liu}, {Liu}, {Liu}, {Liu}, {Lu}, {Luo}, {Lv}, {Ma}, {Ma}, {Ma}, {Mao}, {Min}, {Mitthumsiri}, {Mu}, {Nan}, {Neronov}, {Ou}, {Pang}, {Pattarakijwanich}, {Pei}, {Qi}, {Qi}, {Qiao}, {Qin}, {Ruffolo}, {S{\'a}iz}, {Semikoz}, {Shao}, {Shao}, {Shchegolev}, {Sheng}, {Shu}, {Song}, {Stenkin}, {Stepanov}, {Su}, {Sun}, {Sun}, {Sun}, {Tam}, {Tang}, {Tang}, {Tian}, {Wang}, {Wang}, {Wang}, {Wang}, {Wang}, {Wang}, {Wang}, {Wang}, {Wang}, {Wang}, {Wang}, {Wang}, {Wang}, {Wang}, {Wang}, {Wang}, {Wang}, {Wang}, {Wang}, {Wang}, {Wang}, {Wei}, {Wei}, {Wei}, {Wen}, {Wu}, {Wu}, {Wu}, {Wu}, {Wu}, {Xi}, {Xia}, {Xia}, {Xiang}, {Xiao}, {Xiao}, {Xin}, {Xin}, {Xing}, {Xiong}, {Xu}, {Xu}, {Xu}, {Xu}, {Xue}, {Yan}, {Yan}, {Yan}, {Yang}, {Yang}, {Yang}, {Yang}, {Yang}, {Yang}, {Yang}, {Yang}, {Yang}, {Yao}, {Yao}, {Ye}, {Yin}, {Yin}, {You}, {You}, {Yu}, {Yuan}, {Yue}, {Zeng}, {Zeng}, {Zeng}, {Zha}, {Zhang}, {Zhang}, {Zhang}, {Zhang}, {Zhang}, {Zhang}, {Zhang}, {Zhang}, {Zhang}, {Zhang}, {Zhang}, {Zhang}, {Zhang}, {Zhang}, {Zhang},
  {Zhang}, {Zhang}, {Zhang}, {Zhao}, {Zhao}, {Zhao}, {Zhao}, {Zhao}, {Zheng}, {Zhou}, {Zhou}, {Zhou}, {Zhou}, {Zhou}, {Zhou}, {Zhou}, {Zhu}, {Zhu}, {Zhu}, {Zhu}, {Zuo}, \& {(The Lhaaso Collaboration)}}]{lhaasocat}
---. 2024{\natexlab{b}}, \apjs, 271, 25, \dodoi{10.3847/1538-4365/acfd29}

\bibitem[{{Caraveo} {et~al.}(2003){Caraveo}, {Bignami}, {De Luca}, {Mereghetti}, {Pellizzoni}, {Mignani}, {Tur}, \& {Becker}}]{geminga-pwn3}
{Caraveo}, P.~A., {Bignami}, G.~F., {De Luca}, A., {et~al.} 2003, Science, 301, 1345, \dodoi{10.1126/science.1086973}

\bibitem[{{Cash}(1979)}]{cash}
{Cash}, W. 1979, \apj, 228, 939, \dodoi{10.1086/156922}

\bibitem[{{Chang} \& {Bildsten}(2004)}]{Chang2004}
{Chang}, P., \& {Bildsten}, L. 2004, \apj, 605, 830, \dodoi{10.1086/382271}

\bibitem[{{Chang} {et~al.}(2010){Chang}, {Bildsten}, \& {Arras}}]{Chang2010}
{Chang}, P., {Bildsten}, L., \& {Arras}, P. 2010, \apj, 723, 719, \dodoi{10.1088/0004-637X/723/1/719}

\bibitem[{{De Luca} \& {Molendi}(2004)}]{cxb}
{De Luca}, A., \& {Molendi}, S. 2004, \aap, 419, 837, \dodoi{10.1051/0004-6361:20034421}

\bibitem[{Di~Mauro {et~al.}(2019)Di~Mauro, Manconi, \& Donato}]{DiMauro:2019yvh}
Di~Mauro, M., Manconi, S., \& Donato, F. 2019, Phys. Rev. D, 100, 123015, \dodoi{10.1103/PhysRevD.104.089903}

\bibitem[{Di~Mauro {et~al.}(2020)Di~Mauro, Manconi, \& Donato}]{DiMauro:2019hwn}
---. 2020, Phys. Rev. D, 101, 103035, \dodoi{10.1103/PhysRevD.101.103035}

\bibitem[{Di~Mauro {et~al.}(2021)Di~Mauro, Manconi, Negro, \& Donato}]{DiMauro:2020jbz}
Di~Mauro, M., Manconi, S., Negro, M., \& Donato, F. 2021, Phys. Rev. D, 104, 103002, \dodoi{10.1103/PhysRevD.104.103002}

\bibitem[{{Donath} {et~al.}(2023){Donath}, {Terrier}, {Remy}, {Sinha}, {Nigro}, {Pintore}, {Kh\'elifi}, {Olivera-Nieto}, {Ruiz}, {Br\"ugge}, {Linhoff}, {Contreras}, {Acero}, {Aguasca-Cabot}, {Berge}, {Bhattacharjee}, {Buchner}, {Boisson}, {Carreto Fidalgo}, {Chen}, {de Bony de Lavergne}, {de Miranda Cardoso}, {Deil}, {F\"u\ss{}ling}, {Funk}, {Giunti}, {Hinton}, {Jouvin}, {King}, {Lefaucheur}, {Lemoine-Goumard}, {Lenain}, {L\'opez-Coto}, {Mohrmann}, {Morcuende}, {Panny}, {Regeard}, {Saha}, {Siejkowski}, {Siemiginowska}, {Sip"ocz}, {Unbehaun}, {van Eldik}, {Vuillaume}, \& {Zanin}}]{gammapy}
{Donath}, A., {Terrier}, R., {Remy}, Q., {et~al.} 2023, Astronomy and Astrophysics, 678, A157, \dodoi{10.1051/0004-6361/202346488}

\bibitem[{Fang(2022)}]{Fang:2022fof}
Fang, K. 2022, Front. Astron. Space Sci., 9, 1022100, \dodoi{10.3389/fspas.2022.1022100}

\bibitem[{Fang {et~al.}(2021)Fang, Xi, \& Bi}]{Fang:2021qon}
Fang, K., Xi, S.-Q., \& Bi, X.-J. 2021, Phys. Rev. D, 104, 103024, \dodoi{10.1103/PhysRevD.104.103024}

\bibitem[{{Fomin} {et~al.}(1994){Fomin}, {Stepanian}, {Lamb}, {Lewis}, {Punch}, \& {Weekes}}]{Fomin1994:wobble}
{Fomin}, V.~P., {Stepanian}, A.~A., {Lamb}, R.~C., {et~al.} 1994, Astroparticle Physics, 2, 137, \dodoi{10.1016/0927-6505(94)90036-1}

\bibitem[{{Giacinti} {et~al.}(2020){Giacinti}, {Mitchell}, {L{\'o}pez-Coto}, {Joshi}, {Parsons}, \& {Hinton}}]{Giacinti2020:pulsar-halo}
{Giacinti}, G., {Mitchell}, A.~M.~W., {L{\'o}pez-Coto}, R., {et~al.} 2020, \aap, 636, A113, \dodoi{10.1051/0004-6361/201936505}

\bibitem[{{H.~E.~S.~S. Collaboration} {et~al.}(2023){H.~E.~S.~S. Collaboration}, {Aharonian}, {Ait Benkhali}, {Aschersleben}, {Ashkar}, {Backes}, {Barbosa Martins}, {Batzofin}, {Becherini}, {Berge}, {Bernl{\"o}hr}, {Bi}, {B{\"o}ttcher}, {Boisson}, {Bolmont}, {Borowska}, {Bouyahiaoui}, {Bradascio}, {Brose}, {Brun}, {Bruno}, {Bulik}, {Burger-Scheidlin}, {Cangemi}, {Caroff}, {Casanova}, {Celic}, {Cerruti}, {Chambery}, {Chand}, {Chandra}, {Chen}, {Chibueze}, {Chibueze}, {Cotter}, {Mbarubucyeye}, {Devin}, {Djannati-Ata{\"\i}}, {Dmytriiev}, {Egberts}, {Einecke}, {Ernenwein}, {Feijen}, {Fichet de Clairfontaine}, {Filipovic}, {Fontaine}, {F{\"u}{\ss}ling}, {Funk}, {Gabici}, {Gallant}, {Ghafourizadeh}, {Giavitto}, {Giunti}, {Glawion}, {Glicenstein}, {Goswami}, {Grolleron}, {Grondin}, {Haerer}, {Haupt}, {Hermann}, {Hinton}, {Hofmann}, {Holch}, {Holler}, {Horns}, {Huang}, {Jamrozy}, {Jankowsky}, {Joshi}, {Jung-Richardt}, {Kasai}, {Katarzy{\'n}ski}, {Kh{\'e}lifi}, {Klu{\'z}niak}, {Komin}, {Kosack}, {Kostunin}, {Lang},
  {Le Stum}, {Leitl}, {Lemi{\`e}re}, {Lemoine-Goumard}, {Lenain}, {Leuschner}, {Lohse}, {Luashvili}, {Lypova}, {Mackey}, {Malyshev}, {Marandon}, {Marchegiani}, {Marcowith}, {Marinos}, {Mart{\'\i}-Devesa}, {Marx}, {Maurin}, {Meintjes}, {Meyer}, {Mitchell}, {Moderski}, {Mohrmann}, {Montanari}, {Moulin}, {Muller}, {Nakashima}, {de Naurois}, {Niemiec}, {Noel}, {O'Brien}, {Ohm}, {Olivera-Nieto}, {de Ona Wilhelmi}, {Ostrowski}, {Panny}, {Panter}, {Parsons}, {Peron}, {Prokhorov}, {P{\"u}hlhofer}, {Quirrenbach}, {Reimer}, {Reimer}, {Renaud}, {Reville}, {Rieger}, {Rowell}, {Rudak}, {Ricarte}, {Ruiz-Velasco}, {Sahakian}, {Salzmann}, {Santangelo}, {Sasaki}, {Sch{\"u}ssler}, {Schutte}, {Schwanke}, {Shapopi}, {Sinha}, {Sol}, {Specovius}, {Spencer}, {Stawarz}, {Steinmassl}, {Sushch}, {Suzuki}, {Takahashi}, {Tanaka}, {Tavernier}, {Taylor}, {Terrier}, {Thorpe-Morgan}, {Tsirou}, {Tsuji}, {Vecchi}, {Venter}, {Vink}, {Wagner}, {White}, {Wierzcholska}, {Wong}, {Zacharias}, {Zargaryan}, {Zdziarski}, {Zech}, {Zouari}, \&
  {{\.Z}ywucka}}]{geminga-hess}
{H.~E.~S.~S. Collaboration}, {Aharonian}, F., {Ait Benkhali}, F., {et~al.} 2023, \aap, 673, A148, \dodoi{10.1051/0004-6361/202245776}

\bibitem[{{Halpern} \& {Wang}(1997)}]{Halpern1997}
{Halpern}, J.~P., \& {Wang}, F.~Y.~H. 1997, \apj, 477, 905, \dodoi{10.1086/303743}

\bibitem[{Harris {et~al.}(2020)Harris, Millman, van~der Walt, Gommers, Virtanen, Cournapeau, Wieser, Taylor, Berg, Smith, Kern, Picus, Hoyer, van Kerkwijk, Brett, Haldane, del R{\'{i}}o, Wiebe, Peterson, G{\'{e}}rard-Marchant, Sheppard, Reddy, Weckesser, Abbasi, Gohlke, \& Oliphant}]{numpy}
Harris, C.~R., Millman, K.~J., van~der Walt, S.~J., {et~al.} 2020, Nature, 585, 357, \dodoi{10.1038/s41586-020-2649-2}

\bibitem[{{Ho} \& {Heinke}(2009)}]{Ho2009}
{Ho}, W. C.~G., \& {Heinke}, C.~O. 2009, \nat, 462, 71, \dodoi{10.1038/nature08525}

\bibitem[{{Ho} {et~al.}(2008){Ho}, {Potekhin}, \& {Chabrier}}]{Ho2008}
{Ho}, W. C.~G., {Potekhin}, A.~Y., \& {Chabrier}, G. 2008, \apjs, 178, 102, \dodoi{10.1086/589238}

\bibitem[{{Holder} {et~al.}(2006){Holder}, {Atkins}, {Badran}, {Blaylock}, {Bradbury}, {Buckley}, {Byrum}, {Carter-Lewis}, {Celik}, {Chow}, {Cogan}, {Cui}, {Daniel}, {de la Calle Perez}, {Dowdall}, {Dowkontt}, {Duke}, {Falcone}, {Fegan}, {Finley}, {Fortin}, {Fortson}, {Gibbs}, {Gillanders}, {Glidewell}, {Grube}, {Gutierrez}, {Gyuk}, {Hall}, {Hanna}, {Hays}, {Horan}, {Hughes}, {Humensky}, {Imran}, {Jung}, {Kaaret}, {Kenny}, {Kieda}, {Kildea}, {Knapp}, {Krawczynski}, {Krennrich}, {Lang}, {LeBohec}, {Linton}, {Little}, {Maier}, {Manseri}, {Milovanovic}, {Moriarty}, {Mukherjee}, {Ogden}, {Ong}, {Petry}, {Perkins}, {Pizlo}, {Pohl}, {Quinn}, {Ragan}, {Reynolds}, {Roache}, {Rose}, {Schroedter}, {Sembroski}, {Sleege}, {Steele}, {Swordy}, {Syson}, {Toner}, {Valcarcel}, {Vassiliev}, {Wakely}, {Weekes}, {White}, {Williams}, \& {Wagner}}]{veritas}
{Holder}, J., {Atkins}, R.~W., {Badran}, H.~M., {et~al.} 2006, Astroparticle Physics, 25, 391, \dodoi{10.1016/j.astropartphys.2006.04.002}

\bibitem[{Hunter(2007)}]{matplotlib}
Hunter, J.~D. 2007, Computing in Science \& Engineering, 9, 90, \dodoi{10.1109/MCSE.2007.55}

\bibitem[{{Jansen} {et~al.}(2001){Jansen}, {Lumb}, {Altieri}, {Clavel}, {Ehle}, {Erd}, {Gabriel}, {Guainazzi}, {Gondoin}, {Much}, {Munoz}, {Santos}, {Schartel}, {Texier}, \& {Vacanti}}]{Jansen2001:xmm-orbit}
{Jansen}, F., {Lumb}, D., {Altieri}, B., {et~al.} 2001, \aap, 365, L1, \dodoi{10.1051/0004-6361:20000036}

\bibitem[{Kappl {et~al.}(2015)Kappl, Reinert, \& Winkler}]{Kappl:2015bqa}
Kappl, R., Reinert, A., \& Winkler, M.~W. 2015, JCAP, 10, 034, \dodoi{10.1088/1475-7516/2015/10/034}

\bibitem[{{Khokhriakova} {et~al.}(2024){Khokhriakova}, {Becker}, {Ponti}, {Sasaki}, {Li}, \& {Liu}}]{erosita1}
{Khokhriakova}, A., {Becker}, W., {Ponti}, G., {et~al.} 2024, \aap, 683, A180, \dodoi{10.1051/0004-6361/202347311}

\bibitem[{{Kieda}(2011)}]{Kieda2011:pmt-upgrade}
{Kieda}, D. 2011, in International Cosmic Ray Conference, Vol.~9, International Cosmic Ray Conference, 14, \dodoi{10.7529/ICRC2011/V09/0343}

\bibitem[{{Krause} {et~al.}(2017){Krause}, {Pueschel}, \& {Maier}}]{bdt}
{Krause}, M., {Pueschel}, E., \& {Maier}, G. 2017, Astroparticle Physics, 89, 1, \dodoi{10.1016/j.astropartphys.2017.01.004}

\bibitem[{{Kuntz} \& {Snowden}(2008)}]{Kuntz2008:xmm-particle-bkg}
{Kuntz}, K.~D., \& {Snowden}, S.~L. 2008, \aap, 478, 575, \dodoi{10.1051/0004-6361:20077912}

\bibitem[{Liu(2022)}]{Liu:2022hqf}
Liu, R.-Y. 2022, Int. J. Mod. Phys. A, 37, 2230011, \dodoi{10.1142/S0217751X22300113}

\bibitem[{L\'opez-Coto {et~al.}(2022)L\'opez-Coto, de~O\~na Wilhelmi, Aharonian, Amato, \& Hinton}]{Lopez-Coto:2022igd}
L\'opez-Coto, R., de~O\~na Wilhelmi, E., Aharonian, F., Amato, E., \& Hinton, J. 2022, Nature Astron., 6, 199, \dodoi{10.1038/s41550-021-01580-0}

\bibitem[{{Maier} \& {Holder}(2017)}]{Eventdisplay}
{Maier}, G., \& {Holder}, J. 2017, in International Cosmic Ray Conference, Vol. 301, 35th International Cosmic Ray Conference (ICRC2017), 747, \dodoi{10.22323/1.301.0747}

\bibitem[{Maier {et~al.}(2023)Maier, Holder, McCann, Behera, Duke, Giuri, Skole, Tak, Aliu, Pueschel, Pizlo, Decerpri, Finneagan, Foote, Hughes, Fleischhack, Krawczynski, Prokoph, Grube, Tyler, Berger, Pfrang, Gerard, Beilicke, Kherlakian, Krause, McCutcheon, Nievas-Rosillo, Schroedter, Shayduk, Hakasson, Kelley-Hoskins, Gueta, Ivo, Guenette, Patel, Prado, Welsing, Griffin, Griffiths, O'Brian, Vincent, Vorobiov, Hassan, \& Khassen}]{Eventdisplay_v490p2}
Maier, G., Holder, J., McCann, A., {et~al.} 2023, Eventdisplay: an Analysis and Reconstruction Package for VERITAS, v490.2,  Zenodo, \dodoi{10.5281/zenodo.8154944}

\bibitem[{Manconi {et~al.}(2024)Manconi, Woo, Shang, Krivonos, Tang, Di~Mauro, Donato, Mori, \& Hailey}]{Manconi:2024wlq}
Manconi, S., Woo, J., Shang, R.-Y., {et~al.} 2024, Astron. Astrophys., 689, A326, \dodoi{10.1051/0004-6361/202450242}

\bibitem[{{Mori} \& {Hailey}(2006)}]{Mori2006}
{Mori}, K., \& {Hailey}, C.~J. 2006, \apj, 648, 1139, \dodoi{10.1086/506008}

\bibitem[{{Mori} \& {Heyl}(2007)}]{Mori2007b}
{Mori}, K., \& {Heyl}, J.~S. 2007, \mnras, 376, 895, \dodoi{10.1111/j.1365-2966.2007.11485.x}

\bibitem[{{Mori} \& {Ho}(2007)}]{Mori2007}
{Mori}, K., \& {Ho}, W. C.~G. 2007, \mnras, 377, 905, \dodoi{10.1111/j.1365-2966.2007.11663.x}

\bibitem[{{Mori} {et~al.}(2014){Mori}, {Gotthelf}, {Dufour}, {Kaspi}, {Halpern}, {Beloborodov}, {An}, {Bachetti}, {Boggs}, {Christensen}, {Craig}, {Hailey}, {Harrison}, {Kouveliotou}, {Pivovaroff}, {Stern}, \& {Zhang}}]{Mori2014}
{Mori}, K., {Gotthelf}, E.~V., {Dufour}, F., {et~al.} 2014, \apj, 793, 88, \dodoi{10.1088/0004-637X/793/2/88}

\bibitem[{{Nepomuk Otte}(2011)}]{Otte2011:pmt-upgrade}
{Nepomuk Otte}, A. 2011, arXiv e-prints, arXiv:1110.4702, \dodoi{10.48550/arXiv.1110.4702}

\bibitem[{Osipov {et~al.}(2020)Osipov, Bykov, Petrov, \& Romansky}]{Osipov:2020lty}
Osipov, S.~M., Bykov, A.~M., Petrov, A.~E., \& Romansky, V.~I. 2020, J. Phys. Conf. Ser., 1697, 012009, \dodoi{10.1088/1742-6596/1697/1/012009}

\bibitem[{{Pavlov} {et~al.}(2010){Pavlov}, {Bhattacharyya}, \& {Zavlin}}]{geminga-pwn2}
{Pavlov}, G.~G., {Bhattacharyya}, S., \& {Zavlin}, V.~E. 2010, \apj, 715, 66, \dodoi{10.1088/0004-637X/715/1/66}

\bibitem[{{Pletsch} {et~al.}(2012){Pletsch}, {Guillemot}, {Allen}, {Kramer}, {Aulbert}, {Fehrmann}, {Ray}, {Barr}, {Belfiore}, {Camilo}, {Caraveo}, {{\c{C}}elik}, {Champion}, {Dormody}, {Eatough}, {Ferrara}, {Freire}, {Hessels}, {Keith}, {Kerr}, {de Luca}, {Lyne}, {Marelli}, {McLaughlin}, {Parent}, {Ransom}, {Razzano}, {Reich}, {Saz Parkinson}, {Stappers}, \& {Wolff}}]{fermipsr}
{Pletsch}, H.~J., {Guillemot}, L., {Allen}, B., {et~al.} 2012, \apj, 744, 105, \dodoi{10.1088/0004-637X/744/2/105}

\bibitem[{{Posselt} {et~al.}(2017){Posselt}, {Pavlov}, {Slane}, {Romani}, {Bucciantini}, {Bykov}, {Kargaltsev}, {Weisskopf}, \& {Ng}}]{geminga-pwn1}
{Posselt}, B., {Pavlov}, G.~G., {Slane}, P.~O., {et~al.} 2017, \apj, 835, 66, \dodoi{10.3847/1538-4357/835/1/66}

\bibitem[{Recchia {et~al.}(2021)Recchia, Di~Mauro, Aharonian, Orusa, Donato, Gabici, \& Manconi}]{Recchia:2021kty}
Recchia, S., Di~Mauro, M., Aharonian, F.~A., {et~al.} 2021, Phys. Rev. D, 104, 123017, \dodoi{10.1103/PhysRevD.104.123017}

\bibitem[{{Ruderman} \& {Sutherland}(1975)}]{Ruderman1975:polar-cap}
{Ruderman}, M.~A., \& {Sutherland}, P.~G. 1975, Astrophysical Journal, 196, 51, \dodoi{10.1086/153393}

\bibitem[{{Schroer} {et~al.}(2023){Schroer}, {Evoli}, \& {Blasi}}]{Schroer2023:suppresed-diffusion}
{Schroer}, B., {Evoli}, C., \& {Blasi}, P. 2023, \prd, 107, 123020, \dodoi{10.1103/PhysRevD.107.123020}

\bibitem[{{Smith} {et~al.}(2023){Smith}, {Abdollahi}, {Ajello}, {Bailes}, {Baldini}, {Ballet}, {Baring}, {Bassa}, {Gonzalez}, {Bellazzini}, {Berretta}, {Bhattacharyya}, {Bissaldi}, {Bonino}, {Bottacini}, {Bregeon}, {Bruel}, {Burgay}, {Burnett}, {Cameron}, {Camilo}, {Caputo}, {Caraveo}, {Cavazzuti}, {Chiaro}, {Ciprini}, {Clark}, {Cognard}, {Corongiu}, {Orestano}, {Crnogorcevic}, {Cuoco}, {Cutini}, {D'Ammando}, {de Angelis}, {DeCesar}, {De Gaetano}, {de Menezes}, {Deneva}, {de Palma}, {Di Lalla}, {Dirirsa}, {Di Venere}, {Dom{\'\i}nguez}, {Dumora}, {Fegan}, {Ferrara}, {Fiori}, {Fleischhack}, {Flynn}, {Franckowiak}, {Freire}, {Fukazawa}, {Fusco}, {Galanti}, {Gammaldi}, {Gargano}, {Gasparrini}, {Giacchino}, {Giglietto}, {Giordano}, {Giroletti}, {Green}, {Grenier}, {Guillemot}, {Guiriec}, {Gustafsson}, {Harding}, {Hays}, {Hewitt}, {Horan}, {Hou}, {Jankowski}, {Johnson}, {Johnson}, {Johnston}, {Kataoka}, {Keith}, {Kerr}, {Kramer}, {Kuss}, {Latronico}, {Lee}, {Li}, {Li}, {Limyansky}, {Longo}, {Loparco}, {Lorusso},
  {Lovellette}, {Lower}, {Lubrano}, {Lyne}, {Maan}, {Maldera}, {Manchester}, {Manfreda}, {Marelli}, {Mart{\'\i}-Devesa}, {Mazziotta}, {McEnery}, {Mereu}, {Michelson}, {Mickaliger}, {Mitthumsiri}, {Mizuno}, {Moiseev}, {Monzani}, {Morselli}, {Negro}, {Nemmen}, {Nieder}, {Nuss}, {Omodei}, {Orienti}, {Orlando}, {Ormes}, {Palatiello}, {Paneque}, {Panzarini}, {Parthasarathy}, {Persic}, {Pesce-Rollins}, {Pillera}, {Poon}, {Porter}, {Possenti}, {Principe}, {Rain{\`o}}, {Rando}, {Ransom}, {Ray}, {Razzano}, {Razzaque}, {Reimer}, {Reimer}, {Renault-Tinacci}, {Romani}, {S{\'a}nchez-Conde}, {Parkinson}, {Scotton}, {Serini}, {Sgr{\`o}}, {Shannon}, {Sharma}, {Shen}, {Siskind}, {Spandre}, {Spinelli}, {Stappers}, {Stephens}, {Suson}, {Tabassum}, {Tajima}, {Tak}, {Theureau}, {Thompson}, {Tibolla}, {Torres}, {Valverde}, {Venter}, {Wadiasingh}, {Wang}, {Wang}, {Wang}, {Weltevrede}, {Wood}, {Yan}, {Zaharijas}, {Zhang}, \& {Zhu}}]{3pc}
{Smith}, D.~A., {Abdollahi}, S., {Ajello}, M., {et~al.} 2023, \apj, 958, 191, \dodoi{10.3847/1538-4357/acee67}

\bibitem[{{Snowden} {et~al.}(2004){Snowden}, {Collier}, \& {Kuntz}}]{Snowden2004:xmm-swcx}
{Snowden}, S.~L., {Collier}, M.~R., \& {Kuntz}, K.~D. 2004, \apj, 610, 1182, \dodoi{10.1086/421841}

\bibitem[{Tang \& Piran(2019)}]{Tang:2018wyr}
Tang, X., \& Piran, T. 2019, Mon. Not. Roy. Astron. Soc., 484, 3491, \dodoi{10.1093/mnras/stz268}

\bibitem[{{Verbiest} {et~al.}(2012){Verbiest}, {Weisberg}, {Chael}, {Lee}, \& {Lorimer}}]{geminga-distance}
{Verbiest}, J.~P.~W., {Weisberg}, J.~M., {Chael}, A.~A., {Lee}, K.~J., \& {Lorimer}, D.~R. 2012, \apj, 755, 39, \dodoi{10.1088/0004-637X/755/1/39}

\bibitem[{Vernetto \& Lipari(2016)}]{Vernetto:2016alq}
Vernetto, S., \& Lipari, P. 2016, Phys. Rev. D, 94, 063009, \dodoi{10.1103/PhysRevD.94.063009}

\bibitem[{{Wilms} {et~al.}(2000){Wilms}, {Allen}, \& {McCray}}]{WilmAbund}
{Wilms}, J., {Allen}, A., \& {McCray}, R. 2000, \apj, 542, 914, \dodoi{10.1086/317016}

\bibitem[{{Wood} {et~al.}(2017){Wood}, {Caputo}, {Charles}, {Di Mauro}, {Magill}, {Perkins}, \& {Fermi-LAT Collaboration}}]{fermipy}
{Wood}, M., {Caputo}, R., {Charles}, E., {et~al.} 2017, in International Cosmic Ray Conference, Vol. 301, 35th International Cosmic Ray Conference (ICRC2017), 824, \dodoi{10.22323/1.301.0824}

\bibitem[{{Wu} {et~al.}(2024){Wu}, {Li}, {Liang}, {Ge}, \& {Liu}}]{erosita2}
{Wu}, Q.-Z., {Li}, C.-M., {Liang}, X.-H., {Ge}, C., \& {Liu}, R.-Y. 2024, \apj, 969, 9, \dodoi{10.3847/1538-4357/ad43e1}

\bibitem[{{Yuan} {et~al.}(2017){Yuan}, {Lin}, {Fang}, \& {Bi}}]{bcratio}
{Yuan}, Q., {Lin}, S.-J., {Fang}, K., \& {Bi}, X.-J. 2017, \prd, 95, 083007, \dodoi{10.1103/PhysRevD.95.083007}

\bibitem[{{Zheng} \& {Wang}(2024)}]{pulsar-gating}
{Zheng}, D., \& {Wang}, Z. 2024, \apj, 968, 117, \dodoi{10.3847/1538-4357/ad496d}

\end{thebibliography}
\bibliographystyle{aasjournal}



\end{document}